\RequirePackage[T1]{fontenc}
\documentclass[showkeys,showpacs,superscriptaddress,nofootinbib,longbibliography,preprintnumbers,12pt]{article}

\usepackage{titling}

\usepackage{authblk}
\usepackage{mathtools}

\usepackage{graphicx}
\makeatletter
\let\oldbullet\bullet
\renewcommand{\bullet}{%
  \mathbin{\vcenter{\hbox{\scalebox{1.2}{$\m@th\oldbullet$}}}}% 
}
\makeatother

\usepackage{times}

\usepackage[utf8]{inputenc}
\usepackage{amsmath}
\usepackage{amssymb}
\usepackage{mathtools}
\numberwithin{equation}{section}
\usepackage{slashed}
\usepackage{braket}
\usepackage{bm}
\usepackage{mathrsfs}
\usepackage[svgnames]{xcolor}

\usepackage[colorlinks,citecolor=DarkGreen,linkcolor=FireBrick,linktocpage,pagebackref]{hyperref}
\usepackage{cite}

\usepackage{caption}

\usepackage{times}

\usepackage[utf8]{inputenc}

\usepackage{here}
\usepackage[normalem]{ulem}%取り消し線

\usepackage{url}

\usepackage{dsfont}

\usepackage[top=30truemm,bottom=30truemm,left=25truemm,right=25truemm]{geometry}

\usepackage{spectralsequences}
\usepackage{tikz}
\usetikzlibrary{intersections}
\usetikzlibrary{calc}

\usepackage{xcolor}
\newcommand{\abs}[1]{\left|#1\right|}
\newcommand{\tr}[1]{\mathrm{tr}#1}

\usepackage{tikz-cd}

\usepackage{slashed}

\usepackage{amsthm}
\usepackage{thmtools}
\usepackage{blindtext}
\usepackage{braket}

\declaretheoremstyle[
       shaded={bgcolor=\color{rgb}{0.9,0.9,0.9}}  % comment this line in/out
]{theorem}

\declaretheoremstyle[
       shaded={bgcolor=\color{rgb}{0.9,0.9,0.9}}% comment this line in/out
]{question}

\declaretheoremstyle[
       shaded={bgcolor=\color{rgb}{0.9,0.9,0.9}}  % comment this line in/out
]{remark}

\declaretheoremstyle[
       shaded={bgcolor=\color{rgb}{0.9,0.9,0.9}}  % comment this line in/out
]{proposition}

\declaretheoremstyle[
       shaded={bgcolor=\color{rgb}{0.9,0.9,0.9}}  % comment this line in/out
]{definition}

\declaretheoremstyle[
       shaded={bgcolor=\color{rgb}{0.9,0.9,0.9}}  % comment this line in/out
]{assumption}

\declaretheoremstyle[
       shaded={bgcolor=\color{rgb}{0.9,0.9,0.9}}  % comment this line in/out
]{conjecture}

\declaretheoremstyle[
       shaded={bgcolor=\color{rgb}{0.9,0.9,0.9}}  % comment this line in/out
]{corrorary}

\declaretheoremstyle[
       shaded={bgcolor=\color{rgb}{0.9,0.9,0.9}}  % comment this line in/out
]{axiom}

\declaretheoremstyle[
       shaded={bgcolor=\color{rgb}{0.9,0.9,0.9}}  % comment this line in/out
]{lemma}

\usepackage{mathrsfs}
\usepackage{amsmath,amssymb}

\usepackage{esvect}

\usepackage{adjustbox}

\usepackage[status=draft,inline,nomargin]{fixme}
\fxusetheme{color}
\FXRegisterAuthor{ki}{aki}{\color{magenta}KI}
\FXRegisterAuthor{shu}{shuenv}{\color{red}Shu}

\usetikzlibrary{decorations.pathmorphing,shapes}
\newcounter{sarrow}

\usepackage{xurl}

\usepackage{ulem}

\begin{document}

\title{Parameterized families of 2+1d $G$-cluster states}  

\author[1]{Shuhei Ohyama}

\affil[1]{University of Vienna, Faculty of Physics, Boltzmanngasse 5, A-1090, Vienna, Austria}

\author[2, 3]{Kansei Inamura}
\affil[2]{Mathematical Institute, University of Oxford, Andrew Wiles Building, Woodstock Road, Oxford, OX2 6GG, UK}
\affil[3]{Rudolf Peierls Centre for Theoretical Physics, University of Oxford, Parks Road, Oxford, OX1 3PU, UK}

\date{\today} 

\maketitle

\begin{abstract}
       We construct a $G$-cluster Hamiltonian in 2+1 dimensions and analyze its properties.
       This model exhibits a $G\times2\mathrm{Rep}(G)$ symmetry, where the $2\mathrm{Rep}(G)$ sector realizes a non-invertible symmetry obtained by condensing appropriate algebra objects in $\mathrm{Rep}(G)$.
       Using the symmetry interpolation method, we construct $S^1$- and $S^2$-parameterized families of short-range-entangled (SRE) states by interpolating an either invertible $0$-form or $1$-form symmetry contained in $G\times2\mathrm{Rep}(G)$.
       Applying an adiabatic evolution argument to this family, we analyze the pumped interface mode generated by this adiabatic process.
       We then explicitly construct the symmetry operator acting on the interface and show that the interface mode carries a nontrivial charge under this symmetry, thereby demonstrating the nontriviality of the parameterized family.
\end{abstract}

\setcounter{tocdepth}{3}
\tableofcontents

\section{Introduction and Summary}
\label{sec: Introduction and Summary}

\paragraph{Introduction}

% general background

The notion of topological phases of matter has played a central role in condensed matter physics in recent years.
In particular, symmetry-protected topological (SPT) phases are topological phases of short-range-entangled (SRE) states in the presence of symmetry; although it is one of the most basic classes of topological phases, they are known to exhibit rich physical properties, represented by the bulk-boundary correspondence~\cite{Gu_2009, Pollmann:2009mhk, Pollmann_2010, Chen:2010zpc, Chen_2011_complete, Chen:2011bcp, Chen:2011pg, Schuch_2011}.
Moreover, by considering topological phases realized by one-parameter families of SRE states, one can capture topological pumping phenomena such as the Thouless charge pumping.
More generally, as a generalization of the Thouless pump, one can consider topological phases of families of SRE states parameterized by a general space $X$.
Invariants associated with such families, and the resulting higher pumping phenomena---the so-called generalized Thouless pumps---have recently attracted growing attention~\cite{Thouless83,PhysRevB.82.115120,Kitaev2011SCGP,Kitaev2013SCGP,Kitaev2015IPAM,Kapustin:2020mkl,PhysRevResearch.2.042024,Tantivasadakarn:2021wdv,Shiozaki:2021weu,Hermele2021CMSA,Wen:2021gwc,Bachmann:2022bhx,Spodyneiko:2023vsw,Ohyama:2022cib,Inamura:2024jke,Jones:2025khc,Shiozaki:2025pyo,Li:2025wes}.

In a related but distinct context, there has recently been growing interest across physics in generalizing the notion of symmetry~\cite{Gaiotto:2014kfa}.
In particular, much attention has been focused on non-invertible symmetries, where the usual requirement that symmetry operations be invertible is relaxed.
By taking non-invertible symmetries into account, one can regard self-dualities of theories, such as the Kramers-Wannier self-duality \cite{KW1941}, as symmetries in a generalized sense, even though they cannot be captured by conventional symmetries.
See \cite{Cordova:2022ruw, McGreevy:2022oyu, Schafer-Nameki:2023jdn, Brennan:2023mmt, Bhardwaj:2023kri, Luo:2023ive, Shao:2023gho, Carqueville:2023jhb, Iqbal:2024pee, Costa:2024wks} for recent reviews of non-invertible symmetries.
In 2+1 dimensions, the main focus of this paper, finite non-invertible symmetries are described in terms of fusion 2-categories, defined in~\cite{Douglas:2018qfz} and classified in~\cite{Decoppet:2024htz}.
Recent studies have provided a variety of physical realizations of fusion 2-categorical symmetries~\cite{Kaidi:2021xfk,Bhardwaj:2022lsg,Bhardwaj:2022maz,Bartsch:2022mpm,Bartsch:2022ytj,Delcamp:2023kew,Choi:2024rjm,Inamura:2023qzl,Hsin:2024aqb,Bhardwaj:2024qiv,Bullimore:2024khm,Cordova:2024jlk,Cordova:2024mqg,Cao:2025qhg,Bhardwaj:2025piv,Eck:2025ldx,Vancraeynest-DeCuiper:2025wkh,Hsin:2025ria,Furukawa:2025flp,Inamura:2025cum,KNBalasubramanian:2025vum}.
Motivated by these developments, it is natural to investigate SPT phases protected by non-invertible symmetry and their associated pumping phenomena, and it is becoming clear that such SPT phases exhibit intriguing properties absent in conventional SPT phases~\cite{Thorngren:2019iar,Inamura:2021wuo,Inamura:2021szw,Garre-Rubio:2022uum,Fechisin:2023odt,Seifnashri:2024dsd,Choi:2024rjm,Jia:2024bng,Li:2024fhy,Inamura:2024jke,Pace:2024acq,Meng:2024nxx, Cao:2025qhg,Aksoy:2025rmg,Maeda:2025rxc,Lu:2025rwd,Furukawa:2025flp,ParayilMana:2025nxw,Inamura:2025cum,You:2025uxo,Lu:2025yru}.

% from the perspective of my area of expertise
Tensor networks provide an efficient and diagrammatic framework for describing quantum states in many-body systems; see~\cite{Cirac:2020obd} for a review.
Recent developments have also revealed that tensor networks offer a powerful tool for describing non-invertible symmetries~\cite{Molnar:2022nmh, Lootens:2021tet, Lootens:2022avn, Gorantla:2024ocs}.
In 1+1 dimensions, systematic analyses of SPT phases with non-invertible symmetries have been carried out using a class of tensor networks known as matrix product operators (MPOs)~\cite{Inamura:2021szw,Garre-Rubio:2022uum,Fechisin:2023odt,Jia:2024bng,Inamura:2024jke,Pace:2024acq,Meng:2024nxx, Cao:2025qhg,Lu:2025rwd,Inamura:2025cum,You:2025uxo}.
In 2+1 dimensions, similar studies of non-invertible symmetries are being developed using projected entangled pair operators (PEPOs)~\cite{Delcamp:2023kew,Choi:2024rjm,Inamura:2023qzl, Bullimore:2024khm, Bhardwaj:2024qiv, Cao:2025qhg, Bhardwaj:2025piv, Eck:2025ldx, Vancraeynest-DeCuiper:2025wkh, Inamura:2025cum}.

% comments and summaries on several highly relevant papers
As already mentioned, there has been growing interest in generalized Thouless pumps associated with families of SRE states with symmetry.
However, much of the previous work has focused on analyzing (higher) pumping phenomena for SRE states with invertible symmetries.
In 1+1 dimensions, the generalized Thouless pump phenomenon has been analyzed for families of SRE states with non-invertible symmetries.
In particular, in Ref.~\cite{Inamura:2024jke}, the authors analyzed the $G$-cluster state~\cite{Brell_2015,Fechisin:2023odt}---an SPT phase in 1+1 dimensions protected by $G \times \mathrm{Rep}(G)$ symmetry---and studied the associated pumping phenomena in detail.
The parameterized family is constructed using the so-called symmetry interpolation method.
This method produces families of states by interpolating, in the space of unitary operators, between the identity operator and an invertible symmetry.
In Ref.~\cite{Inamura:2024jke}, the $G$-symmetry sector of $G \times \mathrm{Rep}(G)$-symmetry is interpolated to obtain a parameterized family of $\mathrm{Rep}(G)$-symmetric SRE states.
By continuously modulating this parameter, one can generate a pumped interface mode.
The nontriviality of each family is then systematically characterized by computing the charge of the induced interface mode under the interface symmetry, and it is shown that each family carries a charge labeled by an element of $G$.
From general categorical arguments, one expects that $S^{1}$-families of SRE states with $\mathrm{Rep}(G)$ symmetry are classified by $G$ \cite{Inamura:2024jke, Li:2025wes}, so this lattice result is consistent with the categorical expectation and realizes all sectors predicted by the classification.

% things that have not been clarified in previous research
For the $G$-cluster state in 1+1 dimensions, a simple picture of the Thouless pump has been established.
It is therefore natural to expect that this picture can be generalized to the 2+1-dimensional case by considering the $G$-cluster state in 2+1-dimensions, which is an SPT phase protected by $G\times2\mathrm{Rep}(G)$ symmetry.
However, in 2+1 dimensions, there has not yet been a concrete study of the symmetry structure of the $G$-cluster state and its associated pumping phenomena.
Moreover, the $G\times2\mathrm{Rep}(G)$ symmetry contains, in addition to the $G$-symmetry sector, other invertible symmetries, namely invertible $0$-form and $1$-form symmetries in $2\mathrm{Rep}(G)$.
By interpolating these symmetries, it is expected that one can construct families of 2+1-dimensional $G$-symmetric SRE states and families of $2\mathrm{Rep}(G)$-symmetric SRE states.
However, the symmetry interpolation method developed so far has, in practice, only been applied to onsite $0$-form symmetries, and no analogous construction has been known for these invertible symmetries.

% the objectives and strategies of the research
In this paper, we construct the $G$-cluster state in 2+1 dimensions and, based on this model, build several parameterized families of SRE states.
To this end, we first give a tensor-network construction of the ground state and symmetry operators of the 2+1-dimensional $G$-cluster state explicitly.
Using these representations, we compute the local symmetry action on the ground state and extract the associated (one-dimensionally extended) action tensor, which we refer to as the line-like action tensor.
We then extend the symmetry interpolation method to construct parameterized families by interpolating various invertible symmetries contained in the $G\times2\mathrm{Rep}(G)$ symmetry.
These constructions respectively correspond to interpolating an onsite $0$-form symmetry, a non-onsite $0$-form symmetry, and an onsite $1$-form symmetry.
For the resulting $S^{1}$- and $S^{2}$-families, we apply an adiabatic evolution argument and analyze the interface mode that is pumped in this process.
To confirm the nontriviality of each family, using the line-like action tensor, we construct the symmetry action on the interface, namely the interface symmetry operators\footnote{
       This provides an example of a 2+1-dimensional generalization of the strip algebra, namely the strip 2-algebra~\cite{Cordova:2024iti}.
       In this paper, we do not delve into the strip 2-algebraic aspects of the interface symmetry.
       For a systematic analysis of strip 2-algebras, we refer the reader to Ref.~\cite{Inamura:clusterinterface}.
}, and show that the induced interface mode carries a nontrivial charge under this symmetry.

\paragraph{Outline and Summary}
The structure of the paper is as follows.

In Sec.~\ref{sec: Preliminaries}, we review the properties of the $G$-cluster state in 1+1 dimensions that are relevant to the present work.
At the level of individual tensors, many of those appearing in the 2+1-dimensional $G$-cluster state and its symmetry operators are directly inherited from the 1+1-dimensional $G$-cluster state, so we begin by introducing them in that simpler setting.
Moreover, the conceptual relation between interface symmetries and parameterized families, as well as the symmetry interpolation argument, is essentially the same as in the 1+1-dimensional $G$-cluster state.
For this reason, we begin with a brief overview of the 1+1-dimensional case as a prototype for the 2+1-dimensional analysis.

In Sec.~\ref{sec: 2d G-cluster state}, we construct the $G$-cluster state in 2+1 dimensions and analyze its basic properties.
In particular, the symmetry in 2+1 dimensions includes non-invertible symmetries described by $2\mathrm{Rep}(G)$, and we need to construct the corresponding symmetry operators in the tensor-network language.
From general categorical considerations, these non-invertible symmetries can be realized as condensation surfaces, which arise from the condensation of symmetric special Frobenius algebras in $\mathrm{Rep}(G)$~\cite{Gaiotto:2019xmp,Roumpedakis:2022aik}.
To this end, we first construct a symmetric special Frobenius algebra in $\mathrm{Rep}(G)$, then perform a formally analogous condensation procedure at the level of tensor networks to obtain the symmetry PEPO operators, and verify that they indeed define symmetries and, on states satisfying a flatness condition, obey the fusion rules of $2\mathrm{Rep}(G)$.
See Ref.~\cite{Delcamp:2023kew,Inamura:2023qzl} for similar constructions of condensation surface operators on the lattice.
Furthermore, by examining the local action of these symmetries on the ground state, we construct the one-dimensionally extended action tensor, which we refer to as the line-like action tensor.
To describe the line-like action tensor associated with non-invertible symmetries, we make essential use of matrix product state (MPS) representations of modules over a symmetric special Frobenius algebra, which we refer to as module MPSs.

In Sec.~\ref{sec: Parametrized family of G-symmetric states}, we construct parameterized families of $G$-symmetric SRE states by interpolating invertible symmetries contained in $2\mathrm{Rep}(G)$.
First, the invertible $0$-form symmetries in $2\mathrm{Rep}(G)$ are labeled by an element of the second group cohomology $\mathrm{H}^{2}(G;\mathrm{U}(1))$.
These symmetries are a priori realized as PEPOs with nontrivial bond dimension, but by using the invertibility of the underlying algebra, we can factor them into products of local unitaries.
Applying the symmetry interpolation method to these local unitaries, we construct, for each element of $\mathrm{H}^{2}(G;\mathrm{U}(1))$, an $S^{1}$-parameterized family of $G$-symmetric Hamiltonians.
Next, the invertible $1$-form symmetries in $2\mathrm{Rep}(G)$ are labeled by an element of $\mathrm{H}^{1}(G;\mathrm{U}(1))$.
Although these symmetries are onsite, they are $1$-form symmetries and thus the standard symmetry interpolation method cannot be applied directly.
By suitably combining loops of the $1$-form symmetry operators, however, we construct a two-parameter family of unitary operators, which in turn yields, for each element of $\mathrm{H}^{1}(G;\mathrm{U}(1))$, an $S^{2}$-parameterized family of $G$-symmetric Hamiltonians.
For these models we apply an adiabatic evolution argument and analyze the pumped interface modes.
We then show that the resulting interface modes carry nontrivial charge under the interface symmetry, thereby establishing the nontriviality of the families.
From a categorical point of view, it is expected that $S^{1}$- and $S^{2}$-families of $G$-symmetric Hamiltonians are classified by $\mathrm{H}^{2}(G;\mathrm{U}(1))$ and $\mathrm{H}^{1}(G;\mathrm{U}(1))$ \cite{Thorngren:1612.00846, Hsin:2020cgg,Hermele2021CMSA,Thorngren2021YITP,Inamura:2024jke}, respectively, and other $S^{n}$-families with $n\geq 3$ are expected to be trivial.\footnote{\label{fn: higher Berry phase G}
       More precisely, it is expected that there exist nontrivial $S^4$ families of $G$-symmetric SRE states, as measured by the higher Berry phase.
       However, since these $S^4$ families exist regardless of the $G$-symmetry, we do not address them in this paper.
       See \cite{Qi:2023ysw, Ohyama:2024ytt} for the higher Berry phase in 2+1d SRE states on the lattice.
}
In this sense, our construction is both consistent with, and provides a comprehensive realization of, the general classification.

In Sec.~\ref{sec: Parametrized family of 2Rep(G)-symmetric states}, we construct parameterized families of $2\mathrm{Rep}(G)$-symmetric SRE states by interpolating the $G$ symmetry.
Since the $G$ symmetry is an onsite $0$-form symmetry, the standard symmetry interpolation method can be applied directly.
In this way, for each element of $G$, we obtain an $S^{1}$-parameterized family of $2\mathrm{Rep}(G)$-symmetric Hamiltonians.
For these models, we apply an adiabatic evolution argument and analyze the pumped interface modes.
We then show that the pumped interface mode carries a nontrivial charge under the interface symmetry, thereby establishing the nontriviality of the families.
Although condensation surfaces have not been systematically exploited in earlier studies, we find that they play an essential role in the study of interface modes.
As an additional check of nontriviality from a different viewpoint, we examine the change of basis in the space of line-like action tensors and confirm that the families are indeed nontrivial.
From a categorical point of view, it is expected that $S^{1}$-families of $2\mathrm{Rep}(G)$-symmetric Hamiltonians are classified by $G$ \cite{Inamura:2024jke},
and other $S^{n}$-families with $n\geq 3$ are expected to be trivial.\footnote{
       More precisely, similar to the case of $G$-symmetric SRE states discussed in Footnote~\ref{fn: higher Berry phase G}, it is expected that there exist nontrivial $S^{4}$ families of $2\mathrm{Rep}(G)$-symmetric SRE states, whose nontriviality is measured by the higher Berry phase.
       However, since these $S^4$ families exist regardless of the $2\mathrm{Rep}(G)$-symmetry, we do not address them in this paper.
}
In this sense, our construction is both consistent with, and provides a comprehensive realization of, the general classification.

\section{Preliminaries}
\label{sec: Preliminaries}
As mentioned in Sec.~\ref{sec: Introduction and Summary}, in this paper we construct the $G$-cluster state in 2+1d and, based on it, construct parameterized families and verify their nontriviality.
To construct the parameterized families we employ the so-called symmetry interpolation method, and to verify nontriviality we use the interface symmetry.
These methods have already been used for $G$-cluster states in 1+1d \cite{Inamura:2024jke}.
Accordingly, in this section we outline the $G$-cluster state in 1+1d and review the interface symmetry and the symmetry interpolation method based on it.
Along the way, we also review the basic tensors and their properties that will be used for $G$-cluster states in 2+1d.

\subsection{$G$-cluster state in 1+1d}
\label{sec: G-cluster state in 1d}
The cluster state is an example of an SPT phase with $\mathbb{Z}_{2}\times\mathbb{Z}_{2}$-symmetry in 1+1d\cite{PhysRevLett.86.5188,Raussendorf:0301052}.
As a generalization of the cluster state, the $G$-cluster state in 1+1d for a finite group $G$ is introduced in \cite{Brell_2015,Fechisin:2023odt}.
The $G$-cluster state in 1+1d realizes an SPT phase with $G\times \mathrm{Rep}(G)$-symmetry.
In this section, we review the $G$-cluster state in 1+1d.

We consider a one-dimensional periodic lattice with an even number of sites $2N$. 
At each site $i$, we place the group ring $\mathbb{C}[G]$ of $G$, and define the total Hilbert space as their tensor product:
\begin{equation}
       \mathcal{H}_{1+1d}=\bigotimes_{i=1}^{2N}\mathbb{C}[G]_{i}.
\end{equation}
For each site $i$, we take the basis of $\mathbb{C}[G]_{i}$ as $\{\ket{g}_{i}\}_{g\in G}$.
Let $V$ be the set of odd sites and let $L$ be the set of even sites.
Then, the Hamiltonian of the $G$-cluster state is given by
\begin{equation}
       H_{1+1d}=-\sum_{i\in V} \frac{1}{\abs{G}}\sum_{g\in G}  \overleftarrow{X}_{g}^{(i-1)} \overleftarrow{X}_{g}^{(i)} \overrightarrow{X}_{g^{-1}}^{(i+1)}
       - \sum_{i\in L} \frac{1}{\abs{G}}\sum_{\rho\in \mathrm{Irr}(G)} \mathrm{dim}\rho\cdot\tr_{\rho}\left[Z_{\rho}^{(i-1)}Z_{\rho}^{(i)}Z_{\rho}^{\dagger(i+1)}\right],
\end{equation}
where $\mathrm{Irr}(G)$ is the set of irreducible representations of $G$ and the generalized Pauli $X$ operators $\overrightarrow{X}_{g},\overleftarrow{X}_{g}$ are defined as
\begin{eqnarray}
    \overrightarrow{X}_{g}\coloneq\sum_{h\in G} \ket{gh}\bra{h},\;\overleftarrow{X}_{g}\coloneq\sum_{h\in G} \ket{hg}\bra{h},
\end{eqnarray}
and the generalized Pauli $Z$ operator $Z_{\rho}$ is defined as
\begin{eqnarray}
       [Z_{\rho}]_{ab}:=\sum_{g\in G} \rho(g)_{ab}\ket{g}\bra{g},
\end{eqnarray}
and the superscript $(i)$ means that the operator acts on the site $i$.\footnote{In \cite{Fechisin:2023odt}, the right multiplication operator $\overleftarrow{X}_{g}$ is defined by $\sum_{h\in G} \ket{hg^{-1}}\bra{h}$.}
The commutation relations of these operators are given by
\begin{equation}\label{eq: generalized Pauli com}
       [Z_{\rho}]_{ab}\overrightarrow{X}_{g}=\overrightarrow{X}_{g}[\rho(g)\cdot Z_{\rho}]_{ab},\;\;[Z_{\rho}]_{ab}\overleftarrow{X}_{g}=\overleftarrow{X}_{g}[Z_{\rho}\cdot\rho(g)]_{ab},
\end{equation}
where $\cdot$ denotes the multiplication in the representation $\rho$.
The Hamiltonian $H_{1+1d}$ is gapped and has a unique ground state, which is called the $G$-cluster state in 1+1d.

The $G$-cluster state has $G\times \mathrm{Rep}(G)$-symmetry generated by
\begin{equation}
       U_{g}\coloneq\bigotimes_{i\in V}\overrightarrow{X}_{g}^{(i)},\;\;\mathcal{O}_{\rho}\coloneq\tr_{\rho}[\prod_{i\in L}Z_{\rho}^{(i)}],
\end{equation}
where $g\in G$ and $\rho\in \mathrm{Irr}(G)$.
One can show that the second operators satisfy the fusion rule of $\mathrm{Rep}(G)$:
\begin{equation}
       \mathcal{O}_{\rho}\mathcal{O}_{\sigma}=\sum_{\tau\in \mathrm{Irr}(G)}N_{\rho\sigma}^{\tau}\mathcal{O}_{\tau},
\end{equation}
where $N_{\rho\sigma}^{\tau}$ is the fusion coefficient in $\mathrm{Rep}(G)$.
We note that if $G$ is non-abelian, the symmetry is non-invertible.

The ground state and symmetry have simple MPS and MPO representations.
For the MPS representation, we introduce two tensors:
\begin{equation}\label{eq: G-cluster MPS}
       \adjincludegraphics[scale=1.25,trim={10pt 10pt 10pt 10pt},valign = c]{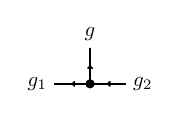}\coloneq\;\frac{1}{\sqrt{\abs{G}}}\;\delta_{g_{1},g}\delta_{g_{2},g},\;\;
       \adjincludegraphics[scale=1.25,trim={10pt 10pt 10pt 10pt},valign = c]{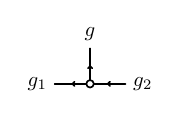}\coloneq\;\delta_{g_{1}^{-1}g_{2},g}.
\end{equation}
The first tensor is called the comultiplication tensor and the second one is called the multiplication tensor.
The ground state is realized by placing the multiplication tensor on each even site and the comultiplication tensor on each odd site and contracting the virtual legs.
Concretely, the MPS representation of the ground state is generated by
\begin{equation}
       A\coloneq \adjincludegraphics[scale=1.25,trim={10pt 10pt 10pt 10pt},valign = c]{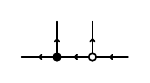}.
\end{equation}
For the MPO representation of the symmetry, we introduce the following MPO tensor for each irreducible representation $\rho\in \mathrm{Irr}(G)$:
\begin{equation}
       \adjincludegraphics[scale=1.25,trim={10pt 10pt 10pt 10pt},valign = c]{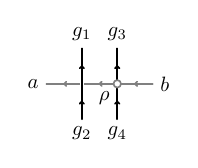} = \rho(g_{3})_{ab}\delta_{g_{1},g_{2}}\delta_{g_{3},g_{4}}.
\end{equation}
Then, the MPO representation of $\mathcal{O}_{\rho}$ is given by placing the above tensor on each even site and contracting the virtual legs.

For conventional symmetry, the SPT invariant is extracted from the projective representation of the symmetry action on the virtual legs of the MPS representation of the ground state\cite{Pollmann_2010, Chen:2010zpc, Schuch_2011}.
The symmetry action of $U_{g}$ on the virtual legs is given by
\begin{equation}\label{eq: pulling through G}
       \adjincludegraphics[scale=1.25,trim={10pt 10pt 10pt 10pt},valign = c]{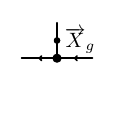}
       \;\;=\;\;
       \adjincludegraphics[scale=1.25,trim={10pt 10pt 10pt 10pt},valign = c]{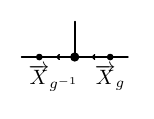}.
\end{equation}
As a generalization of this, we can also push  the MPO symmetry $\mathcal{O}_{\rho}$ through the multiplication tensor as follows:
\begin{equation}\label{eq: pulling through repG}
       \adjincludegraphics[scale=1.25,trim={10pt 10pt 10pt 10pt},valign = c]{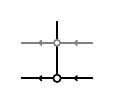}
       \;\;=\;\;
       \adjincludegraphics[scale=1.25,trim={10pt 10pt 10pt 10pt},valign = c]{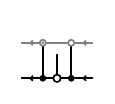},
\end{equation}
where the black dots on the baseline represent the comultiplication tensors and the three-leg gray tensors are defined by
\begin{equation}\label{eq: action tensor basic}
       \adjincludegraphics[scale=1.25,trim={10pt 10pt 10pt 10pt},valign = c]{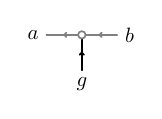}
       \;\;=\;\;\rho(g)_{ab},\;\;
       \adjincludegraphics[scale=1.25,trim={10pt 10pt 10pt 10pt},valign = c]{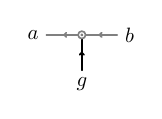}
       \;\;=\;\;\rho(g^{-1})_{ab}.
\end{equation}
Then, the action tensor is defined by inserting the complete set of the basis of the representation $\rho$ on the gray line in between the two gray tensors:
\begin{equation}
       \adjincludegraphics[scale=1.25,trim={10pt 10pt 10pt -5pt},valign = c]{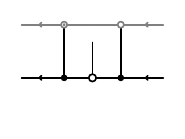}
       \;\;=\;\;\sum_{a=1}^{\dim \rho}\;\;
       \adjincludegraphics[scale=1.25,trim={10pt 10pt 10pt -5pt},valign = c]{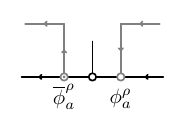},
\end{equation}
where the left and right action tensors $\{\overline{\phi}^{\rho}_{a}\}_{a=1}^{\dim\rho}$ and $\{\phi^{\rho}_{a}\}_{a=1}^{\dim\rho}$ are defined by
\begin{equation}
       \adjincludegraphics[scale=1.25,trim={10pt 10pt 10pt 10pt},valign = c]{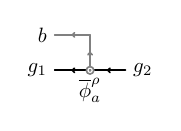}
       =\rho(g^{-1})_{ba} \delta_{g_{1},g_{2}},\;\;
       \adjincludegraphics[scale=1.25,trim={10pt 10pt 10pt 10pt},valign = c]{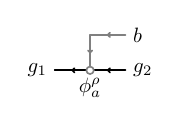}
       =\rho(g)_{ab}\delta_{g_{1},g_{2}}.
\end{equation}
These action tensors satisfy the orthogonality and completeness relations:
\begin{equation}
       \adjincludegraphics[scale=1.25,trim={10pt 10pt 10pt 10pt},valign = c]{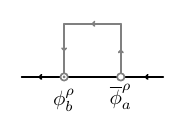}
       \;\;=\delta_{a,b}\;\;
       \adjincludegraphics[scale=1.25,trim={10pt 10pt 10pt 10pt},valign = c]{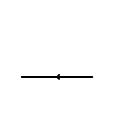}\;\;,\;\;
       \sum_{a = 1}^{\dim \rho} \adjincludegraphics[scale=1.25,trim={10pt 10pt 10pt 10pt},valign = c]{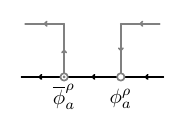}
       \;\;=\;\;
       \adjincludegraphics[scale=1.25,trim={10pt 10pt 10pt 10pt},valign = c]{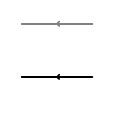}.
\end{equation}
We note that the equation obtained by replacing the multiplication tensor with the unit cell MPS tensor $A$ also holds.
By using the action tensors, one can construct the SPT invariant of the $G$-cluster state in 1+1d.

In Sec.~\ref{sec: 2d G-cluster state}, we generalize the above construction to 2+1d.
In that case, the symmetry structure is described by a fusion 2-category $G\times2\mathrm{Rep}(G)$, and the action tensors become line-like objects.

\subsection{Interface symmetry and parameterized families}
\label{sec: Interface symmetry and parameterized families}

In this section, we review the interface symmetry of the 1+1d $G$-cluster state and its relation to parameterized families.
The interface symmetry is a symmetry acting on the interface between two SPT phases.
In particular, we focus on the self-interface of the $G$-cluster state in 1+1d.

Let us consider the $G$-cluster state on a periodic lattice with defects $\psi_1$ and $\psi_2$
\begin{equation}
       \adjincludegraphics[scale=1.25,trim={10pt 10pt 10pt 10pt},valign = c]{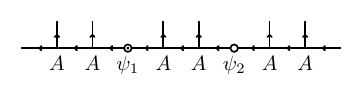},
\end{equation}
and consider the symmetry acting on the state as follows:
\begin{equation}
       \adjincludegraphics[scale=1.25,trim={10pt 10pt 10pt 10pt},valign = c]{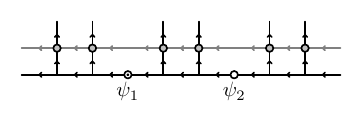}.
\end{equation}
Here, the gray dots represent the symmetry MPO tensors.
Then, by using the orthogonality relation of the action tensors, we find that the symmetry action only affects the defects $\psi_1$ and $\psi_2$ as follows:
\begin{equation}
       (\hat{\mathcal{O}}_{\rho})_{ab}\psi_{1}=\adjincludegraphics[scale=1.25,trim={10pt 10pt 10pt 10pt},valign = c]{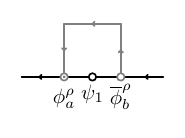},\;\;
       (\hat{\mathcal{O}}_{\rho})_{ba}\psi_{2}=\adjincludegraphics[scale=1.25,trim={10pt 10pt 10pt 10pt},valign = c]{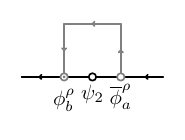},
\end{equation}
where the interface symmetry operators $(\hat{\mathcal{O}}_{\rho})_{ab}$ are defined by
\begin{equation}
       (\hat{\mathcal{O}}_{\rho})_{ab} \coloneq \adjincludegraphics[scale=1.25,trim={10pt 10pt 10pt 10pt},valign = c]{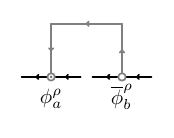}.
\end{equation}
In Ref.\cite{Inamura:2024jke}, the algebra generated by the operators $\{(\hat{\mathcal{O}}_{\rho})_{ab}\left.\right|\rho\in \mathrm{Irr}(G),a,b=1,...,\mathrm{dim}\rho\}$ is called the interface algebra for the self-interface.\footnote{
       These algebraic structures were originally used to study defects at the boundaries of topologically ordered states\cite{Kitaev:2011dxc,Lan:2013wia,Bridgeman:2018jdv, Bridgeman:2019axg, Bridgeman:2019wyu, Barter_2022}, and the same algebraic structures or their generalizations have been investigated in various contexts under various names, including module annular algebras\cite{Bridgeman:2022gdx}, boundary tube algebras\cite{Jia:2024rzr,Choi:2024tri,Choi:2024wfm,Jia:2024zdp,Jia:2025yph}, strip algebras\cite{Cordova:2024iti,Cordova:2024nux,Gagliano:2025gwr}, and interface algebras\cite{Inamura:2024jke,Lu:2025rwd}.
       Recent applications of strip algebras are also discussed in, e.g., \cite{Choi:2023xjw, Cordova:2024vsq, Copetti:2024onh, Copetti:2024dcz, Bhardwaj:2024igy, Heymann:2024vvf, AliAhmad:2025bnd, Benini:2025lav}.
}
In particular, the self-interface modes are classified by the irreducible representations of the interface algebra.

Let us explain the relation between the interface algebra and parameterized families of SPT phases.
It is expected that there is a one-to-one correspondence between the $S^{1}$-families of invertible states and the one-dimensional representations of the self-interface algebra.
Intuitively, this can be understood by considering a Hamiltonian whose parameter $\theta\in[0,2\pi]$ is spatially modulated, i.e., a textured Hamiltonian.
For example, for an $S^{1}$-family on an infinite lattice, we fix the parameter to $\theta=0$ on the left half-space, and on the right half gradually increase $\theta$ from $\theta=0$ to $\theta=2\pi$.
Then the system is homogeneous everywhere except near the center, where a self-interface mode can remain.
This self-interface mode is invertible because it vanishes if we gradually decrease $\theta$ from $2\pi$ to $0$. 
In particular, this self-interface mode is a one-dimensional representation of the interface algebra.
For a more detailed description of this correspondence, see \cite{Inamura:2024jke}.

In the following sections, we generalize the above construction to 2+1d $G$-cluster states.
In that case, the interface symmetry is not an algebra but a (multi)fusion category, and the interface modes are classified by the indecomposable module categories over the interface symmetry category.
We will not discuss the categorical details of the interface symmetry category in this paper, but we will focus on constructing parameterized families of 2+1d $G$-cluster states and analyzing their self-interface modes.
We refer the interested reader to \cite{Inamura:clusterinterface} for more details on the categorical description of the interface symmetry.

\subsection{Symmetry interpolation}
\label{sec: Symmetry interpolation}
There are several ways to construct parameterized families of SPT phases.
In this paper, we use the symmetry interpolation method.
Let us briefly explain the method.
Fist, suppose we have a Hamiltonian $\hat{H}$ with local unitary symmetry $\hat{U}$, i.e., $[\hat{H},\hat{U}]=0$ and $\hat{U}=\prod_{i} \hat{u}_{i}$ for some unitary operator $\hat{u}_{i}$ acting on sites around $i$.
Since $\hat{u}_{i}$ is unitary, we can find a continuous path from the identity operator to $\hat{u}_{i}$, i.e., a continuous family of unitary operators $\hat{u}_{i}(\theta)$ such that $\hat{u}_{i}(0)=\mathrm{id}$ and $\hat{u}_{i}(2\pi)=\hat{u}_{i}$.
Then, we define a continuous family of Hamiltonians by
\begin{equation}
       \hat{H}(\theta)=\hat{U}(\theta)\hat{H}\hat{U}(\theta)^{\dagger},\;\;\hat{U}(\theta)=\prod_{i}\hat{u}_{i}(\theta).
\end{equation}
We note that $\hat{U}(\theta)$ is not $2\pi$-periodic, but $\hat{H}(\theta)$ is $2\pi$-periodic, due to the relation $[\hat{H},\hat{U}]=0$.

For example, in the case of the 1+1d $G$-cluster state, we can take $\hat{U}$ as one of the $G$-symmetry operator $U_{g}$.
As a symmetry interpolation, we can take
\begin{equation}\label{eq: S1 interpolation}
       U_{g}(\theta)=\bigotimes_{i\in V}\overrightarrow{X}_{g}^{(i)}(\theta),\;\;\overrightarrow{X}_{g}^{(i)}(\theta)\coloneq\exp{[\frac{\theta}{2\pi}\log\overrightarrow{X}_{g}^{(i)}]}.
\end{equation}
Here we can take the principal branch of the logarithm as 
\begin{equation}\label{eq: log X}
       \log\overrightarrow{X}_{g}^{(i)}=\sum_{s=0}^{\abs{G}-1}\frac{2\pi i s}{\abs{G}}P_{s}[\overrightarrow{X}_{g}^{(i)}],
\end{equation}
where $P_{s}[\overrightarrow{X}_{g}^{(i)}]$ is the projection operator to the eigenspace of $\overrightarrow{X}_{g}^{(i)}$ with eigenvalue $e^{2\pi i s/\abs{G}}$, i.e.,
\begin{equation}
       P_{s}[X]\coloneq\frac{1}{\abs{G}}\sum_{k=0}^{\abs{G}-1}e^{\frac{-2\pi i s k}{\abs{G}}}X^{k},
\end{equation}
for a unitary operator $X$ satisfying $X^{\abs{G}}=1$.
Indeed, one can check that $\exp{(\log\overrightarrow{X}_{g})}=\overrightarrow{X}_{g}$.
Then, the $S^{1}$-family of the 1+1d $G$-cluster state is given by
\begin{equation}
       H_{1+1d}^{g}(\theta)\coloneq U_{g}(\theta)H_{1+1d}U_{g}(\theta)^{\dagger},
\end{equation}
for each $g\in G$.
This interpolation preserves the $\mathrm{Rep}(G)$-symmetry since $U_{g}(\theta)$ and $\mathcal{O}_{\rho}$ commute for all $\theta$.
Therefore, the $H_{1+1d}^{g}(\theta)$ is a $\mathrm{Rep}(G)$-symmetric $S^{1}$-family. 

       As explained in Sec.~\ref{sec: Interface symmetry and parameterized families}, one can construct a self-interface mode from each $S^{1}$-family $H_{1+1d}^{g}(\theta)$ by considering a textured Hamiltonian.
       In Ref.~\cite{Inamura:2024jke}, it is shown that the associated self-interface algebra has one-dimensional representations labeled by $g \in G$, and the self-interface mode constructed from this $S^{1}$-family corresponds to the one-dimensional representation labeled by $g$.
       This result is consistent with the categorical expectation that $S^{1}$-families of $2\mathrm{Rep}(G)$-symmetric SRE states are classified by $G$.

\section{2+1d $G$-cluster state}
\label{sec: 2d G-cluster state}
In this section, we construct the $G$-cluster state in 2+1d and investigate its properties.
       While in 2+1d the Hamiltonian and the representation of the ground state are given by a natural extension of the 1+1d case, the symmetry structure becomes more complicated because the $\mathrm{Rep}(G)$ symmetry is enhanced to a $2\mathrm{Rep}(G)$ symmetry.
       After representing these symmetries within the tensor-network formalism, we determine the corresponding action tensors.

\subsection{Hamiltonian}
\label{sec: Model}
\begin{figure}[t]
    \centering
    \includegraphics[scale=1]{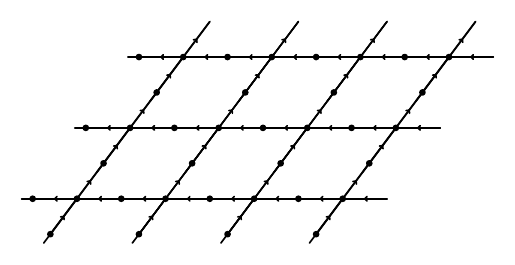}
    \caption{
       A two-dimensional oriented square lattice with periodic boundary conditions.
       The arrows on the links represent the orientation of the lattice.
       We place the group ring $\mathbb{C}[G]$ on each vertex and link of the lattice, denoted by the black dots.
    }
    \label{fig: lattice_ori}
\end{figure}

We consider a two-dimensional oriented square lattice with periodic boundary conditions.
We place the group ring $\mathbb{C}[G]$ of a finite group $G$ on each vertex and link of the lattice, and define the total Hilbert space as their tensor product:
\begin{equation}
       \mathcal{H} \coloneq \left(\bigotimes_{i\in V}\mathbb{C}[G]_{i}\right) \otimes \left(\bigotimes_{\ell\in L}\mathbb{C}[G]_{\ell}\right),
\end{equation}
where $V$ is the set of vertices and $L$ is the set of links.
See Fig.~\ref{fig: lattice_ori}.

Let us define the Hamiltonian of the 2+1d $G$-cluster model.
First, we introduce the following operators:
\begin{eqnarray}
       A_{i,g} &:=& 
           \vcenter{
           \hbox{
               \adjincludegraphics[scale=1.2,trim={10pt 10pt 10pt 10pt},valign = c]{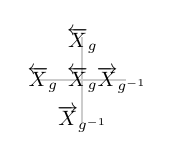}
           }
           },\;
       B_{\ell,\rho} := \mathrm{dim}\rho\cdot\mathrm{tr}_{\rho}[
           \vcenter{
           \hbox{
               \adjincludegraphics[scale=1.2,trim={10pt 10pt 10pt 10pt},valign = c]{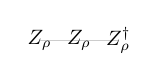}
           }
           }],
\end{eqnarray}
for each vertex $i\in V$ and link $\ell\in L$.
Then, the Hamiltonian of the 2+1d $G$-cluster model is given by
\begin{equation}\label{eq: 2d G-cluster Hamiltonian}
       H\coloneq -\sum_{i\in V}h_{i} - \sum_{\ell\in L}h_{\ell},
\end{equation}
where the local terms are defined by
\begin{equation}
       h_{i}\coloneq \frac{1}{\abs{G}}\sum_{g\in G} A_{i,g},\;\;h_{\ell}\coloneq \frac{1}{\abs{G}}\sum_{\rho\in \mathrm{Irr}(G)} B_{\ell,\rho}.
\end{equation}

We note that the local terms $h_{i}$ and $h_{\ell}$ are projection operators.
In fact, they satisfy
\begin{align}
       \left(h_{i}\right)^2 
       &= \frac{1}{\abs{G}^{2}}\sum_{g_1,g_2 \in G} A_{i, g_1}A_{i, g_2}
       = \frac{1}{\abs{G}^{2}}\sum_{g_1,g_2 \in G} A_{i, g_1 g_2}
       = h_{i}, 
\end{align}
and
\begin{align}\label{eq: projective h_l}
       \begin{aligned}
       \left(h_{\ell}\right)^2 
       &=\frac{1}{\abs{G}^{2}}\sum_{\rho_1, \rho_2 \in \mathrm{Irr}(G)} B_{\ell, \rho_1}B_{\ell, \rho_2}
       =\frac{1}{\abs{G}^{2}}\sum_{\rho_1, \rho_2, \rho_3 \in \mathrm{Irr}(G)} \frac{\mathrm{dim}\rho_1\mathrm{dim}\rho_2}{\mathrm{dim}\rho_3} N_{\rho_1,\rho_2}^{\rho_3}B_{\ell, \rho_3}\\
       &=\frac{1}{\abs{G}^{2}}\sum_{\rho_3 \in \mathrm{Irr}(G)} \abs{G}B_{\ell, \rho_3}
       = h_{\ell}.
       \end{aligned}
\end{align}
Here, in the third equality of Eq.~\eqref{eq: projective h_l}, we used the equation
\begin{equation}
       \sum_{\rho_1,\rho_2}\mathrm{dim}\rho_1\mathrm{dim}\rho_2 N_{\rho_1,\rho_2}^{\rho_3} 
       = \sum_{\rho_1,\rho_2}\mathrm{dim}\rho_1\mathrm{dim}\rho_2 N_{\overline{\rho_1},\rho_3}^{\rho_2} 
       = \sum_{\rho_1}\mathrm{dim}\rho_1\mathrm{dim}\overline{\rho_1}\mathrm{dim}\rho_3
       = \abs{G}\mathrm{dim}\rho_3.
\end{equation}
In addition, one can check that the projection operators commute with each other.
Therefore, the ground state of $H$ is given by the simultaneous eigenstate with eigenvalue $1$ of all the projection operators.

To obtain the ground state explicitly, we diagonalize the projection operators by using the unitary operator
\begin{equation}
       U_{\rm diag}\coloneq\prod_{i\in V}C\overrightarrow{X}_{i,\ell_{r}}C\overrightarrow{X}_{i,\ell_{b}}C\overleftarrow{X}_{i,\ell_{t}}C\overleftarrow{X}_{i,\ell_{l}},
\end{equation}
where $C\overrightarrow{X}_{i,j}$ and $C\overleftarrow{X}_{i,j}$ are the controlled multiplication operators defined by
\begin{equation}
       C\overrightarrow{X}_{i,j}\ket{g_{i},g_{j}} = \ket{g_{i},g_{i}g_{j}},\;\;
       C\overleftarrow{X}_{i,j}\ket{g_{i},g_{j}} = \ket{g_{i},g_{j}g_{i}},
\end{equation}
and $\ell_{r},\ell_{b},\ell_{t},\ell_{l}$ are the right, bottom, top, and left links adjacent to the vertex $i$, respectively.
Then, the local terms are transformed as
\begin{equation}\label{eq: trivialized Ham}
       U_{\rm diag}h_{i}U_{\rm diag}^{\dagger} = \frac{1}{\abs{G}}\sum_{g\in G}\overleftarrow{X}_{g}^{(i)}\;\;\text{and}\;\;
       U_{\rm diag}h_{\ell}U_{\rm diag}^{\dagger} = \frac{1}{\abs{G}}\sum_{\rho\in \mathrm{Irr}(G)}\mathrm{dim}\rho \cdot \tr[Z_{\rho}^{(\ell)}].
\end{equation}
The unique eigenstates of these operators with eigenvalue $1$ are given by
\begin{equation}
       \frac{1}{\sqrt{\abs{G}}}\sum_{g}\ket{g}_{i}\;\;\text{and}\;\;\ket{e}_{\ell},
\end{equation}
respectively, where $e\in G$ is the identity element. 
In fact, we can verify that the second operator in Eq.~\eqref{eq: trivialized Ham} is the projection onto $\ket{e}_{\ell}$:
\begin{equation}
       \frac{1}{\abs{G}}\sum_{\rho\in \mathrm{Irr}(G)}\mathrm{dim}\rho \cdot \tr[Z_{\rho}^{(\ell)}]\ket{g}_{\ell}
       =\frac{1}{\abs{G}}\tr[\rho_{\rm reg}(g)]\ket{g}_{\ell} 
       = \delta_{g,e}\ket{g}.
\end{equation}
Therefore, the ground state of $H$ is unique and given by
\begin{equation}
       \ket{\mathrm{G.S.}} = U_{\rm diag}\left(\bigotimes_{i\in V}\frac{1}{\sqrt{\abs{G}}}\sum_{g}\ket{g}_{i} \otimes \bigotimes_{\ell\in L}\ket{e}_{\ell}\right).
\end{equation}
In particular, $\ket{\mathrm{G.S.}}$ is the simultaneous eigenstate with eigenvalue $1$ of all the stabilizers $h_{i}$ and $h_{\ell}$.
We refer to this ground state as the 2+1d $G$-cluster state.

The Hamiltonian \eqref{eq: 2d G-cluster Hamiltonian} has $0$-form $G$-symmetry and $1$-form $\mathrm{Rep}(G)$-symmetry generated by
\begin{equation}\label{eq: original symmetries G}
    U_{g}:=\prod_{i\in V} \overrightarrow{X}_{g}^{(i)},\;\;\; U_{\rho}(\gamma):= \mathrm{tr}_{\rho}[\prod_{\ell\in\gamma\cap L} [Z_{\rho}^{(\ell)}]^{\mathrm{sgn}(\ell)}],
\end{equation}
for any oriented closed loop $\gamma$, where $\mathrm{sgn}(\ell)$ denotes a function that returns $1$ when the orientation assigned to $l$ is the same as the direction of $\gamma$, and $-1$ when it is opposite.
We note that the order of the product in $U_{\rho}$ is taken along the orientation of $\gamma$.
For example, the smallest loop enclosing a plaquette $p$ with the anti-clockwise orientation is given by
\begin{equation}
       \tr{[Z_{\rho}^{\dagger(\ell_{b})}Z_{\rho}^{\dagger(\ell_{l})}Z_{\rho}^{(\ell_{t})}Z_{\rho}^{(\ell_{r})}]},
\end{equation}
where $\ell_{t},\ell_{b},\ell_{l},\ell_{r}$ are the top, bottom, left, and right links surrounding the plaquette $p$, respectively.
The commutativity of the Hamiltonian and the symmetry operators immediately follows from the commutation relations \eqref{eq: generalized Pauli com} of the generalized Pauli operators.
In addition to this, there exists a $0$-form symmetry generated by the condensation of $\mathrm{Rep}(G)$. 
As a result, the full symmetry becomes $G\times2\mathrm{Rep}(G)$. 
The explicit realization of the condensation surface will be described in the next section.

Let us briefly comment on the topological nature of the $1$-form symmetry.
As an operator acting on the Hilbert space, $U_{\rho}(\gamma)$ depends on the shape of the loop $\gamma$.
From this perspective, the $1$-form symmetry is not topological.
In other words, the topological nature of the $1$-form symmetry in the time direction is guaranteed by the commutativity of the Hamiltonian with the symmetry operators, but there is no such guarantee in the spatial directions.
To ensure the spatial topological nature of the $1$-form symmetry, we impose the following flatness condition on each plaquette $p$:
\begin{align}
       \;\nonumber\\
       \pi_{\rm flat}^{(p)}\Ket{\adjincludegraphics[scale=1.2,trim={10pt 20pt 10pt 20pt},valign = c]{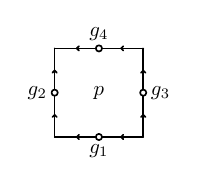}}
       =
       \delta_{g_{2}g_{1},g_{4}g_{3}}\Ket{\adjincludegraphics[scale=1.2,trim={10pt 20pt 10pt 20pt},valign = c]{tikz/out/flatness.pdf}}.\\
       \;\nonumber
\end{align}
On the flat subspace, 
\begin{equation}
       \mathcal{H}_{\rm flat}\coloneq \left(\bigotimes_{p}\pi_{\rm flat}^{(p)}\right) \mathcal{H},
\end{equation}
the $1$-form symmetry operator $U_{\rho}(\gamma)$ depends only on the homotopy class of the loop $\gamma$.
We note that the flatness condition is compatible with the Hamiltonian, i.e., $[\pi_{\rm flat}^{(p)},H]=0$ for all plaquettes $p$.
In the following, we always consider the model restricted to the flat subspace $\mathcal{H}_{\rm flat}$.
For a detailed discussion of topological and non-topological $1$-form symmetries, see Refs.~\cite{Seiberg:2019vrp,Qi:2020jrf,Choi:2024rjm}.

\subsection{Symmetries}
\label{sec: Symmetries}
From a categorical point of view, if there are $1$-form symmetry operators described by a braided fusion category $\mathcal{C}$, then one can obtain $0$-form symmetry operators realized as the condensation of algebra objects $A\in\mathcal{C}$.
The resulting codimension-one defects are referred to as condensation surfaces, and different (Morita classes of) algebra objects in $\mathcal{C}$ generally give rise to different condensation surfaces.
By adding condensation surfaces, the full symmetry category becomes a fusion 2-category called the suspension of $\mathcal{C}$~\cite{Gaiotto:2019xmp,Roumpedakis:2022aik}.
In our case, the $1$-form symmetry is described by $\mathrm{Rep}(G)$ and its suspension is known to be $2\mathrm{Rep}(G)$, the 2-category of 2-representations of $G$~\cite{Decoppet2023Morita}.
In this section, we will explicitly construct the condensation surfaces in our lattice model.

\subsubsection{Frobenius algebra object}
\begin{figure}[t]
       \begin{minipage}{0.5\textwidth}
           \centering
           \begin{equation*}
              \adjincludegraphics[scale=1.25,trim={10pt 10pt 10pt 10pt},valign = c]{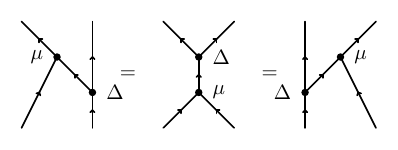}
           \end{equation*}
           \caption*{(a)}    
       \end{minipage}
       \hfill 
       \begin{minipage}{0.2\textwidth}
           \centering
           \begin{equation*}
              \adjincludegraphics[scale=1.25,trim={10pt 10pt 10pt 10pt},valign = c]{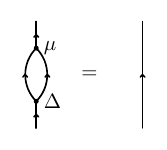}
           \end{equation*}
           \caption*{(b-1)}
       \end{minipage}
       \hfill 
       \begin{minipage}{0.2\textwidth}
           \centering
           \begin{equation*}
              \adjincludegraphics[scale=1.25,trim={10pt 10pt 10pt 10pt},valign = c]{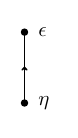}
              =\mathrm{dim}(A)
           \end{equation*}
           \caption*{(b-2)}
       \end{minipage}
       \caption{
       (a) Graphical representation of the Frobenius condition.
       (b) Graphical representation of the speciality condition.
       }
       \label{fig: Frobenius algebra}
\end{figure}

Let us recall the definition of a Frobenius algebra object in a fusion category $\mathcal{C}$.
An object $A$ in a fusion category $\mathcal{C}$ is called a Frobenius algebra object if it is equipped with the following morphisms:
\begin{itemize}
\item multiplication: $\mu: A\otimes A \to A$
\item unit: $\eta: \mathbf{1} \to A$
\item comultiplication: $\Delta: A \to A\otimes A$
\item counit: $\epsilon: A \to \mathbf{1}$
\end{itemize}
satisfying the following conditions:
\begin{itemize}
\item associativity: $\mu\circ (\mu\otimes \mathrm{id}_{A}) = \mu\circ (\mathrm{id}_{A}\otimes \mu)$
\item unitality: $\mu\circ (\eta\otimes \mathrm{id}_{A}) = \mathrm{id}_{A} = \mu\circ (\mathrm{id}_{A}\otimes \eta)$
\item coassociativity: $(\Delta\otimes \mathrm{id}_{A})\circ \Delta = (\mathrm{id}_{A}\otimes \Delta)\circ \Delta$
\item counitality: $(\epsilon\otimes \mathrm{id}_{A})\circ \Delta = \mathrm{id}_{A} = (\mathrm{id}_{A}\otimes \epsilon)\circ \Delta$
\item Frobenius condition: $(\mu\otimes \mathrm{id}_{A})\circ (\mathrm{id}_{A}\otimes \Delta) = \Delta\circ \mu = (\mathrm{id}_{A}\otimes \mu)\circ (\Delta\otimes \mathrm{id}_{A})$
\end{itemize}
Here, $\mathbf{1}$ is the tensor unit of $\mathcal{C}$ and $\mathrm{id}_{A}$ is the identity morphism of $A$.
In addition to this, if $A$ satisfies the following conditions, it is called a special Frobenius algebra object:
\begin{equation}
       \mu\circ \Delta = \mathrm{id}_{A},\;\;\epsilon\circ \eta = \mathrm{dim}(A).
\end{equation}
In addition to this, if $A$ satisfies the following condition, it is called a symmetric special Frobenius algebra object:
\begin{equation}
       (\epsilon\otimes\mathrm{id}_{A^*}) \circ (\mu\otimes\mathrm{id}_{A^*}) \circ (\mathrm{id}_{A}\otimes \mathrm{coev}^{L}) = (\mathrm{id}_{A^*}\otimes\epsilon) \circ (\mathrm{id}_{A^*}\otimes\mu) \circ (\mathrm{coev}^{R}\otimes \mathrm{id}_{A}),
\end{equation}
where $A^*$ denotes the dual of $A$, and $\mathrm{coev}^{L}:\mathbf{1}\to A\otimes A^{*}$ and $\mathrm{coev}^{R}:\mathbf{1}\to A^{*}\otimes A$ are the left and right coevaluation maps, respectively.
See Fig.~\ref{fig: Frobenius algebra} for a graphical representation of the Frobenius condition and the speciality condition.
In the following, we always consider symmetric special Frobenius algebra objects.

\subsubsection{Frobenius algebra objects in $\mathrm{Rep}(G)$}
It is known that (the Morita class of) a symmetric special Frobenius algebra in $\mathrm{Rep}(G)$ is specified by (the equivalence class of) a pair $(H, \omega)$, where $H$ is a subgroup of $G$ and $\omega \in \mathrm{Z}^2(H; \mathrm{U}(1))$ is a 2-cocyle on $H$.\footnote{The equivalence relation is given by the conjugation action of $G$ \cite{Ostrik2003}, that is, the pairs $(H, \omega)$ and $(H^{\prime}, \omega^{\prime})$ are equivalent if and only if there exists an element $g \in G$ such that $H^{\prime} = \mathrm{Ad}_g(H)$ and $[\omega^{\prime}] = [\mathrm{Ad}_g^*(\omega)] \in \mathrm{H}^2(H^{\prime}; \mathrm{U}(1))$.}
To describe the algebra object corresponding to $(H, \omega)$, we  fix a projective unitary representation $(\rho_{\omega},V_{\omega})$ corresponding to $\omega$, where $V_{\omega}$ is a vector space and $\rho_{\omega}:H\to \mathrm{GL}(V_{\omega})$ is a representation map.
Then, the algebra object $A_{H,\omega}^{G}$ is given as an induced representation $\mathrm{Ind}_{H}^{G}(\mathrm{End}(V_{\omega}))$ associated with the represetation $(\mathrm{Ad}\rho_{\omega},\mathrm{End}(V_{\omega}))$ of $H$.
Concretely, $A_{H,\omega}^{G}$ is the space of
$H$-equivariant $\mathrm{GL}(V_{\omega})$-valued function on $G$\cite{Kapustin:2016jqm}:
\begin{equation}
       A_{H,\omega}^{G} := \{f:G\to \mathrm{GL}(V_{\omega}) \left.\right| f(h^{-1}g) = \rho_{\omega}(h) f(g) \rho_{\omega}(h)^{\dagger}\},
\end{equation}
with a representation map $\rho_{H,\omega}^{G}:G\to \mathrm{GL}(A_{H,\omega}^{G})$ given by
\begin{equation}
       \rho_{H,\omega}^{G}(g)f(g') = f(g'g).
\end{equation}
The multiplication map
\begin{equation}
    \mu: A_{H,\omega}^{G}\otimes A_{H,\omega}^{G} \to A_{H,\omega}^{G}
\end{equation}
is given by the point-wise multiplication
\begin{equation}
    (f\cdot f')(g) := f(g)f'(g).
\end{equation}
This multiplication is compatible with the $H$-equivariance.
The counit of $A^G_{H, \omega}$ is given by
\begin{equation}\label{eq: counit}
       \epsilon: f \mapsto \frac{\mathrm{dim}V_{\omega}}{\abs{H}}\sum_{g\in G} \mathrm{tr}_{V_{\omega}}[f(g)].
\end{equation}
Although, the consistent comultiplication is uniquely determined by the above information, we write it down explicitly:
\begin{equation}
       \Delta: A_{H,\omega}^{G} \to A_{H,\omega}^{G}\otimes A_{H,\omega}^{G},\;\; f \mapsto \frac{1}{\mathrm{dim}V_{\omega}}\sum_{i} f\cdot f_{i}\otimes f_{i}^{\dagger},
\end{equation}
where $f_{i}$ is an orthonormal basis of $A_{H,\omega}^{G}$ and ${\dagger}$ implies taking the conjugate transpose in $\mathrm{GL}(A_{H,\omega}^{G})$.
In particular, the multiplication and comultiplication maps are $G$-equivariant, that is,
\begin{equation}\label{eq: g equivariance}
       (\rho_{H,\omega}^{G}(g)\otimes \rho_{H,\omega}^{G}(g))\circ \Delta = \Delta\circ \rho_{H,\omega}^{G}(g),\;\;
       \rho_{H,\omega}^{G}(g)\circ \mu = \mu\circ (\rho_{H,\omega}^{G}(g)\otimes \rho_{H,\omega}^{G}(g)),
\end{equation}
for all $g\in G$.

\subsubsection{Condensation surface}
From a categorical point of view, the condensation of a symmetric special Frobenius algebra object $A$ in a braided fusion category $\mathcal{C}$ gives rise to a $0$-form symmetry in addition to the $1$-form symmetry described by $\mathcal{C}$.
In this paper, we will not discuss the general theory of condensation mathematically, but we will explicitly construct the condensation surfaces in our lattice model.

\begin{figure}[t]
    \centering
    \begin{equation*}
       \mathcal{O}[A] \coloneq \adjincludegraphics[scale=1.5,trim={10pt 10pt 10pt 10pt},valign = c]{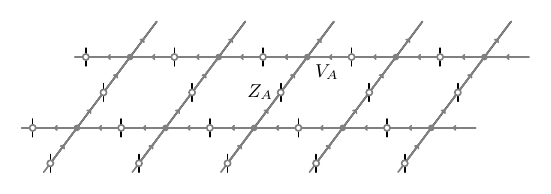}
    \end{equation*}
    \caption{The PEPO representation of the condensation surface. 
    Each white dot represents a generalized Pauli operator $Z_{A}$ associated with the representation $A$ and each black dot represents a 4-valent tensor $V_{A}$ consisting of the multiplication and comultiplication maps.
    }
    \label{fig: condensationPEPO}
\end{figure}

Let $(A=\bigoplus_{i}\rho_{i},\mu,\Delta)$ be an algebra object of $\mathrm{Rep}(G)$, where $\rho_{i}$ are irreducible representations of $G$ and $\mu,\Delta$ are the multiplication and comultiplication maps, respectively.
Since $A$ is a representation of $G$, we can define the generalized Pauli $Z$ operator $Z_{A}$ associated with $A$ by
\begin{equation}
       [Z_{A}]_{ab}\coloneq\sum_{g\in G}\left(\bigoplus_{i}\rho_{i}(g)\right)_{ab}\ket{g}\bra{g}.
\end{equation}
Since $A$ is an algebra object, we can define the operator $\mathcal{O}[A]$ on the lattice as in Fig.~\ref{fig: condensationPEPO}.
We refer to this operator as the condensation surface associated with the algebra object $A$.
This operator obviously commutes with the Hamiltonian \eqref{eq: 2d G-cluster Hamiltonian} due to the $G$-equivariance of the multiplication and comultiplication maps \eqref{eq: g equivariance}.

It is technical but worth mentioning the dependence of the condensation surface on the choice of the representatives $H,\omega$ and $(\rho_{\omega},V_{\omega})$.
From a categorical point of view, a condensation surface for an algebra object $A$ is determined by the Morita class of the algebra object, and does not depend on the choice of the representatives.
In our PEPO realization, the condensation surface can depend on the algebra itself as an operator acting on the total Hilbert space $\mathcal{H}$, but as an operator acting on the flat subspace $\mathcal{H}_{\rm flat}$, it only depends on the Morita class of $A$.
This is because, on the flat subspace, the PEPO labeled by an algebra object $A$ acts as a 2d state sum TQFT partition function associated with $A$ \cite{Kapustin:2016jqm}.
Since the state sum construction provides the same TQFT for Morita equivalent algebras, the PEPOs generated from two Morita equivalent algebras act as the same operator on flat states.
We always consider the flat subspace, and thus we do not need to care about the choice of a representative of the Morita class.
We also note that the flatness condition naturally arises when we construct the $G$-cluster state by gauging\cite{Inamura:clusterinterface}.

\subsubsection{Fusion rules} 
As we have seen, the condensation surfaces are labeled by the Morita classes of symmetric special Frobenius algebra objects in $\mathrm{Rep}(G)$.
Now, we compute the fusion rules of the condensation surfaces and show that they agree with those of $2\mathrm{Rep}(G)$.

First, the tensor product of $G$-equivariant algebras is decomposed into a direct sum of the following induced representations, known as the Mackey decomposition\cite[Theorem 2]{Mackey1951}
\footnote{The authors thank Alex Turzillo for discussions on the decomposition.}:
\begin{equation}\label{eq: Mackey decomposition}
       \mathrm{Ind}_{H}^{G}(\mathrm{End}(V_{\omega}))\otimes \mathrm{Ind}_{K}^{G}(\mathrm{End}(V_{\rho})) \cong \bigoplus_{x\in D} \mathrm{Ind}_{H\cap xKx^{-1}}^{G}(\mathrm{End}(V_{\omega}\otimes V_{\rho^{x}})),
\end{equation}
where $D$ is a set of representatives of the double cosets $H\backslash G/K$ and $\rho^{x}$ is the representation of $K$ obtained by conjugating $\rho$ by $x$, that is, $\rho^x(k) = \rho(x^{-1}kx)$ for $k\in K$.
Accordingly, we can take the projection $p_i$ onto the $i$-th induced representation in the decomposition, and the inclusion $\iota_i$ such that 
\begin{equation}
       \sum_{i} \iota_{i} \circ p_i = \mathrm{id}_{A_{1}\otimes A_{2}},\;\; p_i\circ \iota_{j} = \delta_{ij}\mathrm{id}_{A_{i}}.
\end{equation}
Here, $A_1$ and $A_2$ are defined by $A_1 = \mathrm{Ind}^G_H(\mathrm{End}(V_{\omega}))$ and $A_2 = \mathrm{Ind}^G_K(\mathrm{End}(V_{\rho}))$, and $A_i$ is the $i$th component of the decomposition \eqref{eq: Mackey decomposition}.
Diagrammatically, these are represented as follows:
\begin{align}
       p_{i} = \adjincludegraphics[scale=1.25,trim={10pt 10pt 10pt 10pt},valign = c]{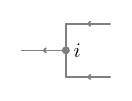},\;\;
       \iota_{i} = \adjincludegraphics[scale=1.25,trim={10pt 10pt 10pt 10pt},valign = c]{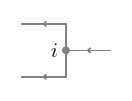},
\end{align}
and
\begin{align}
       \sum_{i}\;\; \adjincludegraphics[scale=1.25,trim={10pt 10pt 10pt 10pt},valign = c]{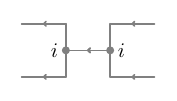}\;\;
       = \adjincludegraphics[scale=1.25,trim={10pt 10pt 10pt 10pt},valign = c]{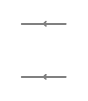},\;\;\;\;
       \adjincludegraphics[scale=1.25,trim={10pt 10pt 10pt 10pt},valign = c]{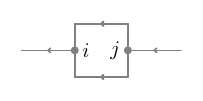}\;\;
       = \delta_{ij}\;\;\adjincludegraphics[scale=1.25,trim={10pt 10pt 10pt 10pt},valign = c]{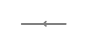}.
\end{align}
The projection and inclusion are compatible with the Frobenius algebra structures, i.e.,
       \begin{align}\label{eq: compatibility of projection and inclusion}
              \begin{aligned}
              &p_{i} \circ (\mu_{1} \otimes \mu_{2}) = \mu_{i} \circ (p_{i}\otimes p_{i}),\;\;
              (p_{i}\otimes p_{i}) \circ (\Delta_{1} \otimes \Delta_{2}) = \Delta_{i} \circ p_{i},\\
              &(\mu_{1} \otimes \mu_{2}) \circ (\iota_{i}\otimes \iota_{i}) = \iota_{i} \circ \mu_{i},\;\;
              (\Delta_{1} \otimes \Delta_{2}) \circ \iota_{i} = (\iota_{i}\otimes \iota_{i}) \circ \Delta_{i}.
              \end{aligned}
       \end{align}
Here, we omitted the trivial braiding between $A_1$ and $A_2$.

By using this projection, we can compute the fusion rule of the PEPOs:
Let $A_{1}$ and $A_{2}$ be two algebra objects in $\mathrm{Rep}(G)$.
Let $P_{12}=\{p_{i}\}$ be a set of projections in the decomposition of the tensor product $A_{1}\otimes A_{2}$.
Then, the product of the two PEPOs is locally decomposed as follows:
\begin{align}\label{eq: fusion rule PEPO}
       \adjincludegraphics[scale=1.5,trim={10pt 10pt 10pt 10pt},valign = c]{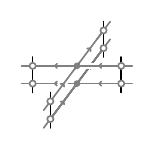}
       \;\;=\;\;\sum_{i_{1},i_{2},i_{3},i_{4}}\;\;
       \adjincludegraphics[scale=1.5,trim={10pt 10pt 10pt 10pt},valign = c]{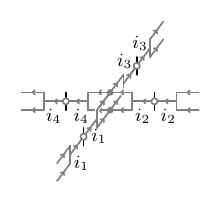},
\end{align}
where $i_{1},i_{2},i_{3},i_{4}$ run over the index of the projections in $P_{12}$.
We can show that, on the right-hand side, the configurations of the projection operators with different indices vanish.
To see this, let us focus on the central part of the right-hand side.
By definition of the 4-valent tensor, this part is written as
\begin{equation}
       \adjincludegraphics[scale=1.25,trim={10pt 10pt 10pt 10pt},valign = c]{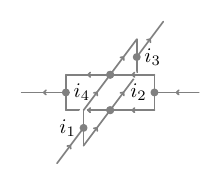}
       =
       \adjincludegraphics[scale=1.25,trim={10pt 10pt 10pt 10pt},valign = c]{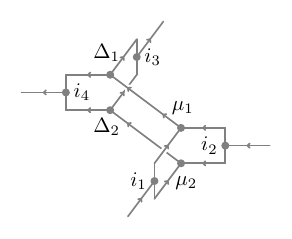}
       =
       \prod_{j=1}^{4}\delta_{i_{j},i}\;\;
       \adjincludegraphics[scale=1.25,trim={10pt 10pt 10pt 10pt},valign = c]{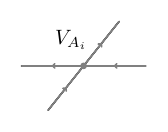},
\end{equation}
where $i$ on the right-hand side is any of $i_1$, $i_2$, $i_3$, and $i_4$.
Here, for the second equality, since the decomposition in Eq.~\eqref{eq: Mackey decomposition} is a direct-sum decomposition as Frobenius algebras, the multiplication and comultiplication between different direct-sum components vanish, which gives rise to the delta functions.
When all the indices are equal, the equality follows from the compatibility condition \eqref{eq: compatibility of projection and inclusion} between the projections and inclusions and the multiplication and comultiplication.
By plugging this into Eq.~\eqref{eq: fusion rule PEPO}, we obtain
\begin{equation}
       \adjincludegraphics[scale=1.5,trim={10pt 10pt 10pt 10pt},valign = c]{tikz/out/fusion_rule_PEPO1.pdf}
       \;\;=\;\;\sum_{i}\;\;
       \adjincludegraphics[scale=1.5,trim={10pt 10pt 10pt 10pt},valign = c]{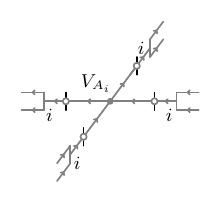}.
\end{equation}
By applying this local decomposition, we obtain the following fusion rule for the PEPO:
\begin{equation}\label{eq: PEPO decomp}
       \mathcal{O}[A_{H,\omega}^{G}]\cdot \mathcal{O}[A^{G}_{K,\rho}] = \sum_{x \in D} \mathcal{O}[A_{H\cap xKx^{-1},\omega\cdot \rho^{x}}^{G}],
\end{equation}
where $D$ is a set of representatives of the double cosets $H\backslash G/K$.
This agrees with the fusion rule of $2\mathrm{Rep}(G)$\cite[Corollary 8.10]{Greenough_2010}.

As an example, we compute two special cases of the decomposition \eqref{eq: Mackey decomposition}.
First, let us consider the case $H=K=G$. 
In this case, $G\backslash G/G=\{e\}$, so the decomposition is
\begin{equation}
       \mathrm{Ind}_{G}^{G}(\mathrm{End}(V_{\omega}))\otimes \mathrm{Ind}_{G}^{G}(\mathrm{End}(V_{\rho})) \cong \mathrm{Ind}_{G}^{G}(\mathrm{End}(V_{\omega}\otimes V_{\rho}))\sim\mathrm{Ind}_{G}^{G}(\mathrm{End}(V_{\omega\cdot\rho})).
\end{equation}
Here, the last symbol $\sim$ denotes Morita equivalence.
Therefore, these condensation surfaces satisfy the fusion rule
\begin{equation}
       \mathcal{O}[A_{G,\omega}^{G}]\cdot \mathcal{O}[A_{G,\rho}^{G}] = \mathcal{O}[A_{G,\omega\cdot\rho}^{G}].
\end{equation}
       We note that the condensation surface $\mathcal{O}[A_{G,1}^{G}]$ acts as the identity operator on $\mathcal{H}_{\text{flat}}$.
       Thus, the condensation surface $\mathcal{O}[A_{G,\omega}^{G}]$ is invertible with its inverse given by $\mathcal{O}[A_{G,\omega^{-1}}^{G}]$.
Next, let us consider the case $H=K=\{e\}$.
In this case, $\{e\}\backslash G/\{e\}=G$ and the cocycles are trivial, so the decomposition is
\begin{equation}\label{eq: full condensation}
       \mathrm{Ind}_{\{e\}}^{G}(\mathrm{End}(\mathbb{C}))\otimes \mathrm{Ind}_{\{e\}}^{G}(\mathrm{End}(\mathbb{C})) \cong \bigoplus_{g\in G} \mathrm{Ind}_{\{e\}}^{G}(\mathrm{End}(\mathbb{C})).
\end{equation}
Therefore, the surface operator $\mathcal{O}[A_{\{e\},1}^{G}]$ satisfies the fusion rule
\begin{equation}
       \mathcal{O}[A_{\{e\},1}^{G}]\cdot \mathcal{O}[A_{\{e\},1}^{G}] = |G|\cdot \mathcal{O}[A_{\{e\},1}^{G}].
\end{equation}
The operator $\mathcal{O}[A_{\{e\},1}^{G}]$ will be referred to as the full-condensation surface because it is physically obtained by condensing all lines on the identity surface. 
In the next two paragraphs, we provide more detailed computations of the fusion rules of the full-condensation surfaces.

\paragraph{Example: $G=\mathbb{Z}_{2}$}

As a concrete example, let us consider the case of $G=\mathbb{Z}_{2}=\{\left.e,a\right|a^2=e\}$ and compute its symmetric special Frobenius algebra objects and their fusion rules.
First, there are two types of algebra objects in $\mathrm{Rep}(\mathbb{Z}_{2})$:

\noindent
(1) $H=\{e\}, \omega=1$,

We take $V_{\omega}$ to be a one-dimensional vector space $\mathbb{C}$ with the trivial representation.
As a basis of $A_{\{e\},1}^{\mathbb{Z}_{2}}$, we can choose 
\begin{align}\label{eq: basis zmod2}
    f_{g}(h) &= \delta_{g,h},
\end{align}
for $g,h\in \mathbb{Z}_{2}$. 
This is the dual group algebra $\mathbb{C}[\mathbb{Z}_{2}]^{\ast}$ as an algebra, and the irreducible decomposition is 
\begin{equation}\label{eq: non_triv alg}
    A_{\{e\},1}^{\mathbb{Z}_{2}} \cong {\bf 1} \oplus \mathrm{sign}.
\end{equation}

\noindent
(2) $H=\mathbb{Z}_{2}, \omega=1$,

We take $V_{\omega}$ to be a one-dimensional vector space $\mathbb{C}$ with the trivial representation.
As a basis of $A_{\mathbb{Z}_{2},1}^{\mathbb{Z}_{2}}$, we can choose
\begin{align}
    f(e) = f(a) = 1.
\end{align}
Therefore,
\begin{equation}
       A_{\mathbb{Z}_{2},1}^{\mathbb{Z}_{2}} \cong {\bf 1}.
\end{equation}
Let us compute the fusion rule between these two algebra objects.
The only nontrivial fusion rule is the decomposition of $A_{\{e\},1}^{\mathbb{Z}_{2}} \otimes A_{\{e\},1}^{\mathbb{Z}_{2}}$.
The product $A_{\{e\},1}^{\mathbb{Z}_{2}}\otimes A_{\{e\},1}^{\mathbb{Z}_{2}}$ is spanned by $f_{e}\otimes f_{e}, f_{e}\otimes f_{a}, f_{a}\otimes f_{e}, f_{a}\otimes f_{a}$.
One can check that 
\begin{equation}
       \left< f_{e} \otimes f_{e}, f_{a} \otimes f_{a} \right> \cong \left< f_{e} \otimes f_{a}, f_{a} \otimes f_{e} \right> \cong A_{\{e\},1}^{\mathbb{Z}_{2}},
\end{equation}
as a $\mathbb{Z}_{2}$-equivariant algebras.
Therefore, we obtain the fusion rule
\begin{equation}
    \mathbf{1} \otimes \mathbf{1} \cong \mathbf{1},\; A_{\{e\},1}^{\mathbb{Z}_{2}} \otimes \mathbf{1} \cong \mathbf{1} \otimes A_{\{e\},1}^{\mathbb{Z}_{2}} \cong A_{\{e\},1}^{\mathbb{Z}_{2}},\; A_{\{e\},1}^{\mathbb{Z}_{2}} \otimes A_{\{e\},1}^{\mathbb{Z}_{2}} \cong A_{\{e\},1}^{\mathbb{Z}_{2}} \oplus A_{\{e\},1}^{\mathbb{Z}_{2}}.
\end{equation}
This agrees with the Mackey decomposition \eqref{eq: Mackey decomposition}.

\paragraph{Example: full-condensation surface for general $G$}
Let us consider the full-condensation surface for general $G$, i.e. $H=\{e\}$ and $\omega=1$.
We take $V_{\omega}$ to be a one-dimensional vector space $\mathbb{C}$ with the trivial representation.
As a basis of $A_{\{e\},1}^{G}$, we can choose
\begin{align}\label{eq: basis G}
    f_{g}(h) &= \delta_{g,h},
\end{align}
for $g,h\in G$. 
Thus, the algebra object is the dual group algebra $\mathbb{C}[G]^{\ast}$.
Let us compute the fusion rule of $A_{\{e\},1}^{G}$.
In our basis $\left<f_g\right>_{g \in G}$, the product in $A_{\{e\},1}^{G}\otimes A_{\{e\},1}^{G}$ is decomposed as
\begin{equation}
       A_{\{e\},1}^{G}\otimes A_{\{e\},1}^{G}\cong\bigoplus_{g\in G}\;\left< f_{g'} \otimes f_{g'g} \right>_{g'\in G},
\end{equation}
where $g$ labels the direct sum component and $g'$ labels the basis of the component.
Let $p_{g}$ be a projection onto the $g$-th component.
Each component is isomorphic to $A_{\{e\},1}^{G}$ as a $G$-equivariant algebra.
Therefore, the fusion rule is
\begin{equation}
       A_{\{e\},1}^{G}\otimes A_{\{e\},1}^{G} \cong  [A_{\{e\},1}^{G}]^{\oplus\abs{G}}.
\end{equation}

\subsection{Tensor network representation}
\label{sec: Tensor network representation}

Let us construct a tensor network representation of the ground state of the Hamiltonian \eqref{eq: 2d G-cluster Hamiltonian} and the symmetry action on it.
A tensor network representation of the ground state is given by the following PEPS tensor:
\begin{equation}\label{eq: G-cluster unit cell}
       \adjincludegraphics[scale=1.25,trim={10pt 10pt 10pt 10pt},valign = c]{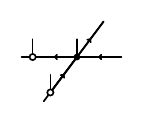},
\end{equation}
where the 4-valent tensor is defined by
\begin{equation}
       \adjincludegraphics[scale=1.25,trim={10pt 10pt 10pt 10pt},valign = c]{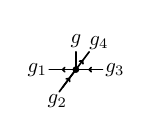}
       =
       \prod_{i=1}^{4}\delta_{g,g_{i}},\;\;
       \adjincludegraphics[scale=1.25,trim={10pt 10pt 10pt 10pt},valign = c]{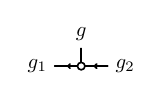}
       =
       \delta_{g_{1}^{-1}g_{2},g}.
\end{equation}
We note that the PEPS tensor is a natural generalization of Eq.~\eqref{eq: G-cluster MPS}.
       One can verify that this PEPS is indeed the unique ground state of the Hamiltonian~\eqref{eq: 2d G-cluster Hamiltonian} by checking that it is stabilized by all the terms in the Hamiltonian.
       To describe elements of $\mathcal{H}$, we introduce the following basis of $\mathcal{H}$:
       \begin{equation}
              \ket{\{g_{i}\},\{g_{ij}\}} \in \mathcal{H},
       \end{equation}
       where $g_{i}$ is the group element on the vertex $i$ and $g_{ij}$ is that on the edge connecting vertices $i$ and $j$.
       Then, one can check that the PEPS generated by the tensor in Eq.~\eqref{eq: G-cluster unit cell} is given by
       \begin{equation}
              \sum_{\{g_{i}\}}\ket{\{g_{i}\},\{g_{i}^{-1}g_{j}\}} \in \mathcal{H}.
       \end{equation}
       This state is clearly invariant under the vertex term $h_{i}$ and the edge term $h_{\ell}$ in the Hamiltonian~\eqref{eq: 2d G-cluster Hamiltonian}.

Let us confirm that this PEPS is invariant under the $G\times2\mathrm{Rep}(G)$ symmetry.
For the $G$-symmetry, the symmetry operators are fractionalized on the virtual legs as follows:
\begin{equation}\label{eq: pulling through G 2d}
       \adjincludegraphics[scale=1.25,trim={10pt 10pt 10pt 10pt},valign = c]{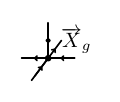}
       =
       \adjincludegraphics[scale=1.25,trim={10pt 10pt 10pt 10pt},valign = c]{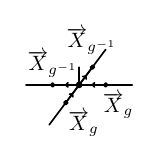}.
\end{equation}
Since the left multiplication on the virtual legs commutes with the white tensor in Eq.~\eqref{eq: G-cluster unit cell}, the symmetry action on the unit cell reduces to the gauge transformation of the PEPS tensor.
Thus, the ground state is invariant under the $G$-symmetry.
For the $2\mathrm{Rep}(G)$-symmetry, due to Eq.~\eqref{eq: pulling through repG}, the symmetry action of the condensation surface on the virtual legs is given by
\begin{equation}\label{eq: sym frac condensation 2d}
       \adjincludegraphics[scale=1.25,trim={10pt 10pt 10pt 10pt},valign = c]{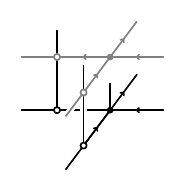}
       \;=\;
       \adjincludegraphics[scale=1.25,trim={10pt 10pt 10pt 10pt},valign = c]{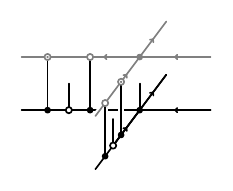}
       \;=\;
       \adjincludegraphics[scale=1.25,trim={10pt 10pt 10pt 10pt},valign = c]{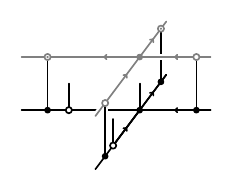}.
\end{equation}
Here, in the first equality, we used Eq.(\ref{eq: pulling through repG}) and the second equality follows from the $G$-equivariance of the multiplication and comultiplication maps in Eq.\eqref{eq: g equivariance}.
Therefore, the action of $\mathcal{O}[A]$ on the ground state $\ket{\mathrm{G.S.}}$ on a torus is given by
\begin{align}
       \begin{aligned}
       \mathcal{O}[A]\ket{\mathrm{G.S.}}
       \;&=\;
       \adjincludegraphics[scale=1.25,trim={10pt 10pt 10pt 10pt},valign = c]{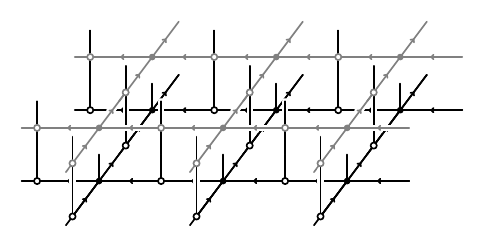}\\
       \;&=\;
       \adjincludegraphics[scale=1.25,trim={10pt 10pt 10pt 10pt},valign = c]{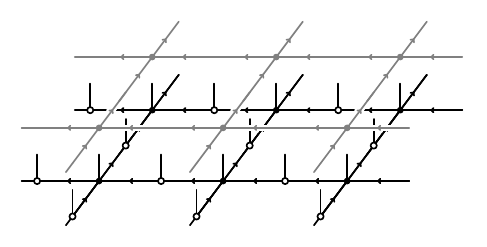}\\
       \;&=\;n(A)\ket{\mathrm{G.S.}}.
       \end{aligned}
\end{align}
Here,  $n(A)$ is defined by
\begin{equation}\label{eq: torus partition function}
       n(A) \coloneq \adjincludegraphics[scale=1.25,trim={10pt 10pt 10pt 10pt},valign = c]{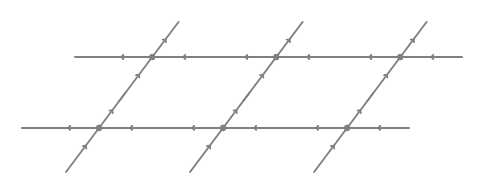} \in\mathbb{N}.
\end{equation}
We note that $n(A)$ does not depend on the system size and for an algebra object labeled by $([H],\omega)$, $n(A_{H,\omega}^{G}) = \abs{G/H}$.
In fact, $n(A)$ is the torus partition function of the TQFT associated with the symmetric special Frobenius algebra $A$ and it is given by $\abs{G/H}$ \cite{Kapustin:2016jqm}.

\subsection{Line-like action tensors}
\label{sec: line-like action tensors}

In the 1+1d case, we defined action tensors by acting the MPO symmetry on the local MPS as in Eq.~(\ref{eq: pulling through repG}).
In the 2+1d case, by acting the onsite $G$-symmetry operator and the symmetry PEPO tensor of condensation surface $\mathcal{O}[A]$ on the local PEPS tensor, we can define action tensors that are extended one-dimensionally.
We refer to these as line-like action tensors.
To compute line-like action tensors for the $2\mathrm{Rep}(G)$-symmetry, it is convenient to introduce the module MPS over the algebra object $A$.

\subsubsection{Line-like action tensors for $G$-symmetry}

Let us first compute line-like action tensors for the $G$-symmetry.
As we showed in Eq.~\eqref{eq: pulling through G 2d}, $\overrightarrow{X}_{g}^{(i)}$ is fractionalized to the virtual legs of the PEPS tensor as
\begin{equation}
       \adjincludegraphics[scale=1.25,trim={10pt 10pt 10pt 10pt},valign = c]{tikz/out/TN_pulling_through1.pdf}
       =
       \adjincludegraphics[scale=1.25,trim={10pt 10pt 10pt 10pt},valign = c]{tikz/out/TN_pulling_through2.pdf}.
\end{equation}
Therefore, the partial action of the symmetry operator $U_{g}$ on the ground state PEPS is given by
\begin{align}\label{eq: line-like action tensor G}
       \begin{aligned}
       \adjincludegraphics[scale=1.25,trim={10pt 10pt 10pt 10pt},valign = c]{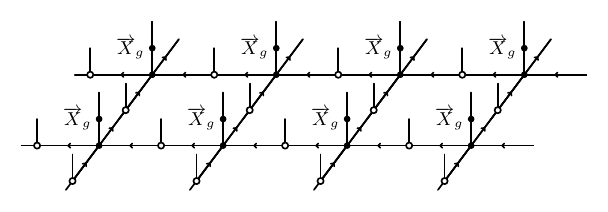}\\
       \;\;=\;\;
       \adjincludegraphics[scale=1.25,trim={10pt 10pt 10pt 10pt},valign = c]{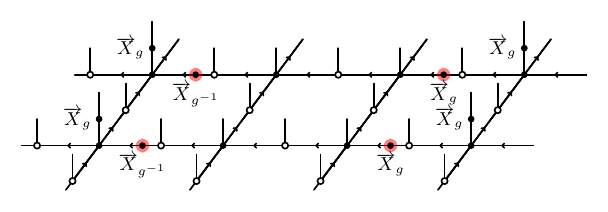}
       \end{aligned}
\end{align}
Here, the middle part of the symmetry operator is absorbed into the ground state, while the left and right parts remain.
Along each interface, an action tensor, which is highlighted in red, appears once per site.
Therefore, the line-like action tensor for the $G$-symmetry is defined by
\begin{equation}\label{eq: line-like action tensor G2}
       \Phi_{g} \coloneq \bigotimes_{v} \overrightarrow{X}_{g},\;\;
       \overline{\Phi}_{g} \coloneq \bigotimes_{v} \overrightarrow{X}_{g^{-1}}.
\end{equation}
Here, the tensor product is taken over all virtual legs along the interface.

\subsubsection{Module MPS}

Let us now introduce the module MPS and derive several important properties of it.
The module MPS will play a key role in computing the line-like action tensors for the $2\mathrm{Rep}(G)$-symmetry.
The computation of the line-like action tensors will be discussed in Sec.~\ref{sec: Line-like action tensors for 2Rep(G)-symmetry}.

\begin{figure}[t]
       \begin{minipage}{0.4\textwidth}
           \centering
           \begin{equation*}
              \adjincludegraphics[scale=1.5,trim={10pt 10pt 10pt 10pt},valign = c]{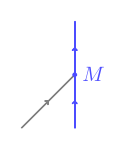}
           \end{equation*}
           \caption*{(a)}    
       \end{minipage}
       \hfill 
       \begin{minipage}{0.6\textwidth}
           \centering
           \begin{equation*}
              \adjincludegraphics[scale=1.5,trim={15pt 10pt 10pt 10pt},valign = c]{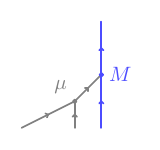}
              \;=              
              \adjincludegraphics[scale=1.5,trim={10pt 10pt 10pt 10pt},valign = c]{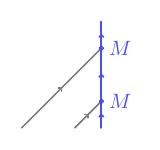}
           \end{equation*}
           \caption*{(b)}
       \end{minipage}
       \caption{
              Fundamental properties of the module action.
              (a) The diagrammatic representation of the module action.
              (b) The diagrammatic representation of the consistency condition.
       }
       \label{fig: module action}
\end{figure}

\paragraph{Module over algebra object}
For an algebra object $A_{H,\omega}^{G}$, a left module is obtained by the following way \cite{Kapustin:2016jqm}:
First recall that we take an irreducible projective representation $(\rho_{\omega},V_{\omega})$ of $H$ with cocycle $\omega$ and the algebra is realized as 
\begin{equation}
       A_{H,\omega}^{G} = \{f:G\to \mathrm{GL}(V_{\omega}) \left.\right| f(h^{-1}g) = \rho_{\omega}(h) f(g) \rho_{\omega}^{\dagger}(h)\}.
\end{equation}
Then, a left module $M^G_{H, \omega}$ over $A_{H,\omega}^{G}$ is realized as an induced module defined by
\begin{equation}
       M_{H,\omega}^{G} := \{m:G\to V_{\omega}\left.\right| m(h^{-1}g) = \rho_{\omega}(h) m(g)\}.
\end{equation}
The module action
\begin{equation}
       M: A_{H,\omega}^{G}\otimes M_{H,\omega}^{G} \to M_{H,\omega}^{G}
\end{equation}
is given by
\begin{equation}
       (f\cdot m)(g) = f(g)m(g),
\end{equation}
that is, the point-wise action of $\mathrm{End}(V_{\omega})$.
One can check that 
\begin{equation}
       (f_{1}\cdot f_{2}) \cdot m = f_{1}\cdot (f_{2}\cdot m),
\end{equation}
and thus $M_{H,\omega}^{G}$ is a left module over $A_{H,\omega}^{G}$.
The diagrammatic representation is presented in Fig.~\ref{fig: module action}(a) and (b).
Since $A^G_{H, \omega}$ is a special Frobenius algebra, $M_{H,\omega}^{G}$ naturally admits a comodule structure $\hat{M}:M_{H,\omega}^{G} \to A_{H,\omega}^{G}\otimes M_{H,\omega}^{G}$:
\begin{equation}\label{eq: comodule structure}
       \adjincludegraphics[scale=1.5,trim={10pt 10pt 10pt 10pt},valign = c]{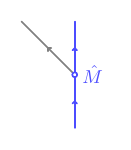}
       \coloneq
       \adjincludegraphics[scale=1.5,trim={10pt 10pt 10pt 10pt},valign = c]{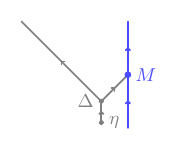}.
\end{equation}
We note that the comodule structure satisfies the following consistency with the module structure:
\begin{equation}
       \adjincludegraphics[scale=1.5,trim={10pt 10pt 10pt 10pt},valign = c]{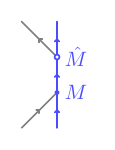}
       =
       \adjincludegraphics[scale=1.5,trim={10pt 10pt 10pt 10pt},valign = c]{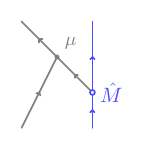}
       =
       \adjincludegraphics[scale=1.5,trim={10pt 10pt 10pt 10pt},valign = c]{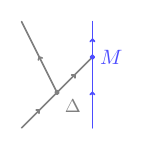}.
\end{equation}
       Let $S_{H\backslash G}$ be a set of representatives of the right $H$-cosets in $G$.
Let us choose bases $\{f^{r}_{ab}\left.\right|r\in S_{H\backslash G},a,b=1,...,\dim V_{\omega}\}$ of $A_{H,\omega}^{G}$ and $\{m^{r}_{a}\left.\right|r\in S_{H\backslash G},a=1,...,\dim V_{\omega}\}$ of $M_{H,\omega}^{G}$ such that
\begin{equation}\label{eq: module MPS basis}
      f_{ab}^{r}(r') = \frac{1}{\sqrt{\mathrm{dim}V_{\omega}}}\;\delta_{rr'}e_{ab}, \;\;\;
      m_{a}^{r}(r') = \delta_{rr'}e_{a},
\end{equation}
where $e_{ab}\in \mathrm{End}(V_{\omega})$ is the matrix unit with $1$ at the $(a,b)$-th entry and $0$ elsewhere, and $e_{a}\in V_{\omega}$ is the vector with $1$ at the $a$-th entry and $0$ elsewhere.
In particular, $f_{ab}^{r}$ and $m_{a}^{r}$ satisfy the following relations:
\begin{equation}\label{eq: basis}
       f_{ab}^{r}(r')\in \mathbb{R},\;\;f_{ab}^{r}\cdot m_{a'}^{r'}=\frac{1}{\sqrt{\mathrm{dim}V_{\omega}}}\;\delta_{rr'}\delta_{ba'}m_{a}^{r}.
\end{equation}
Therefore, as an $A_{H,\omega}^{G}$-module, $M_{H,\omega}^{G}$ decomposes as a direct sum of simple modules indexed by the left $H$-orbits:
\begin{equation}\label{eq: module decomp}
       M_{H,\omega}^{G} \cong \bigoplus_{r\in S_{H\backslash G}} M_{r}, \;\;\; M_{r} := \mathrm{span}\{m_{a}^{r}\left.\right|a=1,...,\dim V_{\omega}\}.
\end{equation}
Physically, each component $M_r$ corresponds to a spontaneously symmetry-broken (SSB) ground state of the $G$-symmetric state sum TQFT associated with the $G$-equivariant algebra $A_{H,\omega}^{G}$\cite{Kapustin:2016jqm}.
For this reason, we will refer to each direct-sum component $M_r$ as an SSB sector of $M_{H,\omega}^{G}$.

For later use, we also define an inner product on $A^G_{H, \omega}$ as
\begin{equation}
       \braket{f|f'} \coloneq \epsilon(f^{\dagger}\cdot f'),
\end{equation}
where $f^{\dagger}$ is the adjoint of $f$ as a function from $G$ to $\mathrm{End}(V_{\omega})$, i.e., $f^{\dagger}(g) \coloneq f(g)^{\dagger}$.
We can check that the basis $\{f^{r}_{ab}\}$ is orthonormal with respect to this inner product:
\begin{equation}
       \braket{f^{r}_{ab}|f^{r'}_{a'b'}} = \delta_{rr'}\delta_{aa'}\delta_{bb'}.
\end{equation}
We also use the ket notation $\ket{f}$ for the same element $f\in A^G_{H,\omega}$ when it is viewed as a vector, and the bra notation $\bra{f}$ to represent the dual vector of $f\in A^G_{H, \omega}$, i.e.,
\begin{equation}
       \bra{f} \coloneq \epsilon(f^{\dagger}\cdot \bullet): A^G_{H, \omega} \to \mathbb{C}.
\end{equation}

\paragraph{Module and comodule MPS}
Module structure maps $M$ and $\hat{M}$ naturally MPS tensors, which we refer to as the module and comodule MPS tensors, respectively.
First, let us consider the module structure map $M$.
The MPS tensor $[M]^{(r;a,b)}_{(r'';d),(r';c)}$ associated with the module structure map is defined by
\begin{equation}
       f_{ab}^{r} \cdot m_{c}^{r'} = \sum_{d,r''}  m_{d}^{r''} [M]^{(r;a,b)}_{(r'';d),(r';c)},
\end{equation}
and Eq.~\eqref{eq: basis} implies that the MPS tensor is given by
\begin{equation}\label{eq: module MPS tensor}
       [M]^{(r;a,b)}_{(r'';d),(r';c)} = \frac{1}{\sqrt{\mathrm{dim}V_{\omega}}}\;\delta_{r,r'}\delta_{r',r''}\delta_{a,d}\delta_{b,c}.
\end{equation}
We define the MPS $\bra{M}_L$ generated by the MPS tensor in Eq.~\eqref{eq: module MPS tensor} as
\begin{align}\label{eq: module MPS}
       \bra{M}_{L}&\coloneq \sum_{\{(r_{k},a_{k},b_{k})\}}\tr\left([M]^{(r_{1};a_{1},b_{1})}\cdots [M]^{(r_{L};a_{L},b_{L})}\right)\bra{f^{r_{1}}_{a_{1}b_{1}},...,f^{r_{L}}_{a_{L}b_{L}}},
\end{align}
where $\bra{f^{r_{1}}_{a_{1}b_{1}},...,f^{r_{L}}_{a_{L}b_{L}}}$ is defined by $\bra{f^{r_1}_{a_1 b_1}}\otimes\cdots\otimes\bra{f^{r_L}_{a_L b_L}}$ and $L$ is the number of sites.
Diagrammatically, it is represented as
\begin{align}\label{eq: module MPS diagram}
       \bra{M}_{L}=\;\;\;\adjincludegraphics[scale=1.25,trim={10pt 10pt 10pt 10pt},valign = c]{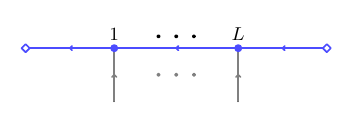}\;\;\;,
\end{align}
where the label above each tensor denotes the site index.
We refer to this MPS $\bra{M}_L$ as the module MPS.
The module MPS is the equal-weight superposition of the SSB ground states corresponding to each component in Eq.~\eqref{eq: module decomp}.
Indeed, the MPS tensor in Eq.~\eqref{eq: module MPS tensor} can be decomposed as $[M]^{(r;a,b)}_{(r'';d),(r';c)} = \delta_{r,r'}\delta_{r',r''} [M_{r}]^{(a,b)}_{d,c}$ where
\begin{equation}\label{eq: simple module MPS tensor}
[M_{r}]^{(a,b)}_{d,c} = \frac{1}{\sqrt{\mathrm{dim}V_{\omega}}}\;\delta_{a,d}\delta_{b,c},
\end{equation}
for $r \in S_{H\backslash G}$, or equivalently,
\begin{equation}
       [M_{r}]^{(a,b)} = \frac{1}{\sqrt{\mathrm{dim}V_{\omega}}}\;e_{ab},
\end{equation}
as a set of $\dim V_{\omega}\times\dim V_{\omega}$ matrices, and the module MPS decomposes as
\begin{equation}
\bra{M}_L = \sum_{r\in S_{H\backslash G}} \bra{M_{r}}_L,
\end{equation}
where $\bra{M_{r}}_L$ is the MPS generated by the MPS tensor in Eq.~\eqref{eq: simple module MPS tensor}:
\begin{align}\label{eq: simple module MPS}
       \bra{M_{r}}_{L}&\coloneq \sum_{\{(a_{k},b_{k})\}}\tr\left([M_{r}]^{(a_{1},b_{1})}\cdots [M_{r}]^{(a_{L},b_{L})}\right)\bra{f^{r}_{a_{1}b_{1}},...,f^{r}_{a_{L}b_{L}}}.
\end{align}
We refer to each MPS $\bra{M_{r}}_L$ as the simple module MPS.
We note that each MPS tensor $\{[M_{r}]^{(a,b)}\}$ is injective as a set of $\dim V_{\omega}\times\dim V_{\omega}$ matrices.

Similarly to the module MPS, the comodule structure map $\hat{M}$ also defines an MPS tensor, which we refer to as the comodule MPS tensor.
The MPS tensor $[\hat{M}]^{(r;a,b)}_{(d;r''),(c;r')}$ is defined by
\begin{equation}
       \hat{M}(m_{c}^{r'}) = \sum_{a,b,r}\sum_{d,r''} f_{ab}^{r} \otimes m_{d}^{r''} [\hat{M}]^{(r;a,b)}_{(d;r''),(c;r')}.
\end{equation}
By using Eq.~\eqref{eq: comodule structure} and Eq.~\eqref{eq: basis}, one can check that the comodule MPS tensor is given by
\begin{equation}\label{eq: comodule MPS tensor}
       [\hat{M}]^{(r;a,b)}_{(r'';d),(r';c)} = \frac{1}{\sqrt{\mathrm{dim}V_{\omega}}}\;\delta_{r,r'}\delta_{r',r''}\delta_{a,c}\delta_{b,d},
\end{equation}
and the MPS $\ket{M}_L$ generated by the comodule MPS tensor in Eq.~\eqref{eq: comodule MPS tensor} is defined as
\begin{align}
       \ket{M}_{L} &\coloneq \sum_{\{(r_{k},a_{k},b_{k})\}}\tr\left([\hat{M}]^{(r_{L};a_{L},b_{L})}\cdots [\hat{M}]^{(r_{1};a_{1},b_{1})}\right)\ket{f^{r_{1}}_{a_{1}b_{1}},...,f^{r_{L}}_{a_{L}b_{L}}},
\end{align}
where $\ket{f^{r_{1}}_{a_{1}b_{1}},...,f^{r_{L}}_{a_{L}b_{L}}}$ is defined by $\ket{f^{r_{1}}_{a_{1}b_{1}}}\otimes\cdots\otimes \ket{f^{r_{L}}_{a_{L}b_{L}}}$.
Diagrammatically, it is represented as
\begin{align}
       \ket{M}_{L}=\;\;\;\adjincludegraphics[scale=1.25,trim={10pt 10pt 10pt 10pt},valign = c]{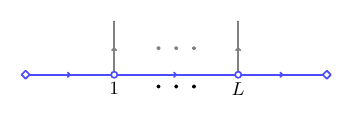}\;\;\;,
\end{align}
where the label below each tensor denotes the site index.
We refer to this MPS $\ket{M}_L$ as the comodule MPS.
The comodule MPS is also the equal-weight superposition of the SSB ground states corresponding to each component in Eq.~\eqref{eq: module decomp}.
Indeed, the comodule MPS tensor in Eq.~\eqref{eq: comodule MPS tensor} can be decomposed as $[\hat{M}]^{(r;a,b)}_{(r'';d),(r';c)} = \delta_{r,r'}\delta_{r',r''} [\hat{M}_{r}]^{(a,b)}_{d,c}$ where
\begin{equation}\label{eq: simple comodule MPS tensor}
[\hat{M}_{r}]^{(a,b)}_{d,c} = \frac{1}{\sqrt{\mathrm{dim}V_{\omega}}}\;\delta_{a,c}\delta_{b,d},
\end{equation}
for $r \in S_{H\backslash G}$, or equivalently,
\begin{equation}
       [\hat{M}_{r}]^{(a,b)} = \frac{1}{\sqrt{\mathrm{dim}V_{\omega}}}\;e_{ba},
\end{equation}
and the comodule MPS decomposes as
\begin{equation}
\ket{M}_{L} = \sum_{r\in S_{H\backslash G}} \ket{M_{r}}_{L},
\end{equation}
where $\ket{M_{r}}_{L}$ is the MPS generated by the comodule MPS tensor in Eq.~\eqref{eq: simple comodule MPS tensor}:
\begin{equation}\label{eq: simple comodule MPS}
       \ket{M_{r}}_{L}\coloneq \sum_{\{(a_{k},b_{k})\}}\tr\left([\hat{M}_{r}]^{(a_{L},b_{L})}\cdots [\hat{M}_{r}]^{(a_{1},b_{1})}\right)\ket{f^{r}_{a_{1}b_{1}},...,f^{r}_{a_{L}b_{L}}}.
\end{equation}
We refer to each MPS $\ket{M_{r}}_{L}$ as the simple comodule MPS.
We note that the comodule MPS tensor $\{[\hat{M}_{r}]^{(a,b)}\}$ is equal to the complex conjugate of the module MPS tensor $\{[M_{r}^\dagger]^{(a,b)}\}$. 
Thus we see that $\ket{M_r}_L$ is the dual vector of $\bra{M_r}_L$:
\begin{align}
       \begin{aligned}
       [\bra{M_{r}}_{L}]^{\dagger} 
       & = \sum_{\{(a_k, b_k)\}} \tr\left( [M_r]^{(a_1, b_1)} \cdots [M_r]^{(a_L, b_L)} \right)^{\ast} \ket{f^r_{a_1 b_1} \cdots f^r_{a_L b_L}} \\
       & = \sum_{\{(a_k, b_k)\}} \tr\left( [M_r^{\dagger}]^{(a_L, b_L)} \cdots [M_r^{\dagger}]^{(a_1, b_1)} \right) \ket{f^r_{a_1 b_1} \cdots f^r_{a_L b_L}}\\
       & = \sum_{\{(a_k, b_k)\}} \tr\left( [\hat{M}_r]^{(a_L, b_L)} \cdots [\hat{M}_r]^{(a_1, b_1)} \right) \ket{f^r_{a_1 b_1} \cdots f^r_{a_L b_L}} = \ket{M_r}_L.
       \end{aligned}
\end{align}
In what follows, we omit the subscript $L$ for simplicity.
We note that the simple module MPSs are orthonormal to each other:
\begin{equation}\label{eq: orthonormality of simple module MPS}
       \braket{M_{r}|M_{r'}} = \delta_{r,r'}.
\end{equation}

Let us describe the action of the symmetry on the SSB ground state $\ket{M_{r}}$.
One can show that the action is given by
\begin{equation}\label{eq: symmetry action on module MPS}
       \bigotimes \rho^{G}_{H,\omega}(g) \ket{M_{r}} = \ket{M_{r'}},
\end{equation}
where $r'\in S_{H\backslash G}$ is the unique element such that $h r' = r g^{-1}$ for a unique $h\in H$.
See App.~\ref{sec: symmetry action on module MPS} for the derivation.

\paragraph{Two important properties}
We would like to compute the symmetry fractionalization of the condensation surface by using the module and comodule MPSs. In this paragraph, we present two properties that are important for this purpose.

First, let us show that the following bubble-creation relation holds:
\begin{equation}\label{eq: bubble creation}
       \adjincludegraphics[scale=1.5,trim={10pt 10pt 10pt 10pt},valign = c]{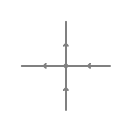}
       \;=\;\mathrm{dim}V_{\omega}\;
       \adjincludegraphics[scale=1.5,trim={10pt 10pt 10pt 10pt},valign = c]{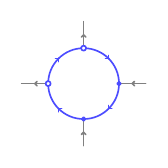}.
\end{equation}
This equality follows from the direct computation using the properties of the module and comodule structures:
\begin{align}\label{eq: bubble creation derivation}
       \adjincludegraphics[scale=1.25,trim={10pt 10pt 10pt 10pt},valign = c]{tikz/out/moduleMPS1.pdf}
       &=
       \adjincludegraphics[scale=1.25,trim={10pt 10pt 10pt 10pt},valign = c]{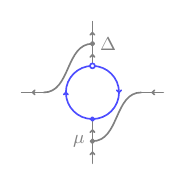}
       =
       \adjincludegraphics[scale=1.25,trim={10pt 10pt 10pt 10pt},valign = c]{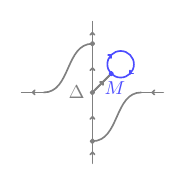}
       =\;\frac{1}{\mathrm{dim}V_{\omega}}\;
       \adjincludegraphics[scale=1.25,trim={10pt 10pt  10pt 10pt},valign = c]{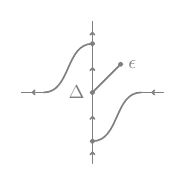},
\end{align}
where, in the first equality, we used the properties of module and comodule structure maps, and the second equality follows from the consistency condition between the module and comodule structures.
To show the last equality, let us compute the partial trace of the structure map $M: A_{H,\omega}^{G}\otimes M_{H,\omega}^{G} \to M_{H,\omega}^{G}$ over $M_{H,\omega}^{G}$.
For an element $f=\sum_{(r;a,b)}X^{r}_{a,b}f^{r}_{a,b} \in A_{H,\omega}^{G}$, the partial trace $\tr_{M^G_{H, \omega}} M: A^G_{H, \omega} \to \mathbb{C}$ is computed as
\begin{align}
       \tr_{M^G_{H, \omega}} M (f) &=\sum_{(r;a)}m_{a}^{r\dagger} f m_{a}^{r}=\sum_{(r';a',b')}\sum_{(r;a)}X_{a',b'}^{r'}m_{a}^{r\dagger} f_{a',b'}^{r'} m_{a}^{r}=\sum_{(r;a)}X_{a,a}^{r}.
\end{align}
On the other hand, the counit $\epsilon: A_{H,\omega}^{G} \to \mathbb{C}$ defined in Eq.~\eqref{eq: counit} is computed as
\begin{align}
       \begin{aligned}
       \epsilon(f) 
       &= \frac{\mathrm{dim}V_{\omega}}{\abs{H}}\sum_{g\in G}\sum_{(r;a,b)}X^{r}_{a,b}\tr_{V_{\omega}}(f^{r}_{a,b}(g))\\
       &= \frac{\mathrm{dim}V_{\omega}}{\abs{H}}\sum_{r'\in S_{H\backslash G}}\sum_{h\in H}\sum_{(r;a,b)}X^{r}_{a,b}\tr_{V_{\omega}}(f_{a,b}^{r}(r'))\\
       &= \mathrm{dim}V_{\omega}\sum_{r'\in S_{H\backslash G}}\sum_{(r;a,b)}X^{r}_{a,b}\delta_{r,r'}\delta_{a,b}\\
       &= \mathrm{dim}V_{\omega}\sum_{(r;a)}X^{r}_{a,a}.
       \end{aligned}
\end{align}
Therefore, the trace $\tr_{M_{H,\omega}^{G}} M: A_{H,\omega}^{G} \to \mathbb{C}$ is given
\begin{equation}
       \tr_{M_{H,\omega}^{G}} M = \frac{1}{\mathrm{dim}V_{\omega}}\;\epsilon,
\end{equation}
and thus we obtain the last equality in Eq.~\eqref{eq: bubble creation derivation}.
By using the counitality condition, we can remove the counit in the last diagram and obtain the desired equality.

Second, we show that the transfer matrix of the module MPS is the sum of projectors onto the SSB sectors.
Let us consider the transfer matrix of the module MPS defined by
\begin{equation}
       T_{M_{H,\omega}^{G}} := \adjincludegraphics[scale=1.5,trim={10pt 10pt 10pt 10pt},valign = c]{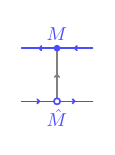}.
\end{equation}
We can show that the transfer matrix satisfies the following equation:
\begin{equation}\label{eq: transfer matrix decomp}
       \adjincludegraphics[scale=1.5,trim={10pt 10pt 10pt 10pt},valign = c]{tikz/out/transfer_mat_module1.pdf}
       \;=\;\frac{1}{\mathrm{dim}V_{\omega}}\;
       \sum_{r\in S_{H\backslash G}}\adjincludegraphics[scale=1.5,trim={10pt 10pt 10pt 10pt},valign = c]{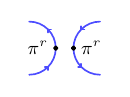}.
\end{equation}
Here, $\pi^{r}$ is the projection onto the SSB sector $M_{r}$ defined in Eq.~\eqref{eq: module decomp}, i.e.,
\begin{equation}
       \pi^{r}(m_{a}^{r'}) = \delta_{r,r'} m_{a}^{r}.
\end{equation}
This equality also follows from the direct computation using the definition of the comodule structure.
Let us evaluate the following diagram:
\begin{equation}
       \adjincludegraphics[scale=1.25,trim={10pt 10pt 10pt 10pt},valign = c]{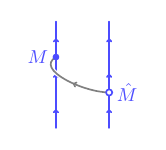}
       =
       \adjincludegraphics[scale=1.25,trim={10pt 10pt 10pt 10pt},valign = c]{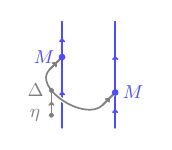}.
\end{equation}
Here, the diagram is the transfer matrix rotated by 90 degrees, with one of the blue lines on left side flipped.
When we substitute $m_{a}^{r}\otimes m_{b}^{r'}$ into the bottom legs, the output of the top legs is computed as
\begin{equation}
       \frac{1}{\mathrm{dim}V_{\omega}}\sum_{r''\in S_{H\backslash G}}\sum_{a',b'=1}^{\dim V_{\omega}} f_{a'b'}^{r''} m_{a}^{r}\otimes (f^{\dagger})_{a'b'}^{r''}m_{b}^{r'} = \frac{1}{\mathrm{dim}V_{\omega}}\;\delta_{r,r'}m_{b}^{r}\otimes m_{a}^{r}.
\end{equation}
Therefore, this diagram acts by swapping the two blue lines and projecting onto the same SSB sector.
By rotating the diagram back, we obtain the desired equality.

\;\\
\noindent
{\bf Example: $G=\mathbb{Z}_{2}$}

When $G=\mathbb{Z}_{2}=\{\left.e,a\right|a^{2}=e\}$, there is only one nontrivial algebra object $A_{H,\omega}^{G}=\bf{1}\oplus \mathrm{sign}$ where $H$ and $\omega$ are both trivial.
Let us compute the module MPS for this algebra object.
Since $H$ and $\omega$ are trivial, $M_{H,\omega}^{G}$ is $A_{H,\omega}^{G}$ itself.
An orthonormal basis of $M_{H,\omega}^{G}$ is given by
\begin{align}
       m^g (h) &= \delta_{g,h},
\end{align}
and the module action is given by\footnote{Here, we omitted the subscript of $m^g$ because it is unique due to $\dim V_{\omega} = 1$.}
\begin{equation}
       f_{g}\cdot m^h = \delta_{g,h} m^h.
\end{equation}
Therefore, $M_{H,\omega}^{G}\simeq\mathbb{C} m^e \oplus \mathbb{C} m^a$ gives the irreducible decomposition.
Then, the MPS matrices are given by
\begin{align}
       \begin{aligned}
       M_{e}= 1,\;\;\;M_{a} = 1.
       \end{aligned}
\end{align}
The MPS matrices $M_{e}$ and $M_{a}$ generate $\bigotimes\ket{e}$ and $\bigotimes\ket{a}$, respectively.
Remark that each MPS matrix is injective as a set of $1\times1$ matrices.

\;\\
\noindent
{\bf Example: Full-condensation surface for $G$}

Let us compute the module MPS for the full-condensation surface $\mathcal{O}[A_{\{e\},1}^{G}]$.
An Orthonormal basis of $M_{H,\omega}^{G}$ is given by
\begin{align}
       m^g (h) &= \delta_{g,h},
\end{align}
and the module action is given by
\begin{equation}
       f_{g}\cdot m^h = \delta_{g,h} m^h.
\end{equation}
Therefore, $M_{H,\omega}^{G}\simeq\bigoplus_{g\in G}\mathbb{C}m_{g}$ gives the irreducible decomposition.
Then, the MPS matrices are given by
\begin{align}
       M_{g} = 1.
\end{align}
The MPS generated by $M_{g}$ is $\bigotimes\ket{g}$, i.e., a $G$-SSB state.
Remark that each MPS matrix is injective as a set of $1\times1$ matrices.

\subsubsection{Line-like action tensors for $2\mathrm{Rep}(G)$-symmetry}
\label{sec: Line-like action tensors for 2Rep(G)-symmetry}
Let us compute the local symmetry action of the condensation surface $\mathcal{O}[A]$ on the ground state, where $A = A^G_{H, \omega}$.
This local symmetry action is described by the following tensors $\phi_{r}^{A}$ and $\overline{\phi}_{r}^{A}$ satisfying
\begin{align}\label{eq: defining property of action tensor}
       \begin{aligned}
       \adjincludegraphics[scale=1.25,trim={10pt 10pt 10pt 10pt},valign = c]{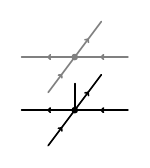}
       &=\sum_{r\in S_{H\backslash G}}\;\mathrm{dim}V_{\omega}\;
       \adjincludegraphics[scale=1.25,trim={10pt 10pt 10pt 10pt},valign = c]{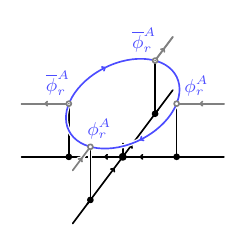}\;,\\
       \adjincludegraphics[scale=1.25,trim={30pt 10pt 10pt 10pt},valign = c]{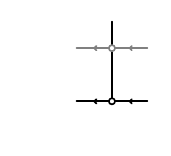}
       &=\sum_{r\in S_{H\backslash G}}\;\mathrm{dim}V_{\omega}\;
       \adjincludegraphics[scale=1.25,trim={10pt 10pt 10pt 10pt},valign = c]{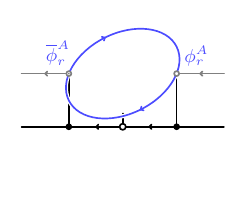}\;,
       \end{aligned}
\end{align}
and the orthogonality condition
\begin{equation}\label{eq: defining property of action tensor2}
       \adjincludegraphics[scale=1.25,trim={10pt 10pt 10pt 10pt},valign = c]{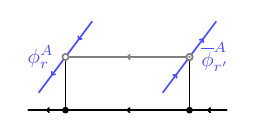}\;=\;\frac{1}{\mathrm{dim}V_{\omega}}\;\delta_{r,r'}\;
       \adjincludegraphics[scale=1.25,trim={10pt 10pt 10pt 10pt},valign = c]{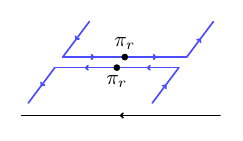}\;.
\end{equation}
By using the module MPS, we can take the action tensors $\phi_{r}^{A}$ and $\overline{\phi}_{r}^{A}$ as follows:
\begin{equation}
       \adjincludegraphics[scale=1.25,trim={10pt 10pt 10pt 10pt},valign = c]{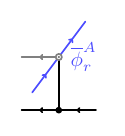}
       \;\coloneq\;
       \adjincludegraphics[scale=1.25,trim={10pt 10pt 10pt 10pt},valign = c]{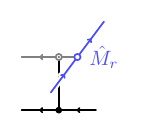},\;\;
       \adjincludegraphics[scale=1.25,trim={10pt 10pt 10pt 10pt},valign = c]{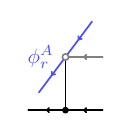}
       \;\coloneq\;
       \adjincludegraphics[scale=1.25,trim={10pt 10pt 10pt 10pt},valign = c]{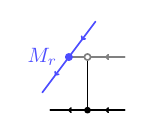},
\end{equation}
where $M_{r}$ and $\hat{M}_{r}$ are the module and comodule structure maps restricted to the SSB sector labeled by $r\in S_{H\backslash G}$, and the gray dots on the right-hand sides represent the tensors defined in Eq.~\eqref{eq: action tensor basic}.
In fact, one can verify that the first defining property in Eq,~\eqref{eq: defining property of action tensor} holds by using the property \eqref{eq: bubble creation} and the second defining property holds by using a similar argument to Eq.~\eqref{eq: bubble creation derivation}.
The orthogonality condition also holds due to Eq.~\eqref{eq: transfer matrix decomp}.

By using these tensors, the symmetry action on the unit cell can be described as follows:
\begin{equation}
       \adjincludegraphics[scale=1.25,trim={10pt 10pt 10pt 10pt},valign = c]{tikz/out/symmetry_action1.pdf}
       \;=\;
       \sum_{r} \mathrm{dim}V_{\omega}\;
       \adjincludegraphics[scale=1.25,trim={10pt 10pt 10pt 10pt},valign = c]{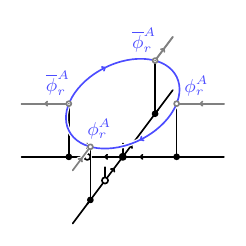}.
\end{equation}
Moreover, the two blue circles can be merged into one:
\begin{align}
       \begin{aligned}
       &\sum_{r,r'}\;(\mathrm{dim}V_{\omega})^{2}\;\;
       \adjincludegraphics[scale=1.25,trim={10pt 10pt 10pt 10pt},valign = c]{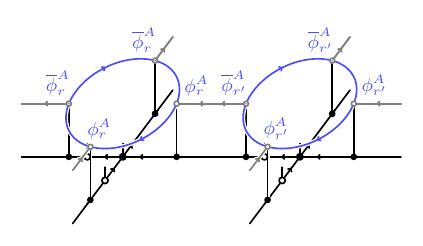}\\
       \;=\;
       &\sum_{r=1}\;\mathrm{dim}V_{\omega}\;\;
       \adjincludegraphics[scale=1.25,trim={10pt 10pt 10pt 10pt},valign = c]{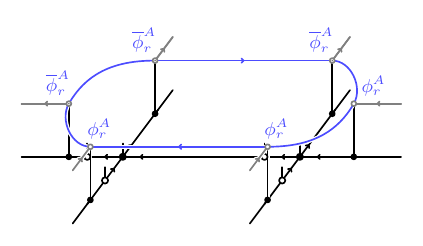}.
       \end{aligned}
\end{align}
For later use, we consider the following symmetry action: 
for the ground state on a torus, we apply a partial symmetry action supported on a strip wrapping around one of the cycles.
The tensor network representation of this partial symmetry action is given by
\begin{align}\label{eq: partial symmetry action repG}
       \begin{aligned}
              &\adjincludegraphics[scale=1.25,trim={10pt 10pt 10pt 10pt},valign = c]{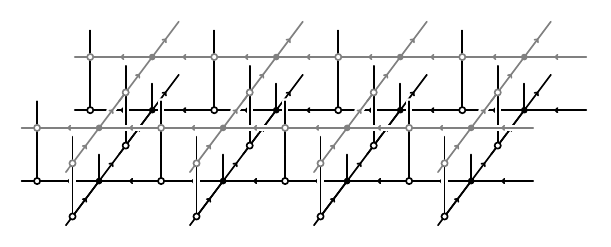}\\
              \;=\;
              \sum_{r}\;&\adjincludegraphics[scale=1.25,trim={10pt 10pt 10pt 10pt},valign = c]{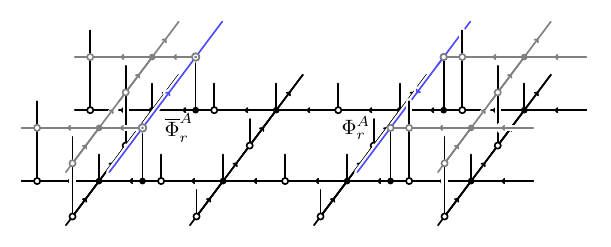},
       \end{aligned}
\end{align}
where periodic boundary conditions are imposed in the two directions, and the line-like tensors $\Phi_{r}^{A}$ and $\overline{\Phi}_{r}^{A}$ are defined as follows:
\begin{align}\label{eq: line-like action tensor RepG}
       \overline{\Phi}^{A}_{r}
       \coloneq\adjincludegraphics[scale=1.25,trim={20pt 10pt 10pt 10pt},valign = c]{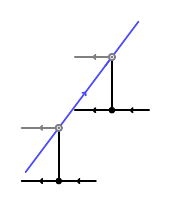}\;
       =\adjincludegraphics[scale=1.25,trim={20pt 10pt 10pt 10pt},valign = c]{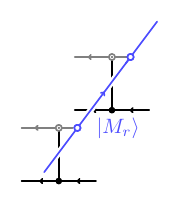},\;\;
       \Phi^{A}_{r}
       \coloneq\adjincludegraphics[scale=1.25,trim={30pt 10pt 10pt 10pt},valign = c]{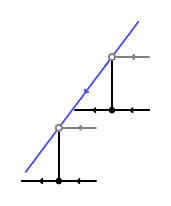}
       =\adjincludegraphics[scale=1.25,trim={10pt 10pt 10pt 10pt},valign = c]{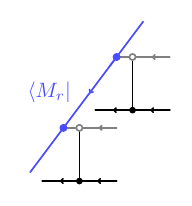}\;.
\end{align}
Here, Eq.~\eqref{eq: partial symmetry action repG} can be derived using Eq.~\eqref{eq: defining property of action tensor} and Eq.~\eqref{eq: defining property of action tensor2}.
We will refer to $\Phi^{A}_{r}$ and $\overline{\Phi}^{A}_{r}$ as the line-like action tensors for the condensation surface $\mathcal{O}[A]$.

\section{Parametrized family of $G$-symmetric states}
\label{sec: Parametrized family of G-symmetric states}
       From general considerations, it is expected that $S^n$-parameterized families of $G$-symmetric SRE states are classified by $\mathrm{H}^{2}(G;\mathrm{U}(1))$ and $\mathrm{H}^{1}(G;\mathrm{U}(1))$ for $n=1$ and $n=2$, respectively, and that the classification is trivial for all $n\geq3$ up to the higher Berry phase \cite{Thorngren:1612.00846, Hsin:2020cgg, Hermele2021CMSA,Thorngren2021YITP}.
In this section, we construct $S^{1}$- and $S^{2}$-parameterized families of $G$-symmetric states labeled by elements in $\mathrm{H}^{2}(G;\mathrm{U}(1))$ and $\mathrm{H}^{1}(G;\mathrm{U}(1))$, respectively, by using the symmetry interpolation.

\subsection{$S^{1}$-parameterized families}
\label{sec: S1-parameterized families G}
In this section, we construct $S^{1}$-parameterized families of $G$-symmetric states by using the symmetry interpolation method.
From a categorical point of view, $S^{1}$-parameterized families are classified by $\mathrm{H}^{2}(G;\mathrm{U}(1))$ \cite{Thorngren:1612.00846, Hsin:2020cgg, Hermele2021CMSA,Thorngren2021YITP}.
Since each invertible $0$-form symmetries in $2\mathrm{Rep}(G)$ is labeled by an element in $\mathrm{H}^{2}(G;\mathrm{U}(1))$, it is natural to expect that an $S^{1}$-parameterized family corresponding to $[\omega]\in\mathrm{H}^{2}(G;\mathrm{U}(1))$ is constructed by interpolating the symmetry labeled by $[\omega]$.
However, we cannot directly apply the symmetry interpolation method discussed in Sec.~\ref{sec: Symmetry interpolation} to an invertible condensation surface in $2\mathrm{Rep}(G)$. This is because a condensation surface defined in Fig.~\ref{fig: condensationPEPO} is not written as a product of local unitary operators even when it is invertible.
To resolve this issue, we first construct a local unitary decomposition of an invertible condensation surface in $2\mathrm{Rep}(G)$.
After that, we apply the symmetry interpolation method to construct $S^{1}$-parameterized families of $G$-symmetric states.
To confirm that the constructed families are nontrivial, we compute the self-interface mode pumped by an adiabatic evolution along the $S^{1}$-parameter and show that it realizes a nontrivial $G$-SPT phase in 1+1d.

\subsubsection{Invertible condensation surface in $2\mathrm{Rep}(G)$}
As we showed in Sec.\ref{sec: Symmetries}, invertible $0$-form symmetries in $2\mathrm{Rep}(G)$ are labeled by cohomology classes $[\omega]$ in $\mathrm{H}^{2}(G,\mathrm{U}(1))$.
In what follows, we will show that the symmetry operator labeled by $[\omega]$ can be decomposed into the product of local unitary operators.
We will also compute its local action on the $G$-cluster state for later use.

We first describe the Frobenius algebra object corresponding to the invertible condensation surface labeled by $[\omega]$.
To this end, let us take a representative cocycle $\omega$.
The corresponding special symmetric Frobenius algebra object is given by
\begin{equation}
       A^{G}_{G,\omega} = \mathrm{End}(V_{\omega}),
\end{equation}
where $(\rho_{\omega},V_{\omega})$ is a projective representation of $G$ with cocycle $\omega$.
As a projective representation, we take the twisted regular representation $(\rho_{\omega}^{\rm reg},\mathbb{C}[G]^{\omega})$:
as a vector space, the representation space is defined by the group algebra
\begin{equation}
       \mathbb{C}[G]^{\omega} \coloneq \mathbb{C}[G],
\end{equation}
and the representation map is defined by
\begin{equation}\label{eq: twisted regular rep}
       \rho_{\omega}^{\rm reg}(g) v_{h} = \omega(g,h)v_{gh},
\end{equation}
where $\{v_{g}\}_{g\in G}$ is the basis of $\mathbb{C}[G]^{\omega}$ labeled by group elements.
In what follows, we will normalize the cocycle as $\omega(e,g)=\omega(g,e)=1$ for all $g\in G$.
We also choose the gauge so that $\omega(g, g^{-1}) = 1$ for all $g \in G$. This normalization and gauge choice imply $\rho_{\omega}^{\rm reg}(e) = \mathrm{id}$ and $\rho_{\omega}^{\rm reg}(g^{-1}) = \rho_{\omega}^{\rm reg}(g)^{-1} = \rho_{\omega^{-1}}^{\rm reg}(g)^{\rm T}$, where the superscript $\rm T$ denotes the transpose.
For convenience, let us write down the special symmetric Frobenius algebra structure explicitly.
Let $\{v^{g}\}_{g\in G}$ be the dual basis of $(\mathbb{C}[G]^{\omega})^{\ast}$ so that $v^{g}(v_{h})=\delta_{g,h}$.
Then, the $G$-action on $\mathrm{End}(\mathbb{C}[G]^{\omega}) \cong (\mathbb{C}[G]^{\omega})^{\ast}\otimes\mathbb{C}[G]^{\omega}$ is given by
\begin{align}
       \rho^{G}_{G,\omega}(g)(v^{h}\otimes v_{l}) = \frac{\omega(g,l)}{\omega(g,h)} v^{gh}\otimes v_{gl},
\end{align}
and the multiplication and comultiplication are given by
\begin{align}
       \begin{aligned}
       \mu&: (v^{g_{1}}\otimes v_{g_{2}})\otimes (v^{h_{1}}\otimes v_{h_{2}}) \mapsto \delta_{g_{2},h_{1}} v^{g_{1}}\otimes v_{h_{2}},\\
       \Delta&: v^{g_{1}}\otimes v_{g_{2}} \mapsto \frac{1}{\abs{G}}\sum_{h\in G} (v^{g_{1}}\otimes v_{h})\otimes (v^{h}\otimes v_{g_{2}}).
       \end{aligned}
\end{align}
We can depict these structure maps diagrammatically as follows:
\begin{align}\label{eq: doubled line notation}
       \adjincludegraphics[scale=1.25,trim={10pt 10pt 10pt 10pt},valign = c]{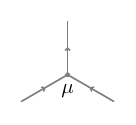}
       =\;
       \adjincludegraphics[scale=1.25,trim={10pt 10pt 10pt 10pt},valign = c]{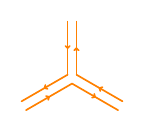},\;\;\;
       \adjincludegraphics[scale=1.25,trim={10pt 10pt 10pt 10pt},valign = c]{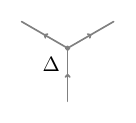}
       =\; \frac{1}{|G|}
       \adjincludegraphics[scale=1.25,trim={10pt 10pt 10pt 10pt},valign = c]{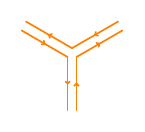}.
\end{align}
Here, the orange lines on the right-hand side oriented in the same direction as the gray line represent the representation space $\mathbb{C}[G]^{\omega}$, while the orange lines oriented opposite to the gray line represent its dual space $(\mathbb{C}[G]^{\omega})^{\ast}$.

\begin{figure}[t]
    \centering
    \adjincludegraphics[scale=1.25,trim={10pt 10pt 10pt 10pt},valign = c]{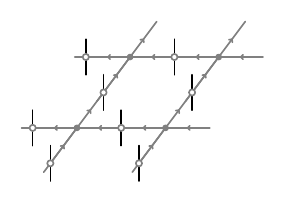}
       \;=
       \adjincludegraphics[scale=1.25,trim={40pt 40pt 40pt 40pt},valign = c]{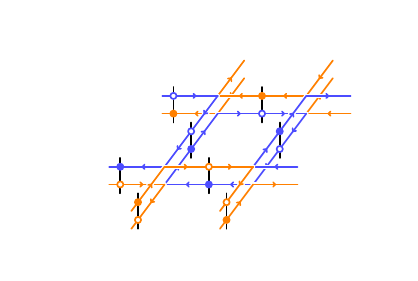}
    \caption{The decomposition of an invertible condensation surface associated with the algebra object $A^{G}_{G,\omega}$.
    To visualize the connectivity of the lines, we use orange and blue colors for the same unitary operators $u^{(p)}_{\omega}$.}
    \label{fig: S1 condensation surface decomposition}
\end{figure}

By using the doubled line notation in Eq.~\eqref{eq: doubled line notation}, we can break up the condensation surface $\mathcal{O}[A^G_{G, \omega}]$ into small pieces.
Recall that the condensation surface consists of the following local tensors:
\begin{equation}\label{eq: invertible condensation surface}
       \adjincludegraphics[scale=1.25,trim={10pt 10pt 10pt 10pt},valign = c]{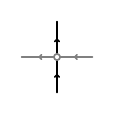}
       \;=\;
       \sum_{g\in G} \rho^{G}_{G,\omega}(g) \otimes \ket{g}\bra{g},\;\;\;\;
       \adjincludegraphics[scale=1.25,trim={10pt 10pt 10pt 10pt},valign = c]{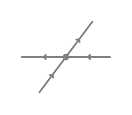}
       \;=\;
       \Delta\circ\mu.
\end{equation}
The first tensor in Eq.~\eqref{eq: invertible condensation surface} can be broken up as
\begin{equation}
       \adjincludegraphics[scale=1.25,trim={10pt 10pt 10pt 10pt},valign = c]{tikz/out/TN_S1_condensation_link1.pdf}
       \;\;=\;\;
       \adjincludegraphics[scale=1.25,trim={10pt 10pt 10pt 10pt},valign = c]{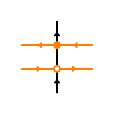},
\end{equation}
where the filled and unfilled dots represent the generalized Pauli $Z$ operators associated with $\rho_{\omega}^{\rm reg}$ and $\rho_{\omega^{-1}}^{\rm reg}$, respectively, i.e.,
\begin{equation}\label{eq: S1 orange tensors}
       \adjincludegraphics[scale=1.25,trim={10pt 10pt 10pt 10pt},valign = c]{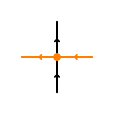}
       \;=\;
       \sum_{g\in G} \rho_{\omega}^{\rm reg}(g)\otimes\ket{g}\bra{g},\;\;
       \adjincludegraphics[scale=1.25,trim={10pt 10pt 10pt 10pt},valign = c]{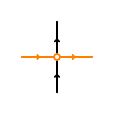}
       \;=\;
       \sum_{g\in G} \rho_{\omega}^{\rm reg}(g^{-1})\otimes\ket{g}\bra{g}.
\end{equation}
In the second equation, we used the relation $\rho_{\omega^{-1}}^{\rm reg}(g) = \rho_{\omega}^{\rm reg}(g^{-1})^{\rm T}$.
The second tensor in Eq.~\eqref{eq: invertible condensation surface} can be broken up as
\begin{equation}
       \adjincludegraphics[scale=1.25,trim={10pt 10pt 10pt 10pt},valign = c]{tikz/out/TN_S1_condensation_vertex1.pdf}
       \;\;=\;\;\frac{1}{\abs{G}}\;\;
       \adjincludegraphics[scale=1.25,trim={10pt 10pt 10pt 10pt},valign = c]{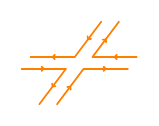}.
\end{equation}
By combining these decompositions, we can break up the condensation surface into small pieces as follows:
\begin{equation}
       \mathcal{O}[A^{G}_{G,\omega}] = \prod_{p} u^{(p)}_{\omega}.
\end{equation}
Here, $u^{(p)}_{\omega}$ is a local unitary operator acting on the physical degrees of freedom around the plaquette $p$ defined by
\begin{equation}
       u^{(p)}_{\omega} \coloneq \frac{1}{\abs{G}}\adjincludegraphics[scale=1.25,trim={10pt 10pt 10pt 10pt},valign = c]{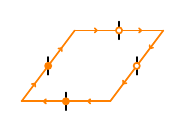},
\end{equation}
where the orange tensors are defined in Eq.~\eqref{eq: S1 orange tensors}.
See Fig.~\ref{fig: S1 condensation surface decomposition} for the graphical representation of the decomposition.

Let us compute the local symmetry action of the invertible condensation surface on the ground state.
The local symmetry action is given by
\begin{equation}\label{eq: S1 symmetry action}
       \adjincludegraphics[scale=1.25,trim={10pt 10pt 10pt 10pt},valign = c]{tikz/out/symmetry_action1.pdf}
       \;=\;
       \adjincludegraphics[scale=1.25,trim={10pt 10pt 10pt 10pt},valign = c]{tikz/out/symmetry_action3.pdf}
       \;=\;\frac{1}{\abs{G}}\;
       \adjincludegraphics[scale=1.25,trim={10pt 10pt 10pt 10pt},valign = c]{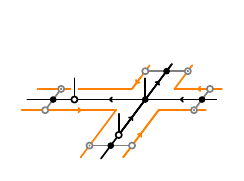},
\end{equation}
where we used the general property of the condensation surface \eqref{eq: sym frac condensation 2d} in the first equality.
The black dots in the last diagram represent the copy tensor and the gray dots are defined by
\begin{equation}\label{eq: S1 gray tensor}
       \adjincludegraphics[scale=1.2,trim={10pt 10pt 10pt 10pt},valign = c]{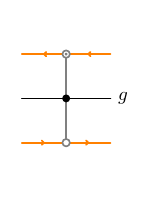}
       =
       \adjincludegraphics[scale=1.2,trim={10pt 10pt 10pt 10pt},valign = c]{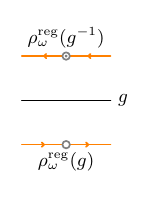}.
\end{equation} 
In particular, we can confirm that the ground state is invariant under the action of the condensation surface.

\subsubsection{Symmetry interpolation of invertible $0$-form symmetry in $2\mathrm{Rep}(G)$}
By using the doubled line notation, the invertible condensation surface can be broken up into local unitary operators $u^{(p)}_{\omega}$ acting around each plaquette $p$.
We can find an interpolation of the invertible condensation surface by interpolating between $u_{\omega}^{(p)}$ and the identity operator.
To find such an interpolation, let us first write down the action of $u_{\omega}^{(p)}$ explicitly.
The action of $u^{(p)}_{\omega}$ on the basis state is computed as 
\begin{align}
       u^{(p)}_{\omega}\Ket{\adjincludegraphics[scale=1,trim={10pt 20pt 10pt 20pt},valign = c]{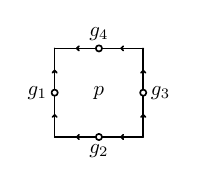}}
       &=\cfrac{1}{\abs{G}}\;
       \tr{[\rho^{\rm reg}_{\omega}(g_{3}^{-1})\rho^{\rm reg}_{\omega}(g_{4}^{-1})\rho^{\rm reg}_{\omega}(g_{1})\rho^{\rm reg}_{\omega}(g_{2})]}\Ket{\adjincludegraphics[scale=1,trim={10pt 20pt 10pt 20pt},valign = c]{tikz/out/flatness2.pdf}}\nonumber\\
       &\;\nonumber\\
       &=\cfrac{1}{\abs{G}}\;
       \frac{\omega(g_{1},g_{2})}{\omega(g_{4},g_{3})}\;\tr{[\rho^{\rm reg}_{\omega}(g_{3}^{-1}g_{4}^{-1})\rho^{\rm reg}_{\omega}(g_{1}g_{2})]}\Ket{\adjincludegraphics[scale=1,trim={10pt 20pt 10pt 20pt},valign = c]{tikz/out/flatness2.pdf}}\nonumber\\
       \;\nonumber\\
       &=
       \frac{\omega(g_{1},g_{2})}{\omega(g_{4},g_{3})}\Ket{\adjincludegraphics[scale=1,trim={10pt 20pt 10pt 20pt},valign = c]{tikz/out/flatness2.pdf}}.
\end{align}
Here, the second equality follows from $\omega(g_3^{-1}, g_4^{-1}) = \omega(g_4, g_3)^{-1}$, which is a consequence of our gauge choice of $\omega$, and the third equality follows from the flatness condition $g_1 g_2 = g_4 g_3$.
By using this expression, we can define an interpolation of the invertible condensation surface as follows:
\begin{equation}
       \mathcal{O}[A^{G}_{G,\omega}](\theta) \coloneq \prod_{p} u^{(p)}_{\omega}(\theta),
\end{equation}
where $u^{(p)}_{\omega}(\theta)$ is defined by
\begin{align}
       &\;\nonumber\\
       u^{(p)}_{\omega}(\theta)\Ket{\adjincludegraphics[scale=1,trim={10pt 20pt 10pt 20pt},valign = c]{tikz/out/flatness2.pdf}}
       &=
       \left[\frac{\omega(g_{1},g_{2})}{\omega(g_{4},g_{3})}\right]^{\frac{\theta}{2\pi}}\Ket{\adjincludegraphics[scale=1,trim={10pt 20pt 10pt 20pt},valign = c]{tikz/out/flatness2.pdf}}.\\
       \;\nonumber
\end{align}
Here, for a complex number $z\in \mathbb{C}$ with $\abs{z}=1$, we take the principal branch of the logarithm as $z=e^{i\phi}$ where $0\leq \phi < 2\pi$ and define $z^{\frac{\theta}{2\pi}} = e^{i\frac{\theta}{2\pi}\phi}$ for $\theta\in [0,2\pi]$.
The parameterized family of Hamiltonians is defined by conjugating the $G$-cluster Hamiltonian $H$ in Eq.~\eqref{eq: 2d G-cluster Hamiltonian} by $U_{\omega}(\theta)$:
\begin{equation}
       H_{\omega}(\theta) = \mathcal{O}[A^{G}_{G,\omega}](\theta) H \mathcal{O}[A^{G}_{G,\omega}](\theta)^{\dagger}.
\end{equation}
We note that the family has the $G$-symmetry for all $\theta\in[0,2\pi]$ since $\mathcal{O}[A^{G}_{G,\omega}](\theta)$ commutes with the generators of the $G$-symmetry.

\subsubsection{Pumped interface mode}
\label{sec: pumped interface mode S1}

Now, let us confirm that the above $S^{1}$-parameterized family is nontrivial by computing the pumped interface mode.
To this end, we use the adiabatic evolution argument.
Let us briefly review the argument here:
Consider an $S^{1}$-parameterized family $H(\theta)$ with $G$-symmetry..
For such a family, we consider a spatially inhomogeneous Hamiltonian, a textured Hamiltonian $H^{\rm text}(\theta)$.
Here, a textured Hamiltonian means a Hamiltonian whose parameters are modulated along one spatial direction (taken to be the $x$-axis below).
More precisely, $H^{\rm text}(\theta)$ is defined by the following properties in the regions near $x=0$, $x \gg 0$, and $x \ll 0$:
\begin{enumerate}
       \item In the region $x \gg 0$, the local terms of $H^{\rm text}(\theta)$ coincide with those of $H(\theta)$.
       \item In the region $x \ll 0$, the local terms of $H^{\rm text}(\theta)$ coincide with those of $H(0)$.
       \item In a neighborhood of $x=0$, the local terms of $H^{\rm text}(\theta)$ are chosen arbitrarily, as long as $H^{\rm text}(\theta)$ remains uniquely gapped and preserves the symmetry $G$.
\end{enumerate}
By adiabatically changing $\theta$ from $0$ to $2\pi$, the ground state of $H^{\rm text}(\theta)$ evolves continuously.
Eventually, the ground state returns to itself except for the region near $x=0$, but the state near $x=0$ may change nontrivially.
This change can be regarded as a pumped interface mode.
If the family $H(\theta)$ is nontrivial, then the pumped interface mode is also nontrivial.
We note that this argument can also be applied to the case where the symmetry is non-invertible.
In Sec.~\ref{sec: S1-parameterized families RepG} we present an analogous analysis for the $S^{1}$-family of $2\mathrm{Rep}(G)$-symmetric states.

Let us apply this argument to our $S^{1}$-parameterized family $H_{\omega}(\theta)$ with the $G$-symmetry.
We consider a textured Hamiltonian $H_{\omega}^{\rm text}(\theta)$ defined by
\begin{equation}
       H_{\omega}^{\rm text}(\theta) \coloneq \mathcal{O}[A^{G}_{G,\omega}]^{x>0}(\theta) H \mathcal{O}[A^{G}_{G,\omega}]^{x>0}(\theta)^{\dagger},
\end{equation}
where $\mathcal{O}[A^{G}_{G,\omega}]^{x>0}(\theta)$ is the truncated condensation surface defined by
\begin{equation}\label{eq: truncated condensation}
       \mathcal{O}[A^{G}_{G,\omega}]^{x>0}(\theta) \coloneq \prod_{p,x_{p}>0} u^{(p)}_{\omega}(\theta),
\end{equation}
where $x_p$ is the $x$-coordinate of the plaquette $p$.
This textured Hamiltonian satisfies the above three properties: the first and second properties are obvious from the definition, and the third property holds since the truncated condensation surface is unitary, and it commutes with the $G$-symmetry.
The ground state of $H_{\omega}^{\rm text}(\theta)$ is given by
\begin{equation}
       \ket{\mathrm{G.S.}_{\omega}^{\rm text}(\theta)} \coloneq \mathcal{O}[A^{G}_{G,\omega}]^{x>0}(\theta)\ket{\mathrm{G.S.}}.
\end{equation}
We refer to this state as the textured ground state.

\begin{figure}[t]
    \centering
    \adjincludegraphics[scale=1.4,trim={10pt 10pt 10pt 10pt},valign = c]{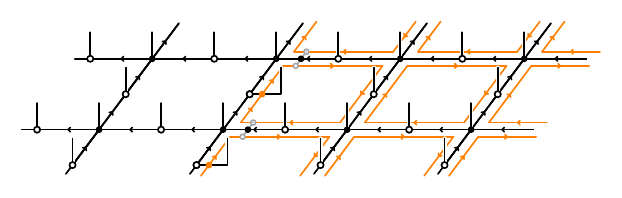}\\
    \caption*{(a)}
    \adjincludegraphics[scale=1.4,trim={10pt 10pt 10pt 10pt},valign = c]{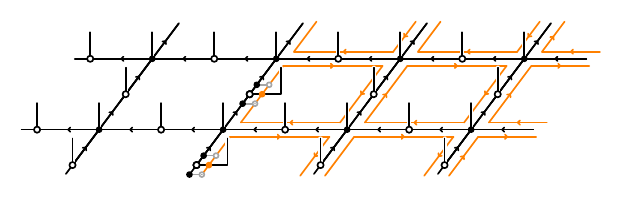}
    \caption*{(b)}
    \adjincludegraphics[scale=1.4,trim={10pt 10pt 10pt 10pt},valign = c]{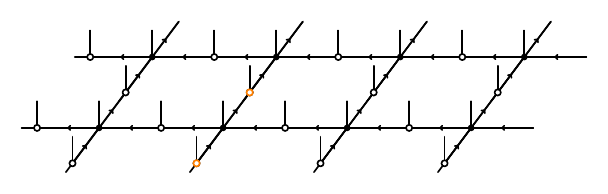}
    \caption*{(c)}
    \caption{
       Graphical calculation of the pumped interface mode for the $S^{1}$-parameterized family $H_{\omega}(\theta)$.
              (a) By using the property \eqref{eq: S1 symmetry action}, an invertible condensation surface is absorbed into the ground state away from the interface at $x=0$.
              The gray and black tensors defined in Eq.~\eqref{eq: S1 gray tensor} remain at each vertex tensor on $x=0$, and the orange tensors defined in Eq.~\eqref{eq: S1 orange tensors} remain on each link tensor on $x=0$.
              (b) Since the black tensor in Eq.~\eqref{eq: S1 gray tensor} is the copy tensor, we can split it into two parts and pull them through the vertex tensors.
              As a result, the nontriviality is concentrated on the link tensors at $x=0$.
              (c) By using the definition of the orange tensors in Eq.~\eqref{eq: S1 pumped interface mode}, we obtain the final form of the pumped interface mode.
    }
    \label{fig: S1 adiabatic evolution}
\end{figure}

To see the pumped interface mode, let us consider the ground state at $\theta=2\pi$.
By using the property \eqref{eq: S1 symmetry action}, one can show that the ground state at $\theta=2\pi$ is obtained from the ground state at $\theta=0$ by replacing the link tensors along the interface at $x=0$ with the following tensors:
\begin{equation}\label{eq: S1 pumped interface mode}
       \adjincludegraphics[scale=1.5,trim={10pt 10pt 10pt 10pt},valign = c]{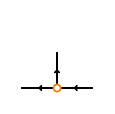}
       \;=\;\frac{1}{\abs{G}}\;
       \adjincludegraphics[scale=1.5,trim={10pt 10pt 10pt 10pt},valign = c]{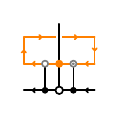}\;.
\end{equation}
See Fig.~\ref{fig: S1 adiabatic evolution} for the detailed graphical calculation.
It follows from a direct calculation that the above orange tensor is the twisted multiplication tensor:
\begin{equation}
       \adjincludegraphics[scale=1.5,trim={10pt 10pt 10pt 10pt},valign = c]{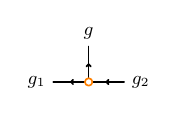}
       =\;
       \omega(g_{2}^{-1},g_{1})\delta_{g_{1}g,g_{2}}.
\end{equation}
Consequently, the pumped interface mode is given by the following MPS:
\begin{equation}
       \adjincludegraphics[scale=1.3,trim={10pt 10pt 10pt 10pt},valign = c]{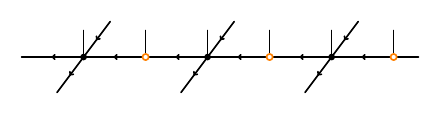}.
\end{equation}
We refer to this MPS as the pumped interface MPS.
In the above diagram, we will refer to the legs extending toward the front and back as the effective physical legs, the vertical leg as the original physical leg, and the horizontal leg as the virtual leg of the MPS.
We note that the pumped interface MPS is injective.

Let us compute the $G$-symmetry action on the pumped interface MPS.
To this end, we consider the action of the $G$-symmetry on the textured ground state at $\theta = 2\pi$ and derive the effective symmetry operators acting on the interface.
We first recall that the $G$-symmetry operator on the Hilbert space is given by
\begin{equation}
       U_{g} = \bigotimes_{i\in V} \overrightarrow{X}_{g}^{(i)}.
\end{equation}
Due to Eq.~\eqref{eq: line-like action tensor G},
the symmetry operator acting on the physical legs away from the interface reduces to the line-like action tensors \eqref{eq: line-like action tensor G2} acting on the effective physical legs of the pumped interface MPS.
On the other hand, for the physical legs at the interface, we keep the on-site symmetry operators $\overrightarrow{X}_g^{(i)}$ as they are.
Consequently, the total action of the $G$-symmetry on the pumped interface MPS is given by the following diagram:
\begin{equation}
       \adjincludegraphics[scale=1.3,trim={10pt 10pt 10pt 10pt},valign = c]{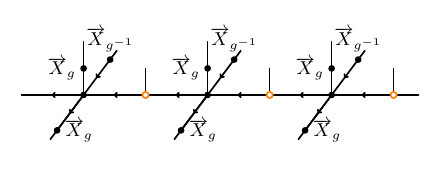}.
\end{equation}

In what follows, we show that the pumped interface MPS is invariant under the above $G$-symmetry action and has a nontrivial SPT invariant as a $G$-symmetric injective MPS.
First, let us show the invariance under the $G$-symmetry action.
To this end, we push the symmetry operators through the MPS tensors:
\begin{align}
       \begin{aligned}
       \adjincludegraphics[scale=1.25,trim={10pt 10pt 10pt 10pt},valign = c]{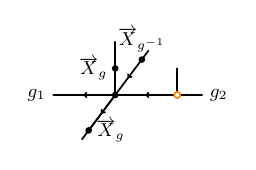}
       \;&=\;
       \adjincludegraphics[scale=1.25,trim={10pt 10pt 10pt 10pt},valign = c]{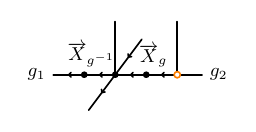}\\
       \;&=\;\omega(g_{2}^{-1},g_{1})\;
       \adjincludegraphics[scale=1.25,trim={10pt 10pt 10pt 10pt},valign = c]{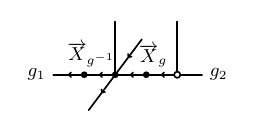}\\
       \;&=\;\omega(g_{2}^{-1},g_{1})\;
       \adjincludegraphics[scale=1.25,trim={10pt 10pt 10pt 10pt},valign = c]{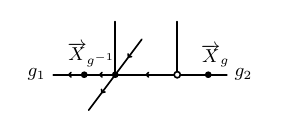}\\
       \;&=\;\frac{\omega(g_{2}^{-1},g_{1})}{\omega((gg_{2})^{-1},gg_{1})}\;
       \adjincludegraphics[scale=1.25,trim={10pt 10pt 10pt 10pt},valign = c]{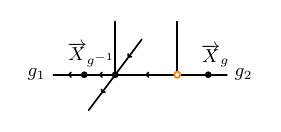}\\
       \;&=\omega(g^{-1},gg_{1})^{-1}\omega(g,g_{2})^{-1}\;
       \adjincludegraphics[scale=1.25,trim={10pt 10pt 10pt 10pt},valign = c]{tikz/out/TN_S1_pumped_interface_pulling_through5.pdf}\\
       \;&=
       \adjincludegraphics[scale=1.25,trim={10pt 10pt 10pt 10pt},valign = c]{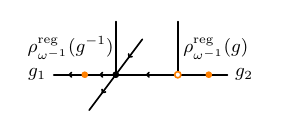},
       \end{aligned}
\end{align}
where $\rho^{\rm reg}_{\omega^{-1}}$ is the twisted regular representation defined by Eq.~\eqref{eq: twisted regular rep} with the twist $\omega^{-1}$. 
In the first equality, we used the $G$-invariance of the copy tensor, in the second and fourth equalities we used the definition of the pumped interface MPS tensor, and in the third equality we used the commutativity of the multiplication tensor with the left multiplication $\overrightarrow{X}_g$.
In the fifth equality, we used the cocycle condition on $\omega$.
Consequently, the pumped interface MPS with periodic boundary condition is invariant under the $G$-symmetry action.
The above equation shows that the symmetry action on the pumped interface MPS is fractionalized into the projective actions $\rho^{\rm reg}_{\omega^{-1}}$ on the virtual leg.
Thus, the SPT invariant of this $G$-symmetric MPS is given by $[\omega^{-1}] \in \mathrm{H}^{2}(G;\mathrm{U}(1))$.
In particular, when $[\omega]$ is nontrivial, the pumped interface mode is nontrivial, which implies that the $S^1$-family of Hamiltonians $H_{\omega}(\theta)$ is nontrivial.

\subsection{$S^{2}$-parameterized families}
\label{sec: S2-parameterized families}
In this section, we construct $S^{2}$-parameterized families of $G$-symmetric states by using the symmetry interpolation method.
From a categorical point of view, $S^{2}$-parameterized families are classified by $\mathrm{H}^{1}(G;\mathrm{U}(1))$ \cite{Thorngren:1612.00846, Hsin:2020cgg, Hermele2021CMSA,Thorngren2021YITP}.
Since each invertible $1$-form symmetry in $2\mathrm{Rep}(G)$ is labeled by an element of $\mathrm{H}^{1}(G;\mathrm{U}(1))$, it is natural to expect that an $S^{2}$-parameterized family corresponding to $\chi\in\mathrm{H}^{1}(G;\mathrm{U}(1))$ is constructed by interpolating the symmetry labeled by $\chi$.
However, the symmetry interpolation method introduced in Sec.~\ref{sec: Symmetry interpolation} is directly applicable to neither $S^2$-families nor 1-form symmetries. To circumvent this issue, we first construct two (trivial) $S^1$-families by applying the symmetry interpolation method to the product of Wilson loops labeled by $\chi$. By further interpolating between these two families, we will obtain an $S^2$-family with $G$-symmetry for each $\chi \in \mathrm{H}^1(G; \mathrm{U}(1))$.

An $S^{2}$-family can be regarded as an $S^{1}$-family of $S^{1}$-families.
Hence, by applying the adiabatic evolution argument to one of the $S^{1}$ parameters, we obtain an $S^{1}$-family of $G$-symmetric pumped interface modes.
The nontriviality of the $S^{2}$-family then reduces to the nontriviality of this $G$-symmetric $S^{1}$-family of pumped interface modes.
Therefore, by applying the adiabatic evolution argument once again to the $S^{1}$-family of interfaces, we confirm the nontriviality of the $S^2$-family of $G$-symmetric states.

\subsubsection{Interpolation of invertible $1$-form symmetry in $2\mathrm{Rep}(G)$}
Each invertible $1$-form symmetry in $2\mathrm{Rep}(G)$ is labeled by a character $\chi: G \to \mathrm{U}(1)$.
For each character $\chi$, we now construct an $S^2$-parameterized family of $G$-symmetric SRE states.
As mentioned above, an $S^2$-family is obtained by interpolating between two trivial $S^1$-families.
As such, we begin with the construction of the two $S^1$-families.

To construct the first $S^1$-family, we apply the symmetry interpolation method to the following operator:
\begin{equation}
\mathrm{id} = \prod_{p} u_{\chi}^{(p)}.
\label{eq: Wilson loop decomposition}
\end{equation}
Here, the tensor product is taken over all plaquettes, and $u_{\chi}^{(p)}$ is the Wilson loop operator around the plaquette $p$, i.e.,
\begin{equation}
u_{\chi}^{(p)} \coloneq Z_{\chi}^{\dagger(\ell_{b})}Z_{\chi}^{\dagger(\ell_{l})}Z_{\chi}^{(\ell_{t})}Z_{\chi}^{(\ell_{r})},
\end{equation}
where $\ell_{t},\ell_{b},\ell_{l},\ell_{r}$ are the top, bottom, left, and right links surrounding the plaquette $p$, respectively.
Since the operator~\eqref{eq: Wilson loop decomposition} is the product of local unitary operators, we can apply the symmetry interpolation method in Sec.~\ref{sec: Symmetry interpolation} to this operator.
Specifically, to obtain an $S^1$-family, we interpolate between the Wilson loop $u_{\chi}^{(p)}$ and the identity operator as
\begin{equation}
v_{\chi}^{(p)}(r) \coloneq [Z_{\chi}^{(\ell_{b})}]^{-r}[Z_{\chi}^{(\ell_{l})}]^{-r}[Z_{\chi}^{(\ell_{t})}]^{r}[Z_{\chi}^{(\ell_{r})}]^{r}, \qquad
r \in [0, 1].
\end{equation}
Here, the operator $[Z_{\chi}^{(\ell)}]^r$ and $[Z_{\chi}^{(\ell)}]^{-r}$ are defined by
\begin{equation}
[Z_{\chi}^{(\ell)}]^{r}\ket{g} \coloneq \chi(g)^{r}\ket{g}, \qquad
[Z_{\chi}^{(\ell)}]^{-r}\ket{g} \coloneq \chi(g)^{-r}\ket{g},
\end{equation}
where $\chi(g)^{r} = e^{i \phi r}$ and $\chi(g)^{-r} = e^{-i \phi r}$ for $\chi(g)=e^{i\phi}$ with $0 \leq \phi < 2 \pi$.
Using the above interpolation, we can construct an $S^1$-family of Hamiltonians as
\begin{equation}
H_{\chi}^{V}(r) \coloneq V(r) H V(r)^{\dagger}, \qquad
V_{\chi}(r) \coloneq \prod_{p} v_{\chi}^{(p)}(r),
\label{eq: HV}
\end{equation}
where $H$ is the Hamiltonian~\eqref{eq: 2d G-cluster Hamiltonian} of the 2+1d $G$-cluster model.
We note that $V_{\chi}(r) = 1$ for all $r \in [0, 1]$ on the lattice with periodic boundary conditions because the contributions from adjacent plaquettes cancel out.
Therefore, the above $S^1$-family is trivial, i.e., $H_{\chi}^V(r) = H$ for all $r \in [0, 1]$.
We will use this trivial family later to construct a nontrivial $S^2$-family.

To construct the second $S^1$-family, we again consider the same decomposition~\eqref{eq: Wilson loop decomposition} of the identity operator, but take a different interpolation.
Specifically, on the plaquette $p$ with $x$-coordinate $x_p$ being even, we interpolate between the Wilson loop $u_{\chi}^{(p)}$ and the identity operator as
\begin{equation}
w_{\chi}^{(p)+}(s) \coloneq \exp(s \log u_{\chi}^{(p)}), \qquad s \in [0, 1].
\label{eq: wp plus}
\end{equation}
On the other hand, on the plaquette $p$ with the $x$-coordinante $x_p$ being odd, we interpolate between $u_{\chi}^{(p)}$ and the identity operator as
\begin{equation}
w_{\chi}^{(p)-}(s) \coloneq \exp(-s \log u_{\chi}^{(p)\dagger}), \qquad s \in [0, 1].
\label{eq: wp minus}
\end{equation}
Here, we recall that the logarithm of an operator $X$ satisfying $X^N = \mathrm{id}$ for a positive integer $N$ is defined by\footnote{In particular, when $X = u_{\chi}^{(p)}$ or $u_{\chi}^{(p) \dagger}$, we take $N = \abs{\mathrm{H}^1(G; \mathrm{U}(1)}$.}
\begin{equation}
\log X \coloneq \sum_{n = 0}^{N-1} i\frac{2\pi n}{N} P_n[X],
\end{equation}
where $P_n[X] \coloneq \frac{1}{N} \sum_{m=0}^{N-1} e^{-i\frac{2\pi nm}{N}} X^m$ is the projection to the eigenspace of $X$ with eigenvalue $e^{i\frac{2\pi n}{N}}$.
The reason to choose different interpolations depending on the $x$-coordinate will be clarified later; see footnote~\ref{fn: wpm}.
By using the interpolations~\eqref{eq: wp plus} and \eqref{eq: wp minus}, we can construct an $S^1$-family of Hamiltonians as
\begin{equation}
H_{\chi}^W(s) \coloneq W_{\chi}(s) H W_{\chi}(s)^{\dagger}, \qquad
W_{\chi}(s) \coloneq \prod_{p, x_p{\rm : even}} w_{\chi}^{(p)+}(s) \prod_{p, x_p{\rm : odd}} w_{\chi}^{(p)-}(s),
\label{eq: HW}
\end{equation}
where $H$ is again the $G$-cluster Hamiltonian~\eqref{eq: 2d G-cluster Hamiltonian}.
We note that the above family is also trivial, i.e., $H_{\chi}^W(s) = H$ for all $s \in [0, 1]$, because $w_{\chi}^{(p)+}(s)$ and $w_{\chi}^{(p)-}(s)$ are polynomials of the Wilson loop $u_{\chi}^{(p)}$ and hence commute with the $G$-cluster Hamiltonian $H$.

Now, we construct an $S^2$-family by interpolating between the above two trivial $S^1$-families.
To this end, we first show that such an interpolation exists. 
More specifically, we show that there exists an interpolation between one-parameter families of local unitary operators $\{v_{\chi}^{(p)}(r) v_{\chi}^{(p^{\prime})}(r)\}_{r \in [0, 1]}$ and $\{w_{\chi}^{(p)+}(s) w_{\chi}^{(p^{\prime})-}(s)\}_{s \in [0, 1]}$.
Here, $p$ and $p^{\prime}$ are adjacent plaquettes whose $x$-coordinates are even and odd, respectively.
An interpolation between these families exists if and only if the following $S^1$-family obtained by concatenating them is contractible in the space of unitary operators:
\begin{equation}
u_{\chi}^{(p, p^{\prime})}(t) = 
\begin{cases}
v_{\chi}^{(p)}(2t) v_{\chi}^{(p^{\prime})}(2t), \qquad & 0 \leq t \leq 1/2, \\
w_{\chi}^{(p)+}(2-2t) w_{\chi}^{(p^{\prime})-}(2-2t), \qquad & 1/2 \leq t \leq 1.
\end{cases}
\end{equation}
In general, an $S^1$-family of unitary operators is contractible if and only if their determinants have no winding number.
For the unitary operators defined above, the winding number is zero because the determinant is constant:
\begin{equation}
\mathrm{det} (u_{\chi}^{(p, p^{\prime})}(t)) = 1, \qquad \forall t \in [0, 1].
\label{eq: det}
\end{equation}
For $t \in [0, 1/2]$, the above equation follows from $\mathrm{det} (v_{\chi}^{(p)}(r)) = 1$, which holds because $Z_{\chi}^{r}$ and $Z_{\chi}^{-r}$ have the opposite determinants. 
On the other hand, for $t \in [1/2, 1]$, the above equation follows from $\mathrm{det} (w_{\chi}^{(p)+}(s)) = (\mathrm{det} (w_{\chi}^{(p)-}(s)))^{-1}$, which holds because $u_{\chi}^{(p)}$ and $u_{\chi}^{(p)\dagger}$ have the same set of eigenvalues.\footnote{We note that each of $\mathrm{det} (w_{\chi}^{(p)+}(s))$ and $\mathrm{det} (w_{\chi}^{(p)-}(s))$ is generally not equal to one and may contribute to the winding number. In particular, each of the one-parameter families $\{w_{\chi}^{(p)+}(s)\}_{s \in [0, 1]}$ and $\{w_{\chi}^{(p)-}(s)\}_{s \in [0, 1]}$ may not be continuously deformed into $\{v_{\chi}^{(p)}(r)\}_{r \in [0, 1]}$. \label{fn: wpm}}
Equation~\eqref{eq: det} shows that the $S^1$-family of unitary operators $\{u_{\chi}^{(p, p^{\prime})}(t)\}_{t \in [0, 1]}$ is contractible.
Thus, there exists an interpolation between one-parameter families of unitary operators $\{v_{\chi}^{(p)}(r) v_{\chi}^{(p^{\prime})}(r)\}_{r \in [0, 1]}$ and $\{w_{\chi}^{(p)+}(s) w_{\chi}^{(p^{\prime})-}(s)\}_{s \in [0, 1]}$.
Such an interpolation is unique up to homotopy because the second homotopy group of $SU(n)$ vanishes for any $n\geq2$.

\begin{figure}[t]
       \begin{minipage}{0.5\textwidth}
           \centering
           \begin{equation*}
              \adjincludegraphics[scale=1.5,trim={10pt 10pt 10pt 10pt},valign = c]{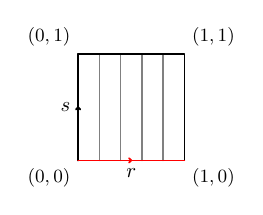}
           \end{equation*}
           \caption*{(a)}    
       \end{minipage}
       \hfill 
       \begin{minipage}{0.5\textwidth}
           \centering
           \begin{equation*}
              \adjincludegraphics[scale=1.5,trim={10pt 10pt 10pt 10pt},valign = c]{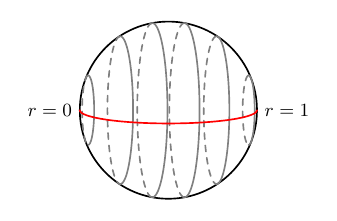}
           \end{equation*}
           \caption*{(b)}
       \end{minipage}
       \caption{
              The parameter spaces of $U_{\chi}(r,s)$ and $H_{\chi}(r,s)$. 
              (a) The parameter space of $U_{\chi}(r,s)$ is a square $[0, 1] \times [0, 1]$. On the bottom and left edges, we have $U_{\chi}(r, 0) = U_{\chi}(0, s) = \mathrm{id}$. On the top and right edges, we have $U_{\chi}(r, 1) = V_{\chi}(r)$ and $U_{\chi}(1, s) = W_{\chi}(s)$, respectively. In particular, $U_{\chi}(r, s)$ on the boundary of the square commutes with the $G$-cluster Hamiltonian $H$.
              (b) The parameter space of $H_{\chi}(r, s)$ is a sphere $S^2$ because the boundary of the square collapses into a single point after conjugating the Hamiltonian $H$ by $U_{\chi}(r, s)$. The gray lines represent the lines with constant $r$, while the red line represents the line with $s=0$. All points on the red line are identified because $U_{\chi}(r, 0) = \mathrm{id}$ for all $r \in [0, 1]$.
       }
       \label{fig: 2-parameter sq}
\end{figure}

The existence of the above interpolation guarantees that there exists a two-parameter family of unitary operators $\{u_{\chi}^{(p, p^{\prime})}(r, s) \mid r, s \in [0, 1]\}$ such that
\begin{equation}
\begin{aligned}
u_{\chi}^{(p, p^{\prime})}(r, 0) &= \mathrm{id}, &\quad
u_{\chi}^{(p, p^{\prime})}(0, s) &= \mathrm{id}, \\
u_{\chi}^{(p, p^{\prime})}(r, 1) &= v_{\chi}^{(p)}(r) v_{\chi}^{(p^{\prime})}(r), &\quad
u_{\chi}^{(p, p^{\prime})} (1, s) &= w_{\chi}^{(p)+}(s) w_{\chi}^{(p^{\prime})-}(s).
\end{aligned}
\label{eq: boundary unitary}
\end{equation}
The above conditions uniquely determine the unitary operators in the interior of the parameter space up to homotopy because the interpolation is unique up to homotopy.
Using the above two-parameter family of unitary operators, we can construct a two-parameter family of Hamiltonians as
\begin{equation}
H_{\chi}(r, s) \coloneq U_{\chi}(r, s) H U_{\chi}(r, s)^{\dagger}, \qquad
U_{\chi}(r, s) \coloneq \prod_{p, x_p{\rm : even}} u_{\chi}^{(p, p^{\prime})}(r, s).
\end{equation}
Due to Eq.~\eqref{eq: boundary unitary}, the unitary operators $U_{\chi}(r, s)$ on the boundary of the parameter space are given by
\begin{equation}
U_{\chi}(r, 0) = U_{\chi}(0, s) = \mathrm{id}, \quad
U_{\chi}(r, 1) = V_{\chi}(r) = \mathrm{id}, \quad U_{\chi}(1, s) = W_{\chi}(s).
\end{equation}
Since all of these operators commute with $H$, the Hamiltonians $H_{\chi}(r, s)$ on the boundary of the parameter space reduce to
\begin{equation}
H_{\chi}(r, 0) = H_{\chi}(0, s) = H_{\chi}(r, 1) = H_{\chi}(1, s) = H.
\end{equation}
Namely, all points on the boundary of the parameter space are identified.
Therefore, the parameter space of the model reduces to a sphere $S^2$; see Fig.~\ref{fig: 2-parameter sq} for a graphical representation of the parameter space.
We note that the Hamiltonian $H_{\chi}(r, s)$ has the $G$-symmetry for all $r$ and $s$ because $U_{\chi}(r, s)$ acts only on the Hilbert space on the edges and thus commutes with the $G$-symmetry.

\subsubsection{Pumped interface mode}

We confirm the nontriviality of the above $S^{2}$-parameterized family $H_{\chi}(r,s)$ by means of the adiabatic evolution argument.
For each fixed $r\in[0,1]$, the family $\{H_{\chi}(r, s) \mid s \in [0, 1]\}$ is an $S^{1}$-parameterized family of $G$-symmetric Hamiltonians.
Thus, by applying the adiabatic evolution argument to $s$ first, we obtain an $S^{1}$-family of $G$-symmetric pumped interface modes parameterized by $r$.
To see this, let us  take a textured Hamiltonian $H_{\chi}^{\rm text}(r,s)$ defined by
\begin{equation}
       H_{\chi}^{\rm text}(r,s) \coloneq U_{\chi}^{x>0}(r,s) H U_{\chi}^{x>0}(r, s)^{\dagger},
\end{equation}
where $U_{\chi}^{x>0}(r,s)$ is defined by
\begin{equation}
       U_{\chi}^{x>0}(r,s) \coloneq \prod_{p; x_{p}>0, {\rm even}} u_{\chi}^{(p,p')}(r,s).
\end{equation}
This textured Hamiltonian satisfies the three properties listed in Sec.~\ref{sec: pumped interface mode S1}.
The ground state of $H_{\chi}^{\rm text}(r,s)$ is given by
\begin{equation}
       \ket{\mathrm{G.S.}_{\chi}^{\rm text}(r,s)} \coloneq U_{\chi}^{x>0}(r,s)\ket{\mathrm{G.S.}}.
\end{equation}
We refer to this state as the textured ground state.

To see the pumped interface mode, let us consider the ground state at $s=1$.
By definition, the truncated unitary operator $U_{\chi}^{x>0}(r, s)$ at $s=1$ is given by the product of the loop operators $v_{\chi}^{(p)}(r)$ on the right-half plane.
Since the product of $v_{\chi}^{(p)}(r)$ and $v_{\chi}^{(p^{\prime})}(r)$ on the adjacent plaquettes $p$ and $p^{\prime}$ cancel out on the shared edge, this operator reduces to the product of $(Z_{\chi}^{(l)})^r$ on the links along the interface at $x=0$.
Therefore, the ground state at $s=1$ is given by
\begin{equation}
       \ket{\mathrm{G.S.}_{\chi}^{\rm text}(r,s=1)} = \bigotimes_{\ell, x_{\ell}=0} (Z_{\chi}^{(\ell)})^{r} \ket{\mathrm{G.S.}}.
\end{equation}
Thus, the pumped interface mode is given by the following injective MPS:
\begin{equation}\label{eq: pumped interface S2}
       \ket{\psi_{\chi,\mathrm{interface}}(r)}\coloneq\adjincludegraphics[scale=1.5,trim={10pt 10pt 10pt 10pt},valign = c]{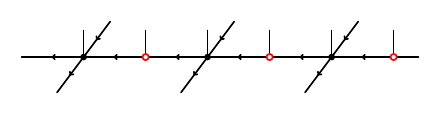},
\end{equation}
where the MPS tensor is defined by
\begin{equation}
       \adjincludegraphics[scale=1.25,trim={10pt 10pt 10pt 10pt},valign = c]{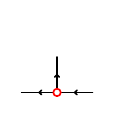}
       =
       \adjincludegraphics[scale=1.25,trim={10pt 10pt 10pt 10pt},valign = c]{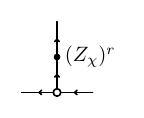}.
\end{equation}
We refer to this MPS as the pumped interface MPS.

We note that the pumped interface MPS \eqref{eq: pumped interface S2} on a periodic chain is periodic in $r\in[0,1]$ and invariant under the $G$-symmetry action for all $r\in[0,1]$.
First, we confirm the periodicity with respect to $r$.
By using Eq.~\eqref{eq: pulling through repG}, one can push the operator $(Z_{\chi}^{(\ell)})^{r}$ at $r=1$ as follows:
\begin{align}\label{eq: S2 pulling through}
       \;\nonumber\\
       \adjincludegraphics[scale=1.25,trim={10pt 10pt 10pt 10pt},valign = c]{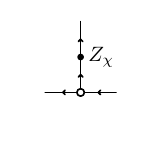}
       =
       \adjincludegraphics[scale=1.25,trim={10pt 10pt 10pt 10pt},valign = c]{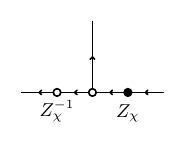}.
\end{align}
This shows that the MPS tensors at $r = 0$ and $r = 1$ are related by the gauge transformation.
Therefore, on a periodic chain, the pumped interface MPS at $r=1$ is equal to that at $r=0$.
Next, we show that the pumped interface MPS is $G$-symmetric.
To this end, let us compute the $G$-symmetry action on the pumped interface MPS.
The action of the symmetry on the interface is given by the following operator, as in the case of the $S^{1}$-family:
\begin{equation}\label{eq: S2 G-action}
       \adjincludegraphics[scale=1.25,trim={10pt 10pt 10pt 10pt},valign = c]{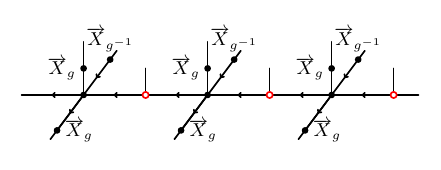}.
\end{equation}
Due to Eq.~\eqref{eq: pulling through G 2d}, the operators $\overrightarrow{X}_g$ and $\overrightarrow{X}_{g^{-1}}$ on the original and effective physical legs can be pushed onto the virtual legs. These operators then cancel out on the virtual legs because they commute with the red tensors in Eq.~\eqref{eq: S2 G-action}. Thus, the pumped interface MPS~\eqref{eq: pumped interface S2} is invariant under the action of the $G$-symmetry.

Now, we have obtained an $S^{1}$-family of $G$-symmetric injective MPSs parameterized by $r\in[0,1]$.
We can confirm the nontriviality of this $S^{1}$-family by applying the adiabatic evolution argument once again.
To see this, we consider the following textured states associated with this $S^{1}$-family\footnote{
       Strictly speaking, to apply the adiabatic evolution argument, we need to construct a family of 1+1-dimensional $G$-symmetric Hamiltonians whose ground state is given by Eq.~\eqref{eq: pumped interface S2}, since in the discussion of the adiabatic evolution, we have assumed the existence of Hamiltonians.
       However, to avoid complicating the discussion, we omit the explicit construction of Hamiltonians here.
       In general, for an injective MPS, there exists a gapped Hamiltonian for which that state is the unique ground state~\cite{Perez-Garcia:2006nqo}.
       Thus, from the above family of injective MPSs \eqref{eq: pumped interface S2}, we can construct a one-parameter family of parent Hamiltonians.
       The state \eqref{eq: textured interface} can be regarded as a textured ground state of this family of parent Hamiltonians.
       Moreover, the nontriviality of this $S^{1}$-family of $G$-symmetric injective MPSs can also be verified without using the adiabatic evolution argument, by employing the method of Ref.~\cite{Shiozaki:2021weu}.
       }:
\begin{equation}\label{eq: textured interface}
       \ket{\psi_{\chi,\mathrm{interface}}^{\rm text}(r)} \coloneq \bigotimes_{\ell,y_{\ell}>0} (Z_{\chi}^{(\ell)})^{r} \ket{\psi_{\chi,\mathrm{interface}}(r)}.
\end{equation}
Diagrammatically, this state is represented as
\begin{equation}
       \ket{\psi_{\chi,\mathrm{interface}}^{\rm text}(r)} = \adjincludegraphics[scale=1.25,trim={10pt 10pt 10pt 10pt},valign = c]{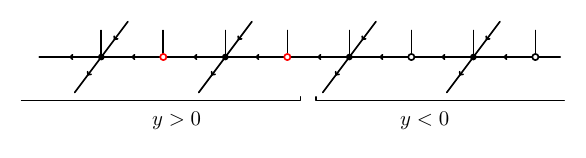}.
\end{equation}
By using Eq.~\eqref{eq: S2 pulling through}, one can see that the operator $Z_{\chi}$ appears on the virtual leg as a pumped interface mode at $r=1$:
\begin{equation}
       \ket{\psi_{\chi,\mathrm{interface}}^{\rm text}(1)} = \adjincludegraphics[scale=1.25,trim={10pt 10pt 10pt 10pt},valign = c]{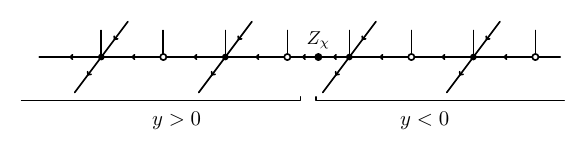}.
\end{equation}
The nontriviality of the $S^{1}$-family with $G$-symmetry can be verified by showing that this pumped interface mode carries a nontrivial charge with respect to the $G$-symmetry action on the interface.
By applying the $G$ symmetry operator defined in Eq.~\eqref{eq: S2 G-action} to $\ket{\psi_{\chi,\mathrm{interface}}^{\rm text}(r=1)}$, we find that the symmetry action on the interface mode $Z_{\chi}$ is given by
\begin{equation}
    Z_{\chi} \longmapsto \overrightarrow{X}_{g}\, Z_{\chi}\, \overrightarrow{X}_{g}^{\dagger}= \chi(g)^{-1} Z_{\chi}.
\end{equation}
This shows that the pumped interface mode at $r=1$ carries the nontrivial charge $\chi^{-1}$ with respect to the $G$-symmetry action on the interface.
Consequently, we have confirmed the nontriviality of the $S^{2}$-parameterized family $H_{\chi}(r, s)$ by applying the adiabatic evolution argument twice.

\section{Parameterized family of $2\mathrm{Rep}(G)$-symmetric states}
\label{sec: Parametrized family of 2Rep(G)-symmetric states}
From general considerations, it is expected that $S^{n}$-parameterized families of $2\mathrm{Rep}(G)$-symmetric SRE states are classified by $G$ for $n=1$, and that the classification is trivial for all $n\geq 2$ up to the higher Berry phase \cite{Inamura:2024jke}.
In this section, we construct all $S^{1}$-parameterized families of $2\mathrm{Rep}(G)$-symmetric SRE states by applying the symmetry interpolation method to the $0$-form symmetry $G$ in the 2+1d $G$-cluster state.

\subsection{$S^{1}$-parameterized families}
\label{sec: S1-parameterized families RepG}
In this subsection, we will construct $S^1$-parameterized families of $2\mathrm{Rep}(G)$-symmetric SRE states using the symmetry interpolation method.
We will also show the nontriviality of these families using the adiabatic evolution argument.

\subsubsection{Model}
We first define an $S^1$-parameterized family of $2\mathrm{Rep}(G)$-symmetric models for each symmetry operator labeled by $g \in G$.
Recall that the $G$-symmetry of the $G$-cluster state is given by 
\begin{equation}
       U_{g} = \bigotimes_{i\in V} \overrightarrow{X}_{g}^{(i)}.
\end{equation}
As an interpolation of the $G$-symmetry, we consider the same interpolation as in Eq.~\eqref{eq: S1 interpolation}:
\begin{equation}\label{eq: S1 interpolation RepG}
       U_{g}(\theta)=\bigotimes_{i\in V}\overrightarrow{X}_{g}^{(i)}(\theta),\;\;\overrightarrow{X}_{g}^{(i)}(\theta)\coloneq\exp{[\frac{\theta}{2\pi}\log\overrightarrow{X}_{g}^{(i)}]}.
\end{equation}
See Eq.~\eqref{eq: log X} for the definition of $\log\overrightarrow{X}_{g}^{(i)}$.
We note that $U_{g}(\theta)$ commutes with the $2 \mathrm{Rep}(G)$ symmetry for all $\theta\in[0,2\pi]$.
Then, we define an $S^{1}$-parameterized family of $2\mathrm{Rep}(G)$-symmetric Hamiltonians by conjugating the $G$-cluster Hamiltonian $H$ in Eq.~\eqref{eq: 2d G-cluster Hamiltonian} by $U_{g}(\theta)$:
\begin{equation}
       H_{g}(\theta) = U_{g}(\theta) H U_{g}(\theta)^{\dagger}.
\end{equation}
The family of unitary operators $U_{g}(\theta)$ is not $2\pi$-periodic, but the Hamiltonian $H_{g}(\theta)$ is $2\pi$-periodic since $U_{g}(2\pi)=U_{g}$ is the $G$-symmetry of $H$.
The ground state of $H_{g}(\theta)$ is given by
\begin{equation}\label{eq: 2Rep(G) GS S1}
       \ket{\mathrm{G.S.}_{g}(\theta)} = U_{g}(\theta)\ket{\mathrm{G.S.}}.
\end{equation}

\subsubsection{Pumped interface mode}
\label{sec: pumped interface mode S1 RepG}

Let us confirm the nontriviality of the above $S^{1}$-parameterized family $H_{g}(\theta)$ by applying the adiabatic evolution argument.
We consider a textured Hamiltonian $H_{g}^{\rm text}(\theta)$ defined by
\begin{equation}
       H_{g}^{\rm text}(\theta) \coloneq U_{g}^{x>0}(\theta) H U_{g}^{x>0}(\theta)^{\dagger},
\end{equation}
where $U_{g}^{x>0}(\theta)$ is defined by
\begin{equation}
       U_{g}^{x>0}(\theta) \coloneq \bigotimes_{i,x_{i}>0} \overrightarrow{X}_{g}^{(i)}(\theta).
\end{equation}
We note that the above textured Hamiltonian satisfies the three properties listed in Sec.~\ref{sec: pumped interface mode S1}, where the $G$-symmetry in the last property is now replaced with the $2\mathrm{Rep}(G)$-symmetry.
The ground state of $H_{g}^{\rm text}(\theta)$ is given by
\begin{equation}
       \ket{\mathrm{G.S.}_{g}^{\rm text}(\theta)} \coloneq U_{g}^{x>0}(\theta)\ket{\mathrm{G.S.}}.
\end{equation}

To see the pumped interface mode, let us consider the ground state at $\theta=2\pi$.
Due to the $G$-invariance~\eqref{eq: pulling through G 2d} of the copy tensor, the on-site operator $\overrightarrow{X}_g^{(i)}(\theta)$ at $\theta = 2\pi$ can be turned into $\overrightarrow{X}_g$ and $\overrightarrow{X}_{g^{-1}}$ acting on the virtual legs. These operators cancel out in pairs except for those on the interface. Thus, there remains the left multiplication operator $\overrightarrow{X}_{g}$ on each virtual leg along the interface at $x=0$:
\begin{equation}\label{eq: 2Rep(G) text GS}
\begin{aligned}
       \ket{\mathrm{G.S.}_{g}^{\rm text}(2\pi)} = 
       \adjincludegraphics[scale=1.25,trim={10pt 10pt 10pt 10pt},valign = c]{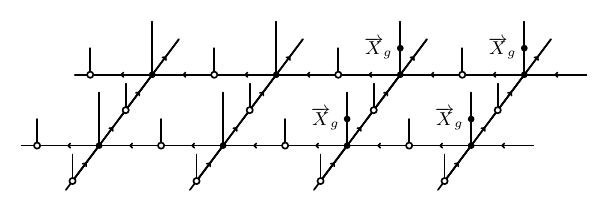}\\
       =
       \adjincludegraphics[scale=1.25,trim={10pt 10pt 10pt 10pt},valign = c]{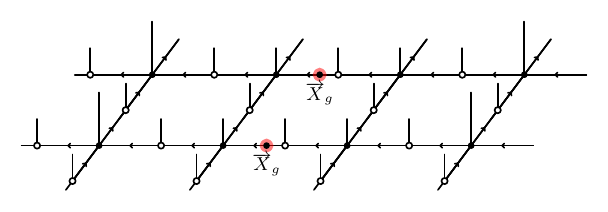}.
\end{aligned}
\end{equation}
The above equation shows that the pumped interface mode at $\theta = 2\pi$ is given by
\begin{equation}\label{eq: 2Rep(G) pumped interface}
\ket{\psi_{g, {\rm interface}}} = \bigotimes \overrightarrow{X}_g = \bigotimes \left( \sum_{h \in G} \ket{gh} \bra{h} \right),
\end{equation}
where the tensor product is taken over all virtual legs along the interface.

Let us consider the action of the $2 \mathrm{Rep}(G)$ symmetry on the interface mode.
The generators of the interface symmetry can be expressed as follows using the line-like action tensor defined in Eq.~\eqref{eq: line-like action tensor RepG}:
\begin{equation}\label{eq: 2Rep(G) interface sym}
       (\mathcal{O}[A])_{rr'} 
       \coloneq 
       \adjincludegraphics[scale=1.25,trim={10pt 10pt 10pt 10pt},valign = c]{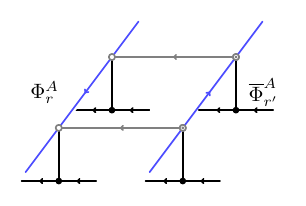}
       =
       \adjincludegraphics[scale=1.25,trim={10pt 10pt 10pt 10pt},valign = c]{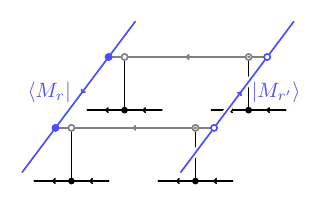}.
\end{equation}
Here, $A$ is a symmetric special Frobenius algebra in $2\mathrm{Rep}(G)$ and $r,r'$ are the labels of the SSB sectors.
These operators can be regarded as the symmetry operators acting on the interface Hilbert space $\bigotimes \mathbb{C}[G]\otimes\mathbb{C}[G]^{\ast}$.
We will refer to the Hilbert space $\mathbb{C}[G]\otimes\mathbb{C}[G]^{\ast}$ as the interface local Hilbert space and denote its basis by $\{\ket{g}\bra{g'}|g,g'\in G\}$.
The symmetry generated by the above operators is an example of the so-called strip 2-algebra discussed in \cite{Cordova:2024iti}.
We do not analyze the full symmetry structure explicitly here; the structure of the interface symmetry, including more general cases, is discussed in \cite{Inamura:clusterinterface}.\footnote{The interface symmetry in this case is described by the multifusion 1-category $\mathrm{Vec}^{\oplus |G|}$ \cite{Inamura:clusterinterface}, which is consistent with the fusion rules~\eqref{eq: interface fusion rules} of the interface symmetry operators.}

Let us show that the pumped interface mode~\eqref{eq: 2Rep(G) pumped interface} is a one-dimensional representation of the interface symmetry.
To this end, we compute the action of the interface symmetry operators~\eqref{eq: 2Rep(G) interface sym} on the pumped interface mode~\eqref{eq: 2Rep(G) pumped interface}.
We first focus on the case where $\mathcal{O}[A]$ is the full condensation surface, i.e., $A = A^G_{\{e\}, 1} = \mathbb{C}[G]^{\ast}$.
In this case, each simple module MPS in Eq.~\eqref{eq: 2Rep(G) interface sym} is labeled by a group element $g \in G$, and the interface symmetry operator $(\mathcal{O}[A^G_{\{e\}, 1}])_{g_1, g_2}$ is the tensor product of operators acting on each interface local Hilbert space:
\begin{equation}\label{eq: interface sym full cond}
       (\mathcal{O}[A^{G}_{\{e\},1}])_{g_{1},g_{2}} = \bigotimes_{i} (\mathcal{O}[A^{G}_{\{e\},1}]^{(i)})_{g_{1},g_{2}}.
\end{equation}
One can easily see that the action of the interface symmetry on each interface local Hilbert space is given by
\begin{equation}\label{eq: onsite interface sym full cond}
       (\mathcal{O}[A^{G}_{\{e\},1}]^{(i)})_{g_{1},g_{2}} \ket{g_{3}}\bra{g_{4}}_{i} = \delta_{g_{3}g_{4}^{-1},g_{1}g_{2}^{-1}}\ket{g_{3}}\bra{g_{4}}_{i}.
\end{equation}
Therefore, the symmetry operator~\eqref{eq: interface sym full cond} acts on the pumped interface mode~\eqref{eq: 2Rep(G) pumped interface} as follows:
\begin{equation}\label{eq: interface rep 1}
(\mathcal{O}[A^G_{\{e\}, 1}])_{g_1, g_2} \ket{\psi_{g, {\rm interface}}} = \delta_{g, g_1g_2^{-1}} \ket{\psi_{g, {\rm interface}}}.
\end{equation}
Similarly, one can also compute the actions of the other interface symmetry operators as
\begin{equation}\label{eq: interface rep 2}
       (\mathcal{O}[A^{G}_{H,\omega}])_{r_{1},r_{2}}\ket{\psi_{g, {\rm interface}}} = \delta_{r_{1},r'}\ket{\psi_{g, {\rm interface}}},
\end{equation}
where $r'\in S_{H\backslash G}$ is the unique element satisfying $hr' = r_{2}g^{-1}$ for a unique $h\in H$.
       Indeed, applying $(\mathcal{O}[A^{G}_{H,\omega}])_{r_{1},r_{2}}$ to $\ket{\psi_{g, {\rm interface}}}$ yields a state proportional to $\ket{\psi_{g, {\rm interface}}}$, with proportionality constant $\bra{M_{r_{1}}}\bigotimes \rho^{G}_{H,\omega}(g)\ket{M_{r_{2}}}$.
       Using the action of the $G$ symmetry on $\ket{M_r}$ given in Eq.~\eqref{eq: symmetry action on module MPS} together with the orthonormality of $\ket{M_r}$ in Eq.~\eqref{eq: orthonormality of simple module MPS}, we find that this factor is $\delta_{r_{1},\,r'}$, where $r'\in S_{H\backslash G}$ is the unique element such that $h r' = r_{2} g^{-1}$ for a unique $h\in H$.
Equations~\eqref{eq: interface rep 1} and \eqref{eq: interface rep 2} show that all symmetry operators on the interface act as either 0 or 1 on the pumped interface mode~\eqref{eq: 2Rep(G) pumped interface}.
In particular, the pumped interface mode~\eqref{eq: 2Rep(G) pumped interface} is a one-dimensional representation of the interface symmetry.

Now, we show that the pumped interface modes of the form~\eqref{eq: 2Rep(G) pumped interface} exhaust all one-dimensional representations of the interface symmetry up to continuous deformation.
To this end, we first enumerate all one-dimensional representations of the interface symmetry.
We start from those of the subsymmetry generated by the operators~\eqref{eq: interface sym full cond}.
As we can see from Eq.~\eqref{eq: onsite interface sym full cond}, the operator $(\mathcal{O}[A^{G}_{\{e\},1}])_{g_{1},g_{2}}$ in Eq.~\eqref{eq: interface sym full cond} depends only on the difference $g_{1}g_{2}^{-1}$.
Thus, it suffices to consider the operators $(\mathcal{O}[A^G_{\{e\}, 1}])_{e, g}$ for $g \in G$.
These symmetry operators obey the following fusion rules due to Eq.~\eqref{eq: onsite interface sym full cond}:
\begin{equation}\label{eq: interface fusion rules}
(\mathcal{O}[A^G_{\{e\}, 1}])_{e, g} (\mathcal{O}[A^G_{\{e\}, 1}])_{e, g^{\prime}} = \delta_{g, g^{\prime}} (\mathcal{O}[A^G_{\{e\}, 1}])_{e, g}.
\end{equation}
We note that each operator $(\mathcal{O}[A^G_{\{e\}, 1}])_{e, g}$ is a projector because it squares to itself.
To write down all one-dimensional representations of the above fusion rules, we let $V_g$ be the image of the on-site projector $(\mathcal{O}[A^{G}_{\{e\},1}]^{(i)})_{e,g}$.
Concretely, $V_g$ is the vector space spanned by the basis vectors $\{\ket{g_{1}}\bra{g_{2}} \mid g_{1}g_{2}^{-1}=g\}_{g_1, g_2 \in G}$. 
The image of the full symmetry operator $(\mathcal{O}[A^G_{\{e\}, 1}])_{e, g}$ is then given by the tensor product $\bigotimes_i V_g$.
By definition, the symmetry operators~\eqref{eq: interface sym full cond} act diagonally on any state $\ket{v_g} \in \bigotimes_i V_g$ as
\begin{equation}\label{eq: one dim rep 1}
(\mathcal{O}[A^G_{\{e\}, 1}])_{e, g^{\prime}} \ket{v_g} = \delta_{g, g^{\prime}} \ket{v_g}.
\end{equation}
Thus, $\ket{v_g}$ is a one-dimensional representation of the fusion rules~\eqref{eq: interface fusion rules}.
Furthermore, one can show that the other symmetry operators also act diagonally on $\ket{v_g}$ as
\begin{equation}\label{eq: one dim rep 2}
       (\mathcal{O}[A^{G}_{H,\omega}])_{r_{1},r_{2}}\ket{v_{g}} = \delta_{r_{1},r'}\ket{v_{g}},
\end{equation}
where $r'\in S_{H\backslash G}$ is the unique element satisfying $hr' = r_{2}g^{-1}$ for a unique $h\in H$.
Therefore, $\ket{v_g}$ is a one-dimensional representation of the full interface symmetry.
Conversely, by construction, any one-dimensional representation of the interface symmetry is of this form.

As is clear from Eqs.~\eqref{eq: one dim rep 1} and \eqref{eq: one dim rep 2}, the symmetry charge of $\ket{v_g} \in \bigotimes_i V_g$ depends only on $g$.\footnote{Here, the symmetry charge of a state refers to the set of eigenvalues of the symmetry operators acting on it.}
In particular, all states in $V_g$ have the same symmetry charge.
Since $V_g$ is connected, any two one-dimensional representations with charge $g$ can be continuously deformed into each other while preserving the symmetry.
Therefore, one-dimensional representations of the interface symmetry are classified by $G$ up to continuous deformations.
The pumped interface mode~\eqref{eq: 2Rep(G) pumped interface} is a representative of the one-dimensional representations carrying charge $g$.

\subsection{Another perspective on nontriviality}
\label{sec: Another perspective on nontriviality}

In 1+1 dimensions, the pump invariant in SPT phases protected by a non-invertible symmetry can be regarded as an automorphism of the fiber functor \cite{Inamura:2024jke}.
More concretely, by choosing an appropriate gauge for the MPS, we can make the MPS tensor $2\pi$-periodic and the associated $L$-symbols\footnote{The $L$-symbols are the data of a fiber functor; see \cite{Garre-Rubio:2022uum} for the definition.} constant with respect to the parameter, and in this gauge, the linear transformation of the space of the action tensors induced by varying the parameter from $0$ to $2\pi$ gives the pump invariant.
In this section, by considering an analogous construction in 2+1 dimensions, we will confirm the nontriviality of the constructed model from another point of view.
More specifically, by choosing a gauge where the PEPS tensor is $2\pi$-periodic, we will see that the nontriviality of the family can be understood as a nontrivial transformation on the space of line-like action tensors.

To find a gauge where the PEPS tensor is $2\pi$-periodic, let us first write down a PEPS representation of the ground state~\eqref{eq: 2Rep(G) GS S1} parameterized by $S^1$.
A PEPS representation of the state~\eqref{eq: 2Rep(G) GS S1} is obtained by applying the unitary operator $\overrightarrow{X}_{g}^{(i)}(\theta)$ defined in Eq.~\eqref{eq: S1 interpolation RepG} to each vertex tensor of the PEPS representation of the $G$-cluster state:
\begin{equation}
       \adjincludegraphics[scale=1.5,trim={10pt 10pt 10pt 10pt},valign = c]{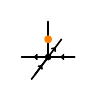}.
\end{equation}
Here, the orange dot represents $\overrightarrow{X}_{g}^{(i)}(\theta)$.
This PEPS tensor is not $2\pi$-periodic with respect to the parameter $\theta$ since it is transformed as in Eq.~\eqref{eq: pulling through G 2d} when $\theta$ is increased by $2\pi$.
To make the PEPS tensor $2\pi$-periodic, we perform a gauge transformation on the virtual legs as follows:
\begin{equation}\label{eq: periodic PEPS}
       \adjincludegraphics[scale=1.5,trim={10pt 10pt 10pt 10pt},valign = c]{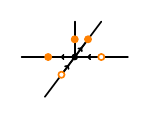},
\end{equation}
where the unfilled orange dots represent $\overrightarrow{X}_{g}(\theta)^{\dagger}$.
Under this gauge transformation, the line-like action tensor defined in Eq.~\eqref{eq: line-like action tensor RepG} is transformed as follows:
\begin{equation}\label{eq: parameterized line-like action tensor}
       \overline{\Phi}^{A}_{r}(\theta)
       \coloneq
       \adjincludegraphics[scale=1.25,trim={20pt 10pt 10pt 10pt},valign = c]{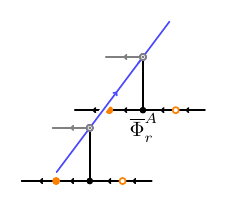}\;\;
       =
       \adjincludegraphics[scale=1.25,trim={20pt 10pt 10pt 10pt},valign = c]{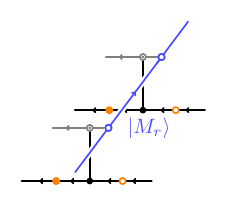}.\;\;
\end{equation}
Here, we recall that $A = A^G_{H, \omega}$.

In this gauge, let us consider how the line-like action tensors change when $\theta$ is increased from $0$ to $2\pi$.\footnote{
       By analogy with the 1+1d case, strictly speaking one should verify that, in this gauge, the $10j$-symbol and related quantities are constant.
       However, since a proper tensor-network formulation of these ingredients involves technical complications, we do not delve into the details here.
       In App.~\ref{sec: Non-abelian 3-cocycle of G-cluster state}, we propose how this information should be incorporated into the tensor network and revisit the pump invariant in this setting.
}
One can easily see that $\overline{\Phi}^{A}_{r}(\theta=2\pi)=\overline{\Phi}^{A}_{r'}(\theta=0)$ where $r'\in S_{H\backslash G}$ is the unique element satisfying $r'=hrg^{-1}$ for a unique $h\in H$.
This is a nontrivial transformation of the space of line-like action tensors if $g\neq e$.
Moreover, if we denote this transformation collectively by $T_g$, it follows that $T_{g_1} \circ T_{g_2} = T_{g_1g_2}$. This is consistent with the fact that $S^1$-parameterized families of $2\mathrm{Rep}(G)$-symmetric SRE states form a group $G$ under the concatenation.

\section*{Acknowledgements}
We would like to thank Andras Molnar and Alex Turzillo for helpful discussions.
S.O. is supported by the European Union's Horizon 2020 research and innovation programme through grant no. 863476 (ERC-CoG SEQUAM). 
S.O. is also supported by JSPS KAKENHI Grant Number 24K00522.
KI is supported in part by the EPSRC Open Fellowship EP/X01276X/1 and by the Leverhulme-Peierls Fellowship funded by the Leverhulme Trust.

\appendix

\section{Symmetry action on module MPS}
\label{sec: symmetry action on module MPS}

Let us compute the symmetry action of $U_{H,\omega}^{G}(g)\coloneq\bigotimes \rho^{G}_{H,\omega}(g)$ on the simple comodule MPS $\ket{M_{r}}$ defined in Eq.~\eqref{eq: simple comodule MPS}.
Since it is easier to compute the action on the simple module MPS defined in Eq.~\eqref{eq: simple module MPS}, we compute it as
\begin{align}
       \begin{aligned}
              U_{H,\omega}^{G}(g)\ket{M_{r}} &= [\bra{M_{r}}U_{H,\omega}^{G}(g)^{\dagger}]^{\dagger}\\
              &= [\sum_{\{(a_{k},b_{k})\}}\tr\left([M_{r}]^{(a_{1},b_{1})}\cdots [M_{r}]^{(a_{L},b_{L})}\right)\bra{g\cdot f^{r}_{a_{1}b_{1}},...,g\cdot f^{r}_{a_{L}b_{L}}}]^{\dagger},
       \end{aligned}
\end{align}
where 
\begin{align}
       \begin{aligned}
              \bra{g\cdot f^{r}_{a_{1}b_{1}},...,g\cdot f^{r}_{a_{L}b_{L}}}\coloneq \bra{\rho^{G}_{H,\omega}(g)\cdot f^{r}_{a_{1}b_{1}}}\otimes \cdots \otimes \bra{\rho^{G}_{H,\omega}(g)\cdot f^{r}_{a_{1}b_{1}}}.
       \end{aligned}
\end{align}
First, let us compute the action of the local unitary operator $\rho^{G}_{H,\omega}(g)$ on the basis $f^{r}_{ab}$ of $A^{G}_{H,\omega}$ defined in Eq.~\eqref{eq: module MPS basis}:
\begin{align}
       \begin{aligned}
       [\rho^{G}_{H,\omega}(g)\cdot f^{r}_{ab}](r') 
       &= f^{r}_{ab}(r'g),\\
       &= \rho_{\omega}(h[r'g]^{-1}) f^{r}_{ab}(r[r'g])\rho_{\omega}(h[r'g]^{-1})^{\dagger},\\
       &= \rho_{\omega}(h[r'g]^{-1}) e_{ab}\rho_{\omega}(h[r'g]^{-1})^{\dagger}\delta_{r,r[r'g]}.
       \end{aligned}
\end{align}
Here, $h[r'g]$ and $r[r'g]$ are defined by the unique decomposition $r'g = h[r'g]^{-1} r[r'g]$ with $h[r'g]\in H$ and $r[r'g]\in S_{H\backslash G}$.
This shows that the action of $\rho^{G}_{H,\omega}(g)$ permutes the basis $\{f^{r}_{ab}\}$ as
\begin{equation}
       [\rho^{G}_{H,\omega}(g)\cdot f^{r}_{ab}]^{\dagger} = \sum_{r'\in S_{H\backslash G}} \rho_{\omega}(h[r'g]^{-1}) (f^{r'}_{ab})^{\dagger} \rho_{\omega}(h[r'g]^{-1})^{\dagger}\delta_{r,r[r'g]}.
\end{equation}
In this basis, the matrix representation of the module map $M$ is computed as
\begin{equation}
\begin{aligned}
       [\rho^{G}_{H,\omega}(g)\cdot f^{r}_{ab}]^{\dagger}m^{r'}_{a'} &= \left[\sum_{r''\in S_{H\backslash G}} \rho_{\omega}(h[r''g]^{-1}) (f^{r''}_{ab})^{\dagger} \rho_{\omega}(h[r''g]^{-1})^{\dagger}\delta_{r,r[r''g]}\right]m^{r'}_{a'},\\
       &= \rho_{\omega}(h[r'g]^{-1}) (f^{r'}_{ab})^{\dagger} \rho_{\omega}(h[r'g]^{-1})^{\dagger}\delta_{r,r[r'g]}m^{r'}_{a'},\\
       &= \sum_{b'} m^{r'}_{b'} \delta_{r,r[r'g]}[m^{r'\dagger}_{b'}\rho_{\omega}(h[r'g]^{-1}) (f^{r'}_{ab})^{\dagger} \rho_{\omega}(h[r'g]^{-1})^{\dagger}m^{r'}_{a'}].
\end{aligned}
\end{equation}
This implies that the symmetry action of $\rho^{G}_{H,\omega}(g)$ on the module MPS is given by
\begin{equation}\label{eq: symmetry action on module MPS tensor}
       [g\cdot M]^{(r;a,b)}_{(s,c),(t,d)} = \delta_{r,r[r'g]}[m^{r'\dagger}_{b'}\rho_{\omega}(h[r'g]^{-1}) (f^{r'}_{ab})^{\dagger} \rho_{\omega}(h[r'g]^{-1})^{\dagger}m^{r'}_{a'}]_{s,t}.
\end{equation}
In particular, each SSB ground state $\ket{M_{r}}$ is mapped to $\ket{M_{r'}}$ for $r'\in S_{H\backslash G}$ such that $r=r[r'g]$, that is, $r'=hrg^{-1}$ for a unique $h\in H$.

\section{Fiber 2-functors and their automorphisms}
\label{sec: Non-abelian 3-cocycle of G-cluster state}
In this paper, we constructed various parameterized families based on the $G$-cluster state and verified their nontriviality.
Although our analysis did not require an explicit description of the SPT invariant of the $G$-cluster state, giving a concrete description of 2+1-dimensional non-invertible symmetric SPT invariants using tensor networks is an interesting problem in its own right, and is also important for formulating pump invariants by methods such as those described in Sec.~\ref{sec: Another perspective on nontriviality}.
A complete description of SPT invariants would require addressing several technical issues in the tensor-network representation, for example, formulating an appropriate notion of injectivity for PEPS, which lies beyond the scope of this paper.
Nevertheless, by analogy with the 1+1-dimensional case, one can to some extent anticipate what form such data should take.
In this section, we propose, based on heuristic considerations, a method for extracting SPT invariants from the PEPS representation of the $G$-cluster state.

From a categorical point of view, SPT phases are expected to be classified by the isomorphism classes of fiber 2-functors, and $S^1$-families are expected to be classified by their automorphisms (up to invertible monoidal modifications).
We briefly explain these data, and then we describe how they should be extracted from the PEPS representation in the case of the $G$-cluster state.
After that, we illustrate our proposal with some examples.
Finally, we revisit the pump invariant described in Sec.~\ref{sec: Another perspective on nontriviality}.

\subsection{Fiber 2-functor of a fusion 2-category}\label{sec: fiber 2-functor}
From the categorical viewpoint, SPT phases protected by a fusion 2-category $\mathcal{C}$ are classified by tensor 2-functors from $\mathcal{C}$ to $2\mathrm{Vec}$, that is, by fiber 2-functors.
Here $2\mathrm{Vec}$ denotes the 2-category of finite semisimple linear categories.
The data of a fiber 2-functor $\mathcal{F}:\mathcal{C}\to 2\mathrm{Vec}$ can be unpacked as follows:
First, a semisimple linear category $\mathcal{F}(x)$ is assigned to each object $x\in \mathcal{C}$.
Next, a linear functor $\mathcal{F}(f):\mathcal{F}(x)\to \mathcal{F}(y)$ is assigned to each 1-morphism $f:x\to y$, and a natural transformation $\mathcal{F}(\alpha):\mathcal{F}(f)\Rightarrow \mathcal{F}(g)$ is assigned to each 2-morphism $\alpha:f\Rightarrow g$.
Then, for each pair of objects $x,y\in \mathcal{C}$, there is given an equivalence
\begin{equation}
       J_{x,y}:\mathcal{F}(x)\otimes \mathcal{F}(y)\to \mathcal{F}(x\otimes y).
\end{equation}
Finally, for each triple of objects $x,y,z\in \mathcal{C}$, there is given a natural isomorphism $\omega_{x,y,z}$ between $J_{x\otimes y,z}\circ [J_{x,y}\otimes\mathrm{id}_{\mathcal{F}(z)}]$ and $J_{x,y\otimes z}\circ [\mathrm{id}_{\mathcal{F}(x)}\otimes J_{y,z}]$ up to associators.\footnote{
       It is known that any weakly monoidal weak 2-category is monoidally equivalent to a semistrict monoidal 2-category in which the associator is trivial \cite[Remark 2.1.4]{Douglas:2018qfz}.
       Thus, one can always take the associator to be trivial without loss of generality.
}
These data are required to be natural in $x,y,z$ and satisfy the usual coherence conditions as a monoidal $2$-functor.

An automorphism $(\alpha,\beta)$ of the fiber 2-functor $\mathcal{F}$ consists of the following data.
First, $\alpha$ is a natural isomorphism from $\mathcal{F}$ to itself.
In particular, to each object $x\in \mathcal{C}$ we assign an autoequivalence $\alpha_{x}:\mathcal{F}(x)\to \mathcal{F}(x)$.
More explicitly, $\alpha_{x}$ sends each simple object $\Phi_{x}^{r}$ in $F(x)$ to another simple object $\Phi_{x}^{\sigma_x(r)}$ for some permutation $\sigma_x$ of the index set of simple objects in $\mathcal{F}(x)$.
Next, $\beta$ is a collection of natural isomorphisms $\beta_{x,y}$ between the functors $\alpha_{x\otimes y}\circ J_{x,y}$ and $J_{x,y}\circ (\alpha_{x}\otimes \alpha_{y})$.
These data should satisfy the suitable coherence conditions.
Finally, an invertible monoidal modification of the automorphism $(\alpha,\beta)$ is given by a collection of natural automorphisms $\mu_x$ of $\alpha_x$ satisfying the suitable coherence conditions.
We refer the reader to Refs.~\cite{gordon1995coherence}\cite[Chaper 2]{schommerpries2014classification} for more detailed explanations of these notions.

\subsection{Fiber 2-functor for the $G$-cluster states}
\label{eq: Fiber 2-functor of the $G$-cluster states}
In this section, we describe how the categorical data introduced in Sec.\ref{sec: fiber 2-functor} should be extracted from the PEPS representation in the case of the $G$-cluster state.

\subsubsection{Point-like action tensors}
First, the objects $\mathcal{O}[A]$ of $2\mathrm{Rep}(G)$ are labeled by symmetric special Frobenius algebras $A$, and the corresponding $\mathcal{F}(\mathcal{O}[A])$ is generated by the line-like action tensors $\{\Phi_{r}^{A}\}_{r=1}^{n(A)}$.
Next, the information of $J_{A_{1},A_{2}}$ is obtained by considering the composition of the line-like action tensors:
The line-like action tensor obtained by composing $\Phi_{r_{1}}^{A_{1}}$ and $\Phi_{r_{2}}^{A_{2}}$ should be equal to the direct sum of line-like action tensors labelled by the symmetric special Frobenius algebras that appear in the direct-sum decomposition of $A_{1}\otimes A_{2}$.
Equivalently, there should exist a topological point-like operator between the object obtained by taking the product of $\Phi_{r_{1}}^{A_{1}}$ and $\Phi_{r_{2}}^{A_{2}}$ and projecting $A_{1}\otimes A_{2}$ onto a direct-sum component $A_{3}$, and the line-like action tensor associated with $A_{3}$.
We call such point-like operators point-like action tensors.
The defining equation of the point-like action tensors can be represented diagrammatically as follows:
\begin{align}\label{eq: point-like action}
       \adjincludegraphics[scale=1.25,trim={10pt 10pt 20pt 10pt},valign = c]{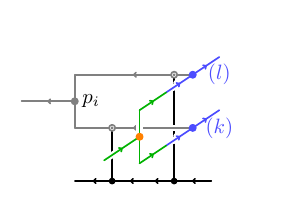}
       =\;\;
       \adjincludegraphics[scale=1.25,trim={10pt 10pt 10pt 10pt},valign = c]{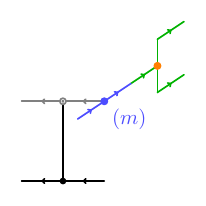},
\end{align}
where $k$, $l$, and $m$ are labels of SSB sectors of $A_{1}$, $A_{2}$, and $A_{3}$ respectively, and $p_{i}$ is a projection from $A_{1}\otimes A_{2}$ onto $A_{3}$.
Let $[V_{A_{1},A_{2}}^{(A_{3};i)}]_{k,l}^{m}$ be the vector space spanned by such point-like action tensors. Then, the natural isomorphism $J_{A_{1},A_{2}}$ can be regarded as a collection of the vector spaces $\{[V_{A_{1},A_{2}}^{(A_{3};i)}]_{r_{1},r_{2}}^{r_{3}}\left.\right|(A_{3};i),r_{a}=1,...,n(A_{a}) \text{ for } a=1,2,3\}$.

To guarantee the existence of a solution to this equation, and to identify what kind of redundancy that solution possesses, it is necessary to understand the redundancy of MPS generated under the fusion of (co)module MPS appearing in the left-hand side of Eq.~\eqref{eq: point-like action}.
We will not go into this point in detail.
Instead, here, we construct a solution for the case in which the algebras $A_{1}$, $A_{2}$ and $A_{3}$ are the algebra that generates the full condensation surface, namely the case $A_i = A^{G}_{\{e\},1}$ for $i=1,2,3$.
In what follows, we focus on an explicit computation for the case $G=\mathbb{Z}_2$. For a general group $G$, the point-like action tensors for the full condensation surfaces can be constructed in the same way.

       \;\\
\noindent
{\bf Example: $G=\mathbb{Z}_{2}$}

Let us compute the point-like action tensors for $G=\mathbb{Z}_{2}$.
In this case, the defining equation of the point-like action tensors becomes
\begin{align}
       \adjincludegraphics[scale=1.25,trim={10pt 10pt 20pt 10pt},valign = c]{tikz/out/fixed_action_tensor1.pdf}
       =\;\;
       \adjincludegraphics[scale=1.25,trim={10pt 10pt 10pt 10pt},valign = c]{tikz/out/fixed_action_tensor2.pdf}
\end{align}
for each $i=1,2$.
To make the equation simpler, we introduce the following notation for the basis of the local Hilbert space:
\begin{align}
       \begin{aligned}
              \ket{+} &\coloneq m_{e},\;\;\;
              \ket{-} \coloneq m_{a},\;\;\;
              \bra{+} &\coloneq m^{\dagger}_{e},\;\;\;
              \bra{-} \coloneq m^{\dagger}_{a},
       \end{aligned}
\end{align}
where $e$ and $a$ are the identity element and the nontrivial element of $\mathbb{Z}_{2}$ respectively.
The solutions for $i=1$ are spanned by
\begin{equation}
    \Lambda_{p_{1}}^{1}
    :=
    \ket{+}
    \begin{array}{c}
        \bra{+} \\
        \otimes \\
        \bra{+}
    \end{array},\;\;\;
    \Lambda_{p_{1}}^{2}
    :=
    \ket{-}
    \begin{array}{c}
        \bra{-} \\
        \otimes \\
        \bra{-}
    \end{array},
\end{equation}
and the solutions for $i=2$ are spanned by
\begin{equation}
    \Lambda_{p_{2}}^{1}
    :=
    \ket{+}
    \begin{array}{c}
        \bra{+} \\
        \otimes \\
        \bra{-}
    \end{array},\;\;\;
    \Lambda_{p_{2}}^{2}
    :=
    \ket{-}
    \begin{array}{c}
        \bra{-} \\
        \otimes \\
        \bra{+}
    \end{array}.
\end{equation}

Let us check this explicitly:
    \begin{equation}
        \adjincludegraphics[scale=1.25,trim={10pt 10pt 20pt 10pt},valign = c]{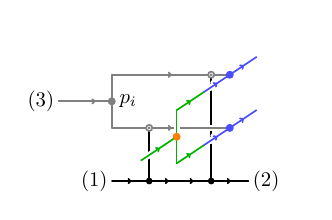}
        =\;\;
        \adjincludegraphics[scale=1.25,trim={10pt 10pt 10pt 10pt},valign = c]{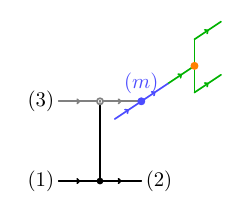}.
    \end{equation}
    First, let us consider the case where $p_{i}=p_{1}$. 
    By contracting the legs $(1)$ and $(2)$ with $\bra{+}$ and $\ket{+}$, the equation becomes
    \begin{equation}
        \adjincludegraphics[scale=1.25,trim={10pt 10pt 10pt 10pt},valign = c]{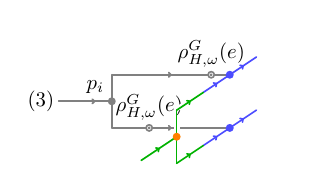}
        =\;\;
        \adjincludegraphics[scale=1.25,trim={10pt 10pt 10pt 10pt},valign = c]{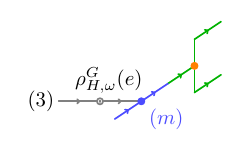}.
    \end{equation}
    Additionally, by contracting the leg $(3)$ with $\ket{+}$ and $\ket{-}$, the equation simplifies to 
    \begin{align}\label{eq: action_tensor_point1}
        \adjincludegraphics[scale=1.25,trim={10pt 10pt 10pt 10pt},valign = c]{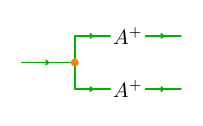}\;\;
        =\;\;
        \adjincludegraphics[scale=1.25,trim={10pt 10pt 10pt 10pt},valign = c]{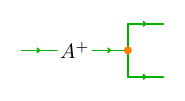},\;\;\;\;
        \adjincludegraphics[scale=1.25,trim={10pt 10pt 10pt 10pt},valign = c]{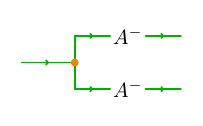}\;\;
        =\;\;
        \adjincludegraphics[scale=1.25,trim={10pt 10pt 10pt 10pt},valign = c]{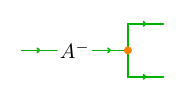},
    \end{align}
    where $A^{\pm}=\ket{\pm}\bra{\pm}$.
    Thus, the solutions are spanned by
    \begin{equation}
        \Lambda_{p_{1}}^{1}
        =
        \ket{+}
        \begin{array}{c}
            \bra{+} \\
            \otimes \\
            \bra{+}
        \end{array},\;\;\;
        \Lambda_{p_{1}}^{2}
        =
        \ket{-}
        \begin{array}{c}
            \bra{-} \\
            \otimes \\
            \bra{-}
        \end{array}.
    \end{equation}
    On the other hand, by contracting the legs $(1)$ and $(2)$ with $\bra{-}$ and $\ket{-}$, the equation becomes
    \begin{equation}
        \adjincludegraphics[scale=1.25,trim={10pt 10pt 20pt 10pt},valign = c]{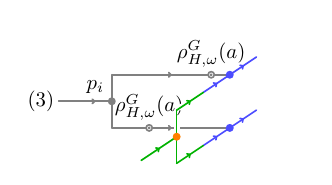}
        =\;\;
        \adjincludegraphics[scale=1.25,trim={10pt 10pt 10pt 10pt},valign = c]{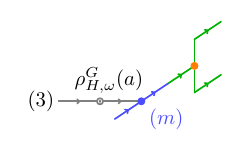}.
    \end{equation}
    This equation is equivalent to Eq.\eqref{eq: action_tensor_point1}.
    Therefore, for the case where $p_{i}=p_{1}$, the solutions are spanned by $\Lambda_{p_{1}}^{1}$ and $\Lambda_{p_{1}}^{2}$.

    Next, let us consider the case where $p_{i}=p_{2}$. 
    By contracting the legs $(1)$, $(2)$ and $(3)$ with $\bra{+}$, $\ket{+}$ and $\ket{\pm}$ respectively, the equation becomes
    \begin{align}\label{eq: action_tensor_point}
        \adjincludegraphics[scale=1.25,trim={10pt 10pt 10pt 10pt},valign = c]{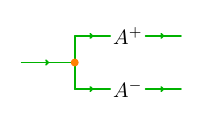}\;\;
        =\;\;
        \adjincludegraphics[scale=1.25,trim={10pt 10pt 10pt 10pt},valign = c]{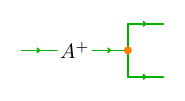},\;\;\;\;
        \adjincludegraphics[scale=1.25,trim={10pt 10pt 10pt 10pt},valign = c]{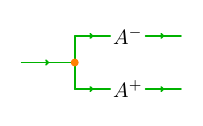}\;\;
        =\;\;
        \adjincludegraphics[scale=1.25,trim={10pt 10pt 10pt 10pt},valign = c]{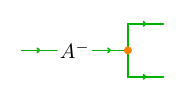}.
    \end{align}
    Furthermore, contracting the legs of $(1)$ and $(2)$ with $\bra{-}$ and $\ket{-}$ reduces to the same equation.
    Therefore, for the case where $p_{i}=p_{2}$, the solutions are spanned by
    \begin{equation}
        \Lambda_{p_{2}}^{1}
        =
        \ket{+}
        \begin{array}{c}
            \bra{+} \\
            \otimes \\
            \bra{-}
        \end{array},\;\;\;
        \Lambda_{p_{2}}^{2}
        =
        \ket{-}
        \begin{array}{c}
            \bra{-} \\
            \otimes \\
            \bra{+}
        \end{array}.
    \end{equation}

\subsubsection{Module 10-j symbol}
Finally, the isomorphism $\omega_{x,y,z}$ can be obtained as a basis transformation of the point-like action tensors between the following vector spaces:
\begin{equation}
       \bigoplus_{(A';i)}\bigoplus_{r_{1},r_{2},r_{3},r_{4}}\bigoplus_{r'} [V_{A_{1},A_{2}}^{(A';i)}]_{r_{1},r_{2}}^{r'} \otimes [V_{A',A_{3}}^{A_{4}}]_{r',r_{3}}^{r_{4}} \to \bigoplus_{(A'';j)}\bigoplus_{r_{1},r_{2},r_{3},r_{4}}\bigoplus_{r''} [V_{A_{2},A_{3}}^{(A'';j)}]_{r_{2},r_{3}}^{r'',r_{4}} \otimes [V_{A_{1},A''}^{A_{4}}]_{r_{1},r''}^{r_{4}}.
\end{equation}
More concretely, the above information can be expressed as the following equation:
\begin{equation}\label{eq: non-abelian 3-cocycle}
    \adjincludegraphics[scale=1.,trim={10pt 10pt 10pt 10pt},valign = c]{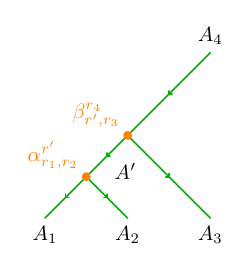}
    =
    \sum_{(A'';\gamma,\delta)} \left[\Omega_{xyz}^{w}\right]_{(A';\alpha,\beta),(A'';\gamma,\delta)}
    \adjincludegraphics[scale=1.,trim={10pt 10pt 10pt 10pt},valign = c]{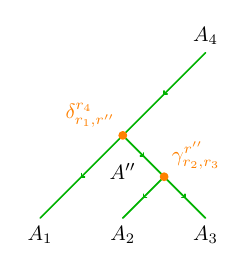},
\end{equation}
where $\alpha_{r_{1},r_{2}}^{r'}$, $\beta_{r',r_{3}}^{r_{4}}$, $\gamma_{r_{2},r_{3}}^{r''}$, and $\delta_{r_{1},r''}^{r_{4}}$ are basis point-like action tensors of $[V_{A_{1},A_{2}}^{(A';i)}]_{r_{1},r_{2}}^{r'}$, $[V_{A',A_{3}}^{A_{4}}]_{r',r_{3}}^{r_{4}}$, $[V_{A_{2},A_{3}}^{(A'';j)}]_{r_{2},r_{3}}^{r'',r_{4}}$, and $[V_{A_{1},A''}^{A_{4}}]_{r_{1},r''}^{r_{4}}$.
We refer to the collection of the coefficients $\left[\Omega_{xyz}^{w}\right]_{(A';\alpha,\beta),(A'';\gamma,\delta)}$ as the module 10-j symbol.
       The SPT invariant of the $G$-cluster state is given by the module 10-j symbol up to gauge transformations.
       However, similar to the case of point-like action tensors, it is difficult in a general setting to guarantee the existence of a solution to this equation and to determine what kind of redundancy such a solution possesses.
       Instead, here we construct a solution for the case of a Frobenius algebra corresponding to a full condensation surface.
       In what follows, we focus on an explicit computation for the case $G=\mathbb{Z}_2$. For a general group $G$, the module 10-j symbol for the full condensation surfaces can be computed in the same way.

\;\\
\noindent
{\bf Example: $G=\mathbb{Z}_{2}$}

To make the equation simpler, we introduce the shorthand notation $\bf{2}\coloneq A_{\{e\},1}^{\mathbb{Z}_{2}}$.
When $G=\mathbb{Z}_{2}$, the left-hand side of the defining equation in Eq.~\eqref{eq: non-abelian 3-cocycle} can be computed as follows:
\begin{align*}
    &\adjincludegraphics[scale=1,trim={10pt 10pt 10pt 10pt},valign = c]{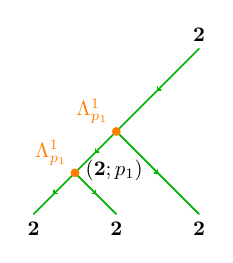}:
    \begin{cases}
        \;\; \ket{+,+,+} &\mapsto \;\; \ket{+},\\
        \;\; \text{otherwise} &\mapsto \;\; 0,
    \end{cases}\;\;
    \adjincludegraphics[scale=1,trim={10pt 10pt 10pt 10pt},valign = c]{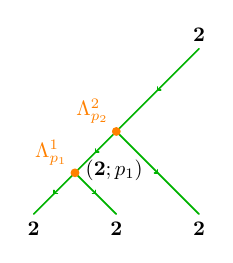}:
    \begin{cases}
        \;\; \ket{+,+,-} &\mapsto \;\; \ket{-},\\
        \;\; \text{otherwise} &\mapsto \;\; 0,
    \end{cases}
\end{align*}
\begin{align*}
    &\adjincludegraphics[scale=1,trim={10pt 10pt 10pt 10pt},valign = c]{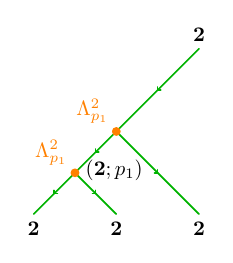}:
    \begin{cases}
        \;\; \ket{-,-,-} &\mapsto \;\; \ket{-},\\
        \;\; \text{otherwise} &\mapsto \;\; 0,
    \end{cases}\;\;
    \adjincludegraphics[scale=1,trim={10pt 10pt 10pt 10pt},valign = c]{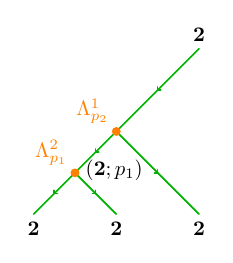}:
    \begin{cases}
        \;\; \ket{-,-,+} &\mapsto \;\; \ket{+},\\
        \;\; \text{otherwise} &\mapsto \;\; 0,
    \end{cases}
\end{align*}
\begin{align*}
    &\adjincludegraphics[scale=1,trim={10pt 10pt 10pt 10pt},valign = c]{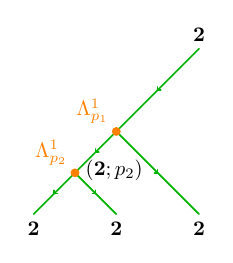}:
    \begin{cases}
        \;\; \ket{-,+,+} &\mapsto \;\; \ket{+},\\
        \;\; \text{otherwise} &\mapsto \;\; 0,
    \end{cases}\;\;
    \adjincludegraphics[scale=1,trim={10pt 10pt 10pt 10pt},valign = c]{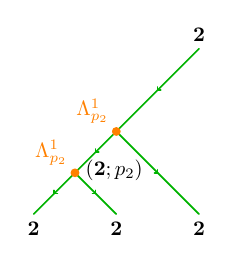}:
    \begin{cases}
        \;\; \ket{-,+,-} &\mapsto \;\; \ket{-},\\
        \;\; \text{otherwise} &\mapsto \;\; 0,
    \end{cases}
\end{align*}
\begin{align*}
    &\adjincludegraphics[scale=1,trim={10pt 10pt 10pt 10pt},valign = c]{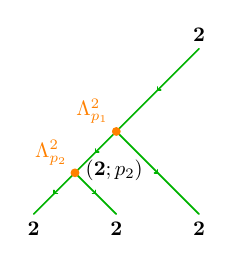}:
    \begin{cases}
        \;\; \ket{+,-,-} &\mapsto \;\; \ket{-},\\
        \;\; \text{otherwise} &\mapsto \;\; 0,
    \end{cases}\;\;
    \adjincludegraphics[scale=1,trim={10pt 10pt 10pt 10pt},valign = c]{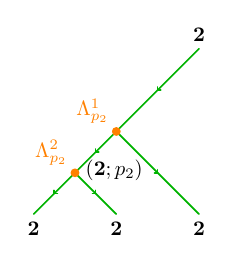}:
    \begin{cases}
        \;\; \ket{+,-,+} &\mapsto \;\; \ket{+},\\
        \;\; \text{otherwise} &\mapsto \;\; 0.
    \end{cases}
\end{align*}
Similarly, the right-hand side of the defining equation in Eq.~\eqref{eq: non-abelian 3-cocycle} can be computed as follows:
\begin{align*}
    &\adjincludegraphics[scale=1,trim={10pt 10pt 10pt 10pt},valign = c]{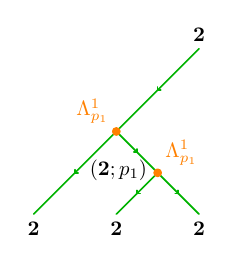}:
    \begin{cases}
        \;\; \ket{+,+,+} &\mapsto \;\; \ket{+},\\
        \;\; \text{otherwise} &\mapsto \;\; 0,
    \end{cases}\;\;
    \adjincludegraphics[scale=1,trim={10pt 10pt 10pt 10pt},valign = c]{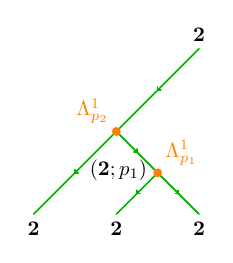}:
    \begin{cases}
        \;\; \ket{-,+,+} &\mapsto \;\; \ket{+},\\
        \;\; \text{otherwise} &\mapsto \;\; 0,
    \end{cases}
\end{align*}
\begin{align*}
    &\adjincludegraphics[scale=1,trim={10pt 10pt 10pt 10pt},valign = c]{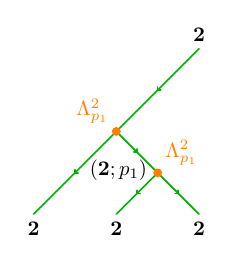}:
    \begin{cases}
        \;\; \ket{-,-,-} &\mapsto \;\; \ket{-},\\
        \;\; \text{otherwise} &\mapsto \;\; 0,
    \end{cases}\;\;
    \adjincludegraphics[scale=1,trim={10pt 10pt 10pt 10pt},valign = c]{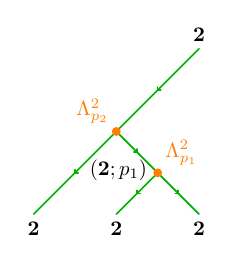}:
    \begin{cases}
        \;\; \ket{+,-,-} &\mapsto \;\; \ket{-},\\
        \;\; \text{otherwise} &\mapsto \;\; 0,
    \end{cases}
\end{align*}
\begin{align*}
    &\adjincludegraphics[scale=1,trim={10pt 10pt 10pt 10pt},valign = c]{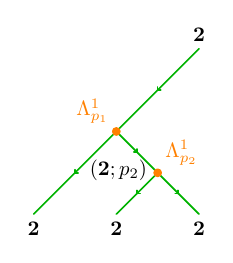}:
    \begin{cases}
        \;\; \ket{+,-,+} &\mapsto \;\; \ket{+},\\
        \;\; \text{otherwise} &\mapsto \;\; 0,
    \end{cases}\;\;
    \adjincludegraphics[scale=1,trim={10pt 10pt 10pt 10pt},valign = c]{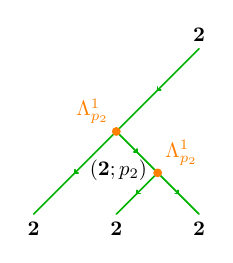}:
    \begin{cases}
        \;\; \ket{-,-,+} &\mapsto \;\; \ket{+},\\
        \;\; \text{otherwise} &\mapsto \;\; 0,
    \end{cases}
\end{align*}
\begin{align*}
    &\adjincludegraphics[scale=1,trim={10pt 10pt 10pt 10pt},valign = c]{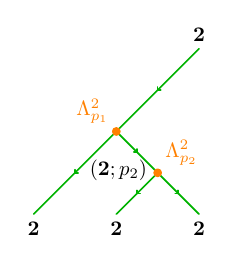}:
    \begin{cases}
        \;\; \ket{-,+,-} &\mapsto \;\; \ket{-},\\
        \;\; \text{otherwise} &\mapsto \;\; 0,
    \end{cases}\;\;
    \adjincludegraphics[scale=1,trim={10pt 10pt 10pt 10pt},valign = c]{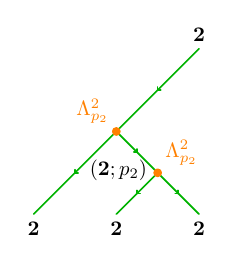}:
    \begin{cases}
        \;\; \ket{+,+,-} &\mapsto \;\; \ket{-},\\
        \;\; \text{otherwise} &\mapsto \;\; 0.
    \end{cases}
\end{align*}
From the above result, we can confirm that the diagrams on the left-hand side and the right-hand side generate the same vector spaces.
By choosing an appropriate ordering of the basis, the module 10-j symbol can be represented as the identity matrix.
This is consistent with the fact that the classification of SPT phases protected only by $2 \mathrm{Rep}(\mathbb{Z}_{2})$ symmetry is trivial.

\subsection{Pump invariant revisited}
\label{eq: Pump invariant revisited}

Based on the proposal in Sec.~\ref{eq: Fiber 2-functor of the $G$-cluster states}, we revisit the pump invariant of the $G$-cluster state discussed in Sec.~\ref{sec: Another perspective on nontriviality}.
In Sec.~\ref{sec: Another perspective on nontriviality}, we fixed a $2\pi$-periodic gauge for the PEPS in Eq.~\eqref{eq: periodic PEPS}, and as a result the line-like action tensor acquired a parameter dependence of the form given in Eq.~\eqref{eq: parameterized line-like action tensor}.
If one uses this parameter-dependent line-like action tensor and considers the defining equation for the point-like action tensor, the parameter dependence drops out of the equation.
Therefore, the point-like action tensor, as well as the module 10-j symbol computed from it, is independent of the parameter.
Consequently, by analogy with the 1+1-dimensional invariant \cite{Inamura:2024jke}, one can regard the transformation induced on the space of line-like action tensors as an invariant of $S^{1}$-family of $2 \mathrm{Rep}(G)$-symmetric SRE states.

From a categorical perspective, it is expected that the classification of $S^1$-families is given by automorphisms of the fiber 2-functor.
The family labeled by $g\in G$ constructed in Sec.~\ref{sec: Parametrized family of 2Rep(G)-symmetric states} induces a transformation on the space of line-like action tensors, which is expected to correspond to an automorphism of the fiber 2-functor characterized by a permutation of $S_{H\backslash G}$, defined as follows:
for $A = A^G_{H, \omega} \in 2 \mathrm{Rep}(G)$ and $r\in S_{H\backslash G}$
\begin{equation}
       \sigma_{A}(r)\coloneqq r', \qquad r' = hrg,
\end{equation}
where $h\in H$ is a unique element such that $hrg\in S_{H\backslash G}$.
This automorphism is nontrivial when $g\neq e$.\footnote{
       More precisely, the classification is given by automorphisms up to invertible monoidal modification. However, an automorphism that nontrivially permutes objects cannot be removed by a monoidal modification.
}

It is further possible to push the parameter dependence down to a lower layer as follows:
While the (co)module MPS in Eq.~\eqref{eq: parameterized line-like action tensor} is $2\pi$-periodic, the (co)module MPS tensor is not $2\pi$-periodic, since at $\theta=2\pi$ it undergoes the gauge transformation specified in Eq.~\eqref{eq: symmetry action on module MPS tensor}.
To cancel this effect, one can perform a gauge transformation using
\begin{equation}
       \hat{V}(\theta)\coloneq [\rho_{H,\omega}^{G}(g)]^{\frac{\theta}{2\pi}}.
\end{equation}
This makes the MPS tensor $2\pi$-periodic.
On the other hand, as a trade-off of this gauge transformation, the defining equation for the point-like action tensor acquires a parameter dependence, and in general the point-like action tensor no longer satisfies $2\pi$-periodicity.
In this gauge, it is expected that the pump invariant can be defined as a linear transformation acting on the space of point-like action tensors.

\bibliography{bibliography}

\providecommand{\href}[2]{#2}\begingroup\raggedright\begin{thebibliography}{100}

\bibitem{Gu_2009}
Z.-C.~Gu and X.-G.~Wen, \emph{{Tensor-entanglement-filtering renormalization approach and symmetry-protected topological order}}, \href{https://doi.org/10.1103/physrevb.80.155131}{\emph{Physical Review B} {\bfseries 80} (2009) } [\href{https://arxiv.org/abs/0903.1069}{{\ttfamily 0903.1069}}].

\bibitem{Pollmann:2009mhk}
F.~Pollmann, E.~Berg, A.M.~Turner and M.~Oshikawa, \emph{{Symmetry protection of topological phases in one-dimensional quantum spin systems}}, \href{https://doi.org/10.1103/PhysRevB.85.075125}{\emph{Phys. Rev. B} {\bfseries 85} (2012) 075125} [\href{https://arxiv.org/abs/0909.4059}{{\ttfamily 0909.4059}}].

\bibitem{Pollmann_2010}
F.~Pollmann, A.M.~Turner, E.~Berg and M.~Oshikawa, \emph{{Entanglement spectrum of a topological phase in one dimension}}, \href{https://doi.org/10.1103/physrevb.81.064439}{\emph{Physical Review B} {\bfseries 81} (2010) } [\href{https://arxiv.org/abs/0910.1811}{{\ttfamily 0910.1811}}].

\bibitem{Chen:2010zpc}
X.~Chen, Z.-C.~Gu and X.-G.~Wen, \emph{{Classification of gapped symmetric phases in one-dimensional spin systems}}, \href{https://doi.org/10.1103/PhysRevB.83.035107}{\emph{Phys. Rev. B} {\bfseries 83} (2011) 035107} [\href{https://arxiv.org/abs/1008.3745}{{\ttfamily 1008.3745}}].

\bibitem{Chen_2011_complete}
X.~Chen, Z.-C.~Gu and X.-G.~Wen, \emph{Complete classification of one-dimensional gapped quantum phases in interacting spin systems}, \href{https://doi.org/10.1103/physrevb.84.235128}{\emph{Physical Review B} {\bfseries 84} (2011) } [\href{https://arxiv.org/abs/1103.3323}{{\ttfamily 1103.3323}}].

\bibitem{Chen:2011bcp}
X.~Chen, Z.-X.~Liu and X.-G.~Wen, \emph{{Two-dimensional symmetry-protected topological orders and their protected gapless edge excitations}}, \href{https://doi.org/10.1103/PhysRevB.84.235141}{\emph{Phys. Rev. B} {\bfseries 84} (2011) 235141} [\href{https://arxiv.org/abs/1106.4752}{{\ttfamily 1106.4752}}].

\bibitem{Chen:2011pg}
X.~Chen, Z.-C.~Gu, Z.-X.~Liu and X.-G.~Wen, \emph{{Symmetry protected topological orders and the group cohomology of their symmetry group}}, \href{https://doi.org/10.1103/PhysRevB.87.155114}{\emph{Phys. Rev. B} {\bfseries 87} (2013) 155114} [\href{https://arxiv.org/abs/1106.4772}{{\ttfamily 1106.4772}}].

\bibitem{Schuch_2011}
N.~Schuch, D.~P\'{e}rez-Garc\'{i}a and I.~Cirac, \emph{{Classifying quantum phases using matrix product states and projected entangled pair states}}, \href{https://doi.org/10.1103/physrevb.84.165139}{\emph{Physical Review B} {\bfseries 84} (2011) } [\href{https://arxiv.org/abs/1010.3732}{{\ttfamily 1010.3732}}].

\bibitem{Thouless83}
D.J.~Thouless, \emph{Quantization of particle transport}, \href{https://doi.org/10.1103/PhysRevB.27.6083}{\emph{Phys. Rev. B} {\bfseries 27} (1983) 6083}.

\bibitem{PhysRevB.82.115120}
J.C.Y.~Teo and C.L.~Kane, \emph{Topological defects and gapless modes in insulators and superconductors}, \href{https://doi.org/10.1103/PhysRevB.82.115120}{\emph{Phys. Rev. B} {\bfseries 82} (2010) 115120}.

\bibitem{Kitaev2011SCGP}
A.~Kitaev, \emph{Toward a topological classification of many-body quantum states with short-range entanglement},  2011.
\newblock \href{http://scgp.stonybrook.edu/video_portal/video.php?id=336}{ talk at Simons Center for Geometry and Physics}.

\bibitem{Kitaev2013SCGP}
A.~Kitaev, \emph{On the classification of short-range entangled states},  2013.
\newblock \href{http://scgp.stonybrook.edu/video_portal/video.php?id=2010}{ talk at Simons Center for Geometry and Physics}.

\bibitem{Kitaev2015IPAM}
A.~Kitaev, \emph{Homotopy-theoretic approach to spt phases in action: $z_{16}$ classification of three-dimensional superconductors},  2015.
\newblock \href{https://www.ipam.ucla.edu/abstract/?tid=12389}{ talk at Institute for Pure and Applied Mathematics}.

\bibitem{Kapustin:2020mkl}
A.~Kapustin and L.~Spodyneiko, \emph{{Higher-dimensional generalizations of the Thouless charge pump}},  \href{https://arxiv.org/abs/2003.09519}{{\ttfamily 2003.09519}}.

\bibitem{PhysRevResearch.2.042024}
Y.~Kuno and Y.~Hatsugai, \emph{Interaction-induced topological charge pump}, \href{https://doi.org/10.1103/PhysRevResearch.2.042024}{\emph{Phys. Rev. Res.} {\bfseries 2} (2020) 042024} [\href{https://arxiv.org/abs/2007.11215}{{\ttfamily 2007.11215}}].

\bibitem{Tantivasadakarn:2021wdv}
N.~Tantivasadakarn, R.~Thorngren, A.~Vishwanath and R.~Verresen, \emph{{Pivot Hamiltonians as generators of symmetry and entanglement}}, \href{https://doi.org/10.21468/SciPostPhys.14.2.012}{\emph{SciPost Phys.} {\bfseries 14} (2023) 012} [\href{https://arxiv.org/abs/2110.07599}{{\ttfamily 2110.07599}}].

\bibitem{Shiozaki:2021weu}
K.~Shiozaki, \emph{{Adiabatic cycles of quantum spin systems}}, \href{https://doi.org/10.1103/PhysRevB.106.125108}{\emph{Phys. Rev. B} {\bfseries 106} (2022) 125108} [\href{https://arxiv.org/abs/2110.10665}{{\ttfamily 2110.10665}}].

\bibitem{Hermele2021CMSA}
M.~Hermele, \emph{Families of gapped systems and quantum pumps},  2021.
\newblock \href{https://www.youtube.com/watch?v=wtaC0tqXZMU}{ talk at Harvard CMSA}.

\bibitem{Wen:2021gwc}
X.~Wen, M.~Qi, A.~Beaudry, J.~Moreno, M.J.~Pflaum, D.~Spiegel et~al., \emph{{Flow of higher Berry curvature and bulk-boundary correspondence in parametrized quantum systems}}, \href{https://doi.org/10.1103/PhysRevB.108.125147}{\emph{Phys. Rev. B} {\bfseries 108} (2023) 125147} [\href{https://arxiv.org/abs/2112.07748}{{\ttfamily 2112.07748}}].

\bibitem{Bachmann:2022bhx}
S.~Bachmann, W.~De~Roeck, M.~Fraas and T.~Jappens, \emph{{A Classification of $G$-Charge Thouless Pumps in 1D Invertible States}}, \href{https://doi.org/10.1007/s00220-024-05010-w}{\emph{Commun. Math. Phys.} {\bfseries 405} (2024) 157} [\href{https://arxiv.org/abs/2204.03763}{{\ttfamily 2204.03763}}].

\bibitem{Spodyneiko:2023vsw}
L.~Spodyneiko, \emph{{Hall conductivity pump}},  \href{https://arxiv.org/abs/2309.14332}{{\ttfamily 2309.14332}}.

\bibitem{Ohyama:2022cib}
S.~Ohyama, K.~Shiozaki and M.~Sato, \emph{{Generalized Thouless pumps in (1+1)-dimensional interacting fermionic systems}}, \href{https://doi.org/10.1103/PhysRevB.106.165115}{\emph{Phys. Rev. B} {\bfseries 106} (2022) 165115} [\href{https://arxiv.org/abs/2206.01110}{{\ttfamily 2206.01110}}].

\bibitem{Inamura:2024jke}
K.~Inamura and S.~Ohyama, \emph{{1+1d SPT phases with fusion category symmetry: interface modes and non-abelian Thouless pump}},  \href{https://arxiv.org/abs/2408.15960}{{\ttfamily 2408.15960}}.

\bibitem{Jones:2025khc}
N.G.~Jones, R.~Thorngren, R.~Verresen and A.~Prakash, \emph{{Charge pumps, pivot Hamiltonians, and symmetry-protected topological phases}}, \href{https://doi.org/10.1103/rtq1-pplf}{\emph{Phys. Rev. B} {\bfseries 112} (2025) 165123} [\href{https://arxiv.org/abs/2507.00995}{{\ttfamily 2507.00995}}].

\bibitem{Shiozaki:2025pyo}
K.~Shiozaki, \emph{{Equivariant Parameter Families of Spin Chains: A Discrete MPS Formulation}},  \href{https://arxiv.org/abs/2507.19932}{{\ttfamily 2507.19932}}.

\bibitem{Li:2025wes}
Y.~Li, M.~Dell'acqua and A.~Mitra, \emph{{Classification of Thouless pumps with non-invertible symmetries and implications for Floquet phases}},  \href{https://arxiv.org/abs/2510.01626}{{\ttfamily 2510.01626}}.

\bibitem{Gaiotto:2014kfa}
D.~Gaiotto, A.~Kapustin, N.~Seiberg and B.~Willett, \emph{{Generalized Global Symmetries}}, \href{https://doi.org/10.1007/JHEP02(2015)172}{\emph{JHEP} {\bfseries 02} (2015) 172} [\href{https://arxiv.org/abs/1412.5148}{{\ttfamily 1412.5148}}].

\bibitem{KW1941}
H.A.~Kramers and G.H.~Wannier, \emph{{Statistics of the Two-Dimensional Ferromagnet. Part I}}, \href{https://doi.org/10.1103/PhysRev.60.252}{\emph{Phys. Rev.} {\bfseries 60} (1941) 252}.

\bibitem{Cordova:2022ruw}
C.~C\'{o}rdova, T.T.~Dumitrescu, K.~Intriligator and S.-H.~Shao, \emph{{Snowmass White Paper: Generalized Symmetries in Quantum Field Theory and Beyond}},  in \emph{{Snowmass 2021}}, 5, 2022 [\href{https://arxiv.org/abs/2205.09545}{{\ttfamily 2205.09545}}].

\bibitem{McGreevy:2022oyu}
J.~McGreevy, \emph{{Generalized Symmetries in Condensed Matter}}, \href{https://doi.org/10.1146/annurev-conmatphys-040721-021029}{\emph{Ann. Rev. Condensed Matter Phys.} {\bfseries 14} (2023) 57} [\href{https://arxiv.org/abs/2204.03045}{{\ttfamily 2204.03045}}].

\bibitem{Schafer-Nameki:2023jdn}
S.~Sch\"{a}fer-Nameki, \emph{{ICTP lectures on (non-)invertible generalized symmetries}}, \href{https://doi.org/10.1016/j.physrep.2024.01.007}{\emph{Phys. Rept.} {\bfseries 1063} (2024) 1} [\href{https://arxiv.org/abs/2305.18296}{{\ttfamily 2305.18296}}].

\bibitem{Brennan:2023mmt}
T.D.~Brennan and S.~Hong, \emph{{Introduction to Generalized Global Symmetries in QFT and Particle Physics}},  \href{https://arxiv.org/abs/2306.00912}{{\ttfamily 2306.00912}}.

\bibitem{Bhardwaj:2023kri}
L.~Bhardwaj, L.E.~Bottini, L.~Fraser-Taliente, L.~Gladden, D.S.W.~Gould, A.~Platschorre et~al., \emph{{Lectures on generalized symmetries}}, \href{https://doi.org/10.1016/j.physrep.2023.11.002}{\emph{Phys. Rept.} {\bfseries 1051} (2024) 1} [\href{https://arxiv.org/abs/2307.07547}{{\ttfamily 2307.07547}}].

\bibitem{Luo:2023ive}
R.~Luo, Q.-R.~Wang and Y.-N.~Wang, \emph{{Lecture notes on generalized symmetries and applications}}, \href{https://doi.org/10.1016/j.physrep.2024.02.002}{\emph{Phys. Rept.} {\bfseries 1065} (2024) 1} [\href{https://arxiv.org/abs/2307.09215}{{\ttfamily 2307.09215}}].

\bibitem{Shao:2023gho}
S.-H.~Shao, \emph{{What's Done Cannot Be Undone: TASI Lectures on Non-Invertible Symmetries}},  \href{https://arxiv.org/abs/2308.00747}{{\ttfamily 2308.00747}}.

\bibitem{Carqueville:2023jhb}
N.~Carqueville, M.~Del~Zotto and I.~Runkel, \emph{{Topological defects}},  \href{https://arxiv.org/abs/2311.02449}{{\ttfamily 2311.02449}}.

\bibitem{Iqbal:2024pee}
N.~Iqbal, \emph{{Jena lectures on generalized global symmetries: principles and applications}},  7, 2024 [\href{https://arxiv.org/abs/2407.20815}{{\ttfamily 2407.20815}}].

\bibitem{Costa:2024wks}
D.~Costa, C.~C{\'o}rdova, M.D.~Zotto, D.~Freed, J.~G{\"o}dicke, A.~Hofer et~al., \emph{Simons lectures on categorical symmetries},  \href{https://arxiv.org/abs/2411.09082}{{\ttfamily 2411.09082}}.

\bibitem{Douglas:2018qfz}
C.L.~Douglas and D.J.~Reutter, \emph{{Fusion 2-categories and a state-sum invariant for 4-manifolds}},  \href{https://arxiv.org/abs/1812.11933}{{\ttfamily 1812.11933}}.

\bibitem{Decoppet:2024htz}
T.D.~D{\'e}coppet, P.~Huston, T.~Johnson-Freyd, D.~Nikshych, D.~Penneys, J.~Plavnik et~al., \emph{{The Classification of Fusion 2-Categories}},  \href{https://arxiv.org/abs/2411.05907}{{\ttfamily 2411.05907}}.

\bibitem{Kaidi:2021xfk}
J.~Kaidi, K.~Ohmori and Y.~Zheng, \emph{{Kramers-Wannier-like Duality Defects in (3+1)D Gauge Theories}}, \href{https://doi.org/10.1103/PhysRevLett.128.111601}{\emph{Phys. Rev. Lett.} {\bfseries 128} (2022) 111601} [\href{https://arxiv.org/abs/2111.01141}{{\ttfamily 2111.01141}}].

\bibitem{Bhardwaj:2022lsg}
L.~Bhardwaj, S.~Sch\"{a}fer-Nameki and J.~Wu, \emph{{Universal Non-Invertible Symmetries}}, \href{https://doi.org/10.1002/prop.202200143}{\emph{Fortsch. Phys.} {\bfseries 70} (2022) 2200143} [\href{https://arxiv.org/abs/2208.05973}{{\ttfamily 2208.05973}}].

\bibitem{Bhardwaj:2022maz}
L.~Bhardwaj, L.E.~Bottini, S.~Schafer-Nameki and A.~Tiwari, \emph{{Non-invertible symmetry webs}}, \href{https://doi.org/10.21468/SciPostPhys.15.4.160}{\emph{SciPost Phys.} {\bfseries 15} (2023) 160} [\href{https://arxiv.org/abs/2212.06842}{{\ttfamily 2212.06842}}].

\bibitem{Bartsch:2022mpm}
T.~Bartsch, M.~Bullimore, A.E.V.~Ferrari and J.~Pearson, \emph{{Non-invertible symmetries and higher representation theory I}}, \href{https://doi.org/10.21468/SciPostPhys.17.1.015}{\emph{SciPost Phys.} {\bfseries 17} (2024) 015} [\href{https://arxiv.org/abs/2208.05993}{{\ttfamily 2208.05993}}].

\bibitem{Bartsch:2022ytj}
T.~Bartsch, M.~Bullimore, A.E.V.~Ferrari and J.~Pearson, \emph{{Non-invertible symmetries and higher representation theory II}}, \href{https://doi.org/10.21468/SciPostPhys.17.2.067}{\emph{SciPost Phys.} {\bfseries 17} (2024) 067} [\href{https://arxiv.org/abs/2212.07393}{{\ttfamily 2212.07393}}].

\bibitem{Delcamp:2023kew}
C.~Delcamp and A.~Tiwari, \emph{{Higher categorical symmetries and gauging in two-dimensional spin systems}}, \href{https://doi.org/10.21468/SciPostPhys.16.4.110}{\emph{SciPost Phys.} {\bfseries 16} (2024) 110} [\href{https://arxiv.org/abs/2301.01259}{{\ttfamily 2301.01259}}].

\bibitem{Choi:2024rjm}
Y.~Choi, Y.~Sanghavi, S.-H.~Shao and Y.~Zheng, \emph{{Non-invertible and higher-form symmetries in 2+1d lattice gauge theories}}, \href{https://doi.org/10.21468/SciPostPhys.18.1.008}{\emph{SciPost Phys.} {\bfseries 18} (2025) 008} [\href{https://arxiv.org/abs/2405.13105}{{\ttfamily 2405.13105}}].

\bibitem{Inamura:2023qzl}
K.~Inamura and K.~Ohmori, \emph{{Fusion Surface Models: 2+1d Lattice Models from Fusion 2-Categories}}, \href{https://doi.org/10.21468/SciPostPhys.16.6.143}{\emph{SciPost Phys.} {\bfseries 16} (2024) 143} [\href{https://arxiv.org/abs/2305.05774}{{\ttfamily 2305.05774}}].

\bibitem{Hsin:2024aqb}
P.-S.~Hsin, R.~Kobayashi and C.~Zhang, \emph{{Fractionalization of coset non-invertible symmetry and exotic Hall conductance}}, \href{https://doi.org/10.21468/SciPostPhys.17.3.095}{\emph{SciPost Phys.} {\bfseries 17} (2024) 095} [\href{https://arxiv.org/abs/2405.20401}{{\ttfamily 2405.20401}}].

\bibitem{Bhardwaj:2024qiv}
L.~Bhardwaj, D.~Pajer, S.~Schafer-Nameki, A.~Tiwari, A.~Warman and J.~Wu, \emph{{Gapped phases in (2+1)d with non-invertible symmetries: Part I}}, \href{https://doi.org/10.21468/SciPostPhys.19.2.056}{\emph{SciPost Phys.} {\bfseries 19} (2025) 056} [\href{https://arxiv.org/abs/2408.05266}{{\ttfamily 2408.05266}}].

\bibitem{Bullimore:2024khm}
M.~Bullimore and J.J.~Pearson, \emph{{Towards All Categorical Symmetries in 2+1 Dimensions}},  \href{https://arxiv.org/abs/2408.13931}{{\ttfamily 2408.13931}}.

\bibitem{Cordova:2024jlk}
C.~Cordova, D.B.~Costa and P.-S.~Hsin, \emph{{Non-invertible symmetries in finite-group gauge theory}}, \href{https://doi.org/10.21468/SciPostPhys.18.1.019}{\emph{SciPost Phys.} {\bfseries 18} (2025) 019} [\href{https://arxiv.org/abs/2407.07964}{{\ttfamily 2407.07964}}].

\bibitem{Cordova:2024mqg}
C.~Cordova, D.B.~Costa and P.-S.~Hsin, \emph{{Non-Invertible Symmetries as Condensation Defects in Finite-Group Gauge Theories}},  \href{https://arxiv.org/abs/2412.16681}{{\ttfamily 2412.16681}}.

\bibitem{Cao:2025qhg}
W.~Cao, M.~Yamazaki and L.~Li, \emph{{Duality viewpoint of noninvertible symmetry protected topological phases}},  \href{https://arxiv.org/abs/2502.20435}{{\ttfamily 2502.20435}}.

\bibitem{Bhardwaj:2025piv}
L.~Bhardwaj, S.~Schafer-Nameki, A.~Tiwari and A.~Warman, \emph{{Gapped Phases in (2+1)d with Non-Invertible Symmetries: Part II}},  \href{https://arxiv.org/abs/2502.20440}{{\ttfamily 2502.20440}}.

\bibitem{Eck:2025ldx}
L.~Eck, \emph{{Dualities between 2+1d fusion surface models from braided fusion categories}}, \href{https://doi.org/10.21468/SciPostPhys.19.6.157}{\emph{SciPost Phys.} {\bfseries 19} (2025) 157} [\href{https://arxiv.org/abs/2501.14722}{{\ttfamily 2501.14722}}].

\bibitem{Vancraeynest-DeCuiper:2025wkh}
B.~Vancraeynest-De~Cuiper and C.~Delcamp, \emph{{Twisted gauging and topological sectors in (2+1)d Abelian lattice gauge theories}}, \href{https://doi.org/10.21468/SciPostPhys.19.2.054}{\emph{SciPost Phys.} {\bfseries 19} (2025) 054} [\href{https://arxiv.org/abs/2501.16301}{{\ttfamily 2501.16301}}].

\bibitem{Hsin:2025ria}
P.-S.~Hsin, R.~Kobayashi and C.~Zhang, \emph{{Anomalies of Coset Non-Invertible Symmetries}},  \href{https://arxiv.org/abs/2503.00105}{{\ttfamily 2503.00105}}.

\bibitem{Furukawa:2025flp}
Y.~Furukawa, \emph{{Lattice models with subsystem/weak non-invertible symmetry-protected topological order}},  \href{https://arxiv.org/abs/2505.11419}{{\ttfamily 2505.11419}}.

\bibitem{Inamura:2025cum}
K.~Inamura, S.-J.~Huang, A.~Tiwari and S.~Schafer-Nameki, \emph{{(2+1)d Lattice Models and Tensor Networks for Gapped Phases with Categorical Symmetry}},  \href{https://arxiv.org/abs/2506.09177}{{\ttfamily 2506.09177}}.

\bibitem{KNBalasubramanian:2025vum}
M.~K.~N.~Balasubramanian, M.~Buican, C.~Delcamp and R.~Radhakrishnan, \emph{{Gauging Non-Invertible Symmetries in (2+1)d Topological Orders}},  \href{https://arxiv.org/abs/2507.01142}{{\ttfamily 2507.01142}}.

\bibitem{Thorngren:2019iar}
R.~Thorngren and Y.~Wang, \emph{{Fusion category symmetry. Part I. Anomaly in-flow and gapped phases}}, \href{https://doi.org/10.1007/JHEP04(2024)132}{\emph{JHEP} {\bfseries 04} (2024) 132} [\href{https://arxiv.org/abs/1912.02817}{{\ttfamily 1912.02817}}].

\bibitem{Inamura:2021wuo}
K.~Inamura, \emph{{Topological field theories and symmetry protected topological phases with fusion category symmetries}}, \href{https://doi.org/10.1007/JHEP05(2021)204}{\emph{JHEP} {\bfseries 05} (2021) 204} [\href{https://arxiv.org/abs/2103.15588}{{\ttfamily 2103.15588}}].

\bibitem{Inamura:2021szw}
K.~Inamura, \emph{{On lattice models of gapped phases with fusion category symmetries}}, \href{https://doi.org/10.1007/JHEP03(2022)036}{\emph{JHEP} {\bfseries 03} (2022) 036} [\href{https://arxiv.org/abs/2110.12882}{{\ttfamily 2110.12882}}].

\bibitem{Garre-Rubio:2022uum}
J.~Garre-Rubio, L.~Lootens and A.~Moln\'ar, \emph{{Classifying phases protected by matrix product operator symmetries using matrix product states}}, \href{https://doi.org/10.22331/q-2023-02-21-927}{\emph{Quantum} {\bfseries 7} (2023) 927} [\href{https://arxiv.org/abs/2203.12563}{{\ttfamily 2203.12563}}].

\bibitem{Fechisin:2023odt}
C.~Fechisin, N.~Tantivasadakarn and V.V.~Albert, \emph{{Noninvertible Symmetry-Protected Topological Order in a Group-Based Cluster State}}, \href{https://doi.org/10.1103/PhysRevX.15.011058}{\emph{Phys. Rev. X} {\bfseries 15} (2025) 011058} [\href{https://arxiv.org/abs/2312.09272}{{\ttfamily 2312.09272}}].

\bibitem{Seifnashri:2024dsd}
S.~Seifnashri and S.-H.~Shao, \emph{{Cluster State as a Noninvertible Symmetry-Protected Topological Phase}}, \href{https://doi.org/10.1103/PhysRevLett.133.116601}{\emph{Phys. Rev. Lett.} {\bfseries 133} (2024) 116601} [\href{https://arxiv.org/abs/2404.01369}{{\ttfamily 2404.01369}}].

\bibitem{Jia:2024bng}
Z.~Jia, \emph{{Generalized cluster states from Hopf algebras: non-invertible symmetry and Hopf tensor network representation}}, \href{https://doi.org/10.1007/JHEP09(2024)147}{\emph{JHEP} {\bfseries 09} (2024) 147} [\href{https://arxiv.org/abs/2405.09277}{{\ttfamily 2405.09277}}].

\bibitem{Li:2024fhy}
Y.~Li and M.~Litvinov, \emph{{Non-invertible SPT, gauging and symmetry fractionalization}},  \href{https://arxiv.org/abs/2405.15951}{{\ttfamily 2405.15951}}.

\bibitem{Pace:2024acq}
S.D.~Pace, H.T.~Lam and {\"O}.M.~Aksoy, \emph{{(SPT-)LSM theorems from projective non-invertible symmetries}}, \href{https://doi.org/10.21468/SciPostPhys.18.1.028}{\emph{SciPost Phys.} {\bfseries 18} (2025) 028} [\href{https://arxiv.org/abs/2409.18113}{{\ttfamily 2409.18113}}].

\bibitem{Meng:2024nxx}
C.~Meng, X.~Yang, T.~Lan and Z.~Gu, \emph{{Non-invertible SPTs: an on-site realization of (1+1)d anomaly-free fusion category symmetry}},  \href{https://arxiv.org/abs/2412.20546}{{\ttfamily 2412.20546}}.

\bibitem{Aksoy:2025rmg}
{\"O}.M.~Aksoy and X.-G.~Wen, \emph{{Phases with non-invertible symmetries in 1+1D -- symmetry protected topological orders as duality automorphisms}},  \href{https://arxiv.org/abs/2503.21764}{{\ttfamily 2503.21764}}.

\bibitem{Maeda:2025rxc}
J.~Maeda and T.~Oishi, \emph{{N-ality symmetry and SPT phases in (1+1)d}}, \href{https://doi.org/10.1007/JHEP12(2025)063}{\emph{JHEP} {\bfseries 12} (2025) 063} [\href{https://arxiv.org/abs/2504.20151}{{\ttfamily 2504.20151}}].

\bibitem{Lu:2025rwd}
D.-C.~Lu, F.~Xu and Y.-Z.~You, \emph{{Strange correlator and string order parameter for non-invertible symmetry protected topological phases in 1+1d}},  \href{https://arxiv.org/abs/2505.00673}{{\ttfamily 2505.00673}}.

\bibitem{ParayilMana:2025nxw}
A.~Parayil~Mana, Y.~Li, H.~Sukeno and T.-C.~Wei, \emph{{Higher-order topological phases protected by non-invertible and subsystem symmetries}},  \href{https://arxiv.org/abs/2505.18119}{{\ttfamily 2505.18119}}.

\bibitem{You:2025uxo}
M.~You, \emph{{Symmetric entanglers for non-invertible SPT phases}},  \href{https://arxiv.org/abs/2509.04581}{{\ttfamily 2509.04581}}.

\bibitem{Lu:2025yru}
D.-C.~Lu and Z.~Sun, \emph{{Intrinsic NISPT Phases, igNISPT Phases, and Mixed Anomalies of Non-Invertible Symmetries}},  \href{https://arxiv.org/abs/2511.01965}{{\ttfamily 2511.01965}}.

\bibitem{Cirac:2020obd}
J.I.~Cirac, D.~P\'erez-Garc\'ia, N.~Schuch and F.~Verstraete, \emph{{Matrix product states and projected entangled pair states: Concepts, symmetries, theorems}}, \href{https://doi.org/10.1103/RevModPhys.93.045003}{\emph{Rev. Mod. Phys.} {\bfseries 93} (2021) 045003} [\href{https://arxiv.org/abs/2011.12127}{{\ttfamily 2011.12127}}].

\bibitem{Molnar:2022nmh}
A.~Molnar, A.R.~de~Alarc\'on, J.~Garre-Rubio, N.~Schuch, J.I.~Cirac and D.~P\'erez-Garc\'\i{}a, \emph{{Matrix product operator algebras I: representations of weak Hopf algebras and projected entangled pair states}},  \href{https://arxiv.org/abs/2204.05940}{{\ttfamily 2204.05940}}.

\bibitem{Lootens:2021tet}
L.~Lootens, C.~Delcamp, G.~Ortiz and F.~Verstraete, \emph{{Dualities in One-Dimensional Quantum Lattice Models: Symmetric Hamiltonians and Matrix Product Operator Intertwiners}}, \href{https://doi.org/10.1103/PRXQuantum.4.020357}{\emph{PRX Quantum} {\bfseries 4} (2023) 020357} [\href{https://arxiv.org/abs/2112.09091}{{\ttfamily 2112.09091}}].

\bibitem{Lootens:2022avn}
L.~Lootens, C.~Delcamp and F.~Verstraete, \emph{{Dualities in One-Dimensional Quantum Lattice Models: Topological Sectors}}, \href{https://doi.org/10.1103/PRXQuantum.5.010338}{\emph{PRX Quantum} {\bfseries 5} (2024) 010338} [\href{https://arxiv.org/abs/2211.03777}{{\ttfamily 2211.03777}}].

\bibitem{Gorantla:2024ocs}
P.~Gorantla, S.-H.~Shao and N.~Tantivasadakarn, \emph{{Tensor Networks for Noninvertible Symmetries in 3+1D and Beyond}}, \href{https://doi.org/10.1103/p32z-v884}{\emph{Phys. Rev. X} {\bfseries 15} (2025) 041006} [\href{https://arxiv.org/abs/2406.12978}{{\ttfamily 2406.12978}}].

\bibitem{Brell_2015}
C.G.~Brell, \emph{Generalized cluster states based on finite groups}, \href{https://doi.org/10.1088/1367-2630/17/2/023029}{\emph{New Journal of Physics} {\bfseries 17} (2015) 023029} [\href{https://arxiv.org/abs/1408.6237}{{\ttfamily 1408.6237}}].

\bibitem{Cordova:2024iti}
C.~Cordova, N.~Holfester and K.~Ohmori, \emph{{Representation theory of solitons}}, \href{https://doi.org/10.1007/JHEP06(2025)001}{\emph{JHEP} {\bfseries 06} (2025) 001} [\href{https://arxiv.org/abs/2408.11045}{{\ttfamily 2408.11045}}].

\bibitem{Inamura:clusterinterface}
K.~Inamura and S.~Ohyama, ``{Generalized cluster states in 2+1d: non-invertible symmetries, interfaces, and parameterized families}.''
\newblock to appear.

\bibitem{Gaiotto:2019xmp}
D.~Gaiotto and T.~Johnson-Freyd, \emph{{Condensations in higher categories}},  \href{https://arxiv.org/abs/1905.09566}{{\ttfamily 1905.09566}}.

\bibitem{Roumpedakis:2022aik}
K.~Roumpedakis, S.~Seifnashri and S.-H.~Shao, \emph{{Higher Gauging and Non-invertible Condensation Defects}}, \href{https://doi.org/10.1007/s00220-023-04706-9}{\emph{Commun. Math. Phys.} {\bfseries 401} (2023) 3043} [\href{https://arxiv.org/abs/2204.02407}{{\ttfamily 2204.02407}}].

\bibitem{Thorngren:1612.00846}
R.~Thorngren and D.V.~Else, \emph{Gauging spatial symmetries and the classification of topological crystalline phases}, \href{https://doi.org/10.1103/physrevx.8.011040}{\emph{Physical Review X} {\bfseries 8} (2018) } [\href{https://arxiv.org/abs/1612.00846}{{\ttfamily 1612.00846}}].

\bibitem{Hsin:2020cgg}
P.-S.~Hsin, A.~Kapustin and R.~Thorngren, \emph{{Berry Phase in Quantum Field Theory: Diabolical Points and Boundary Phenomena}}, \href{https://doi.org/10.1103/PhysRevB.102.245113}{\emph{Phys. Rev. B} {\bfseries 102} (2020) 245113} [\href{https://arxiv.org/abs/2004.10758}{{\ttfamily 2004.10758}}].

\bibitem{Thorngren2021YITP}
R.~Thorngren, \emph{Berry phase, diabolical points, and pivot hamiltonians},  2021.
\newblock \href{https://www.ms.u-tokyo.ac.jp/~yasuyuki/yitp2021.htm}{ talk at Yukawa Institute for Theoretical Physics}.

\bibitem{Qi:2023ysw}
M.~Qi, D.T.~Stephen, X.~Wen, D.~Spiegel, M.J.~Pflaum, A.~Beaudry et~al., \emph{{Charting the space of ground states with tensor networks}}, \href{https://doi.org/10.21468/SciPostPhys.18.5.168}{\emph{SciPost Phys.} {\bfseries 18} (2025) 168} [\href{https://arxiv.org/abs/2305.07700}{{\ttfamily 2305.07700}}].

\bibitem{Ohyama:2024ytt}
S.~Ohyama and S.~Ryu, \emph{{Higher Berry phase from projected entangled pair states in (2+1) dimensions}}, \href{https://doi.org/10.1103/PhysRevB.111.045112}{\emph{Phys. Rev. B} {\bfseries 111} (2025) 045112} [\href{https://arxiv.org/abs/2405.05325}{{\ttfamily 2405.05325}}].

\bibitem{PhysRevLett.86.5188}
R.~Raussendorf and H.J.~Briegel, \emph{{A One-Way Quantum Computer}}, \href{https://doi.org/10.1103/PhysRevLett.86.5188}{\emph{Phys. Rev. Lett.} {\bfseries 86} (2001) 5188}.

\bibitem{Raussendorf:0301052}
R.~Raussendorf, D.E.~Browne and H.J.~Briegel, \emph{{Measurement-based quantum computation on cluster states}}, \href{https://doi.org/10.1103/PhysRevA.68.022312}{\emph{Phys. Rev. A} {\bfseries 68} (2003) 022312} [\href{https://arxiv.org/abs/quant-ph/0301052}{{\ttfamily quant-ph/0301052}}].

\bibitem{Kitaev:2011dxc}
A.~Kitaev and L.~Kong, \emph{{Models for Gapped Boundaries and Domain Walls}}, \href{https://doi.org/10.1007/s00220-012-1500-5}{\emph{Commun. Math. Phys.} {\bfseries 313} (2012) 351} [\href{https://arxiv.org/abs/1104.5047}{{\ttfamily 1104.5047}}].

\bibitem{Lan:2013wia}
T.~Lan and X.-G.~Wen, \emph{{Topological quasiparticles and the holographic bulk-edge relation in (2+1) -dimensional string-net models}}, \href{https://doi.org/10.1103/PhysRevB.90.115119}{\emph{Phys. Rev. B} {\bfseries 90} (2014) 115119} [\href{https://arxiv.org/abs/1311.1784}{{\ttfamily 1311.1784}}].

\bibitem{Bridgeman:2018jdv}
J.C.~Bridgeman, D.~Barter and C.~Jones, \emph{{Fusing Binary Interface Defects in Topological Phases: The $\operatorname{Vec}(\mathbb{Z}/p\mathbb{Z})$ case}}, \href{https://doi.org/10.1063/1.5095941}{\emph{J. Math. Phys.} {\bfseries 60} (2019) 121701} [\href{https://arxiv.org/abs/1810.09469}{{\ttfamily 1810.09469}}].

\bibitem{Bridgeman:2019axg}
J.C.~Bridgeman and D.~Barter, \emph{{Computing defects associated to bounded domain wall structures: the case}}, \href{https://doi.org/10.1088/1751-8121/ab7d60}{\emph{J. Phys. A} {\bfseries 53} (2020) 235206} [\href{https://arxiv.org/abs/1901.08069}{{\ttfamily 1901.08069}}].

\bibitem{Bridgeman:2019wyu}
J.C.~Bridgeman and D.~Barter, \emph{{Computing data for Levin-Wen with defects}}, \href{https://doi.org/10.22331/q-2020-06-04-277}{\emph{Quantum} {\bfseries 4} (2020) 277} [\href{https://arxiv.org/abs/1907.06692}{{\ttfamily 1907.06692}}].

\bibitem{Barter_2022}
D.~Barter, J.~Bridgeman and R.~Wolf, \emph{{Computing associators of endomorphism fusion categories}}, \href{https://doi.org/10.21468/scipostphys.13.2.029}{\emph{SciPost Physics} {\bfseries 13} (2022) } [\href{https://arxiv.org/abs/2110.03644}{{\ttfamily 2110.03644}}].

\bibitem{Bridgeman:2022gdx}
J.C.~Bridgeman, L.~Lootens and F.~Verstraete, \emph{{Invertible Bimodule Categories and Generalized Schur Orthogonality}}, \href{https://doi.org/10.1007/s00220-023-04781-y}{\emph{Commun. Math. Phys.} {\bfseries 402} (2023) 2691} [\href{https://arxiv.org/abs/2211.01947}{{\ttfamily 2211.01947}}].

\bibitem{Jia:2024rzr}
Z.~Jia, S.~Tan and D.~Kaszlikowski, \emph{{Weak Hopf symmetry and tube algebra of the generalized multifusion string-net model}}, \href{https://doi.org/10.1007/JHEP07(2024)207}{\emph{JHEP} {\bfseries 07} (2024) 207} [\href{https://arxiv.org/abs/2403.04446}{{\ttfamily 2403.04446}}].

\bibitem{Choi:2024tri}
Y.~Choi, B.C.~Rayhaun and Y.~Zheng, \emph{{Generalized Tube Algebras, Symmetry-Resolved Partition Functions, and Twisted Boundary States}},  \href{https://arxiv.org/abs/2409.02159}{{\ttfamily 2409.02159}}.

\bibitem{Choi:2024wfm}
Y.~Choi, B.C.~Rayhaun and Y.~Zheng, \emph{{Noninvertible Symmetry-Resolved Affleck-Ludwig-Cardy Formula and Entanglement Entropy from the Boundary Tube Algebra}}, \href{https://doi.org/10.1103/PhysRevLett.133.251602}{\emph{Phys. Rev. Lett.} {\bfseries 133} (2024) 251602} [\href{https://arxiv.org/abs/2409.02806}{{\ttfamily 2409.02806}}].

\bibitem{Jia:2024zdp}
Z.~Jia, \emph{{Weak Hopf non-invertible symmetry-protected topological spin liquid and lattice realization of (1+1)D symmetry topological field theory}},  \href{https://arxiv.org/abs/2412.15336}{{\ttfamily 2412.15336}}.

\bibitem{Jia:2025yph}
Z.~Jia and S.~Tan, \emph{{Weak Hopf tube algebra for domain walls between 2d gapped phases of Turaev-Viro TQFTs}}, \href{https://doi.org/10.1007/JHEP11(2025)018}{\emph{JHEP} {\bfseries 11} (2025) 018} [\href{https://arxiv.org/abs/2507.01515}{{\ttfamily 2507.01515}}].

\bibitem{Cordova:2024nux}
C.~Cordova, D.~Garc{\'\i}a-Sep{\'u}lveda and N.~Holfester, \emph{{Particle-Soliton Degeneracy in 2D Quantum Chromodynamics}},  \href{https://arxiv.org/abs/2412.21153}{{\ttfamily 2412.21153}}.

\bibitem{Gagliano:2025gwr}
F.~Gagliano, A.~Grigoletto and K.~Ohmori, \emph{{Higher Representations and Quark Confinement}},  \href{https://arxiv.org/abs/2501.09069}{{\ttfamily 2501.09069}}.

\bibitem{Choi:2023xjw}
Y.~Choi, B.C.~Rayhaun, Y.~Sanghavi and S.-H.~Shao, \emph{{Remarks on boundaries, anomalies, and noninvertible symmetries}}, \href{https://doi.org/10.1103/PhysRevD.108.125005}{\emph{Phys. Rev. D} {\bfseries 108} (2023) 125005} [\href{https://arxiv.org/abs/2305.09713}{{\ttfamily 2305.09713}}].

\bibitem{Cordova:2024vsq}
C.~C\'{o}rdova, D.~Garc\'\i{}a-Sep\'ulveda and N.~Holfester, \emph{{Particle-soliton degeneracies from spontaneously broken non-invertible symmetry}}, \href{https://doi.org/10.1007/JHEP07(2024)154}{\emph{JHEP} {\bfseries 07} (2024) 154} [\href{https://arxiv.org/abs/2403.08883}{{\ttfamily 2403.08883}}].

\bibitem{Copetti:2024onh}
C.~Copetti, \emph{{Defect Charges, Gapped Boundary Conditions, and the Symmetry TFT}},  \href{https://arxiv.org/abs/2408.01490}{{\ttfamily 2408.01490}}.

\bibitem{Copetti:2024dcz}
C.~Copetti, L.~Cordova and S.~Komatsu, \emph{{S-matrix bootstrap and non-invertible symmetries}}, \href{https://doi.org/10.1007/JHEP03(2025)204}{\emph{JHEP} {\bfseries 03} (2025) 204} [\href{https://arxiv.org/abs/2408.13132}{{\ttfamily 2408.13132}}].

\bibitem{Bhardwaj:2024igy}
L.~Bhardwaj, C.~Copetti, D.~Pajer and S.~Schafer-Nameki, \emph{{Boundary SymTFT}}, \href{https://doi.org/10.21468/SciPostPhys.19.2.061}{\emph{SciPost Phys.} {\bfseries 19} (2025) 061} [\href{https://arxiv.org/abs/2409.02166}{{\ttfamily 2409.02166}}].

\bibitem{Heymann:2024vvf}
J.~Heymann and T.~Quella, \emph{{Revisiting the symmetry-resolved entanglement for noninvertible symmetries in 1+1d conformal field theories}}, \href{https://doi.org/10.1103/lr47-yv3j}{\emph{Phys. Rev. D} {\bfseries 112} (2025) 025004} [\href{https://arxiv.org/abs/2409.02315}{{\ttfamily 2409.02315}}].

\bibitem{AliAhmad:2025bnd}
S.~Ali~Ahmad, M.S.~Klinger and Y.~Wang, \emph{{The Many Faces of Non-invertible Symmetries}},  \href{https://arxiv.org/abs/2509.18072}{{\ttfamily 2509.18072}}.

\bibitem{Benini:2025lav}
F.~Benini, P.~Calabrese, M.~Fossati, A.H.~Singh and M.~Venuti, \emph{{Entanglement Asymmetry for Higher and Noninvertible Symmetries}},  \href{https://arxiv.org/abs/2509.16311}{{\ttfamily 2509.16311}}.

\bibitem{Seiberg:2019vrp}
N.~Seiberg, \emph{{Field Theories With a Vector Global Symmetry}}, \href{https://doi.org/10.21468/SciPostPhys.8.4.050}{\emph{SciPost Phys.} {\bfseries 8} (2020) 050} [\href{https://arxiv.org/abs/1909.10544}{{\ttfamily 1909.10544}}].

\bibitem{Qi:2020jrf}
M.~Qi, L.~Radzihovsky and M.~Hermele, \emph{{Fracton phases via exotic higher-form symmetry-breaking}}, \href{https://doi.org/10.1016/j.aop.2020.168360}{\emph{Annals Phys.} {\bfseries 424} (2021) 168360} [\href{https://arxiv.org/abs/2010.02254}{{\ttfamily 2010.02254}}].

\bibitem{Decoppet2023Morita}
T.D.~D\'{e}coppet, \emph{{The Morita Theory of Fusion 2-Categories}}, \href{https://doi.org/10.21136/hs.2023.07}{\emph{Higher Structures} {\bfseries 7} (2023) 234} [\href{https://arxiv.org/abs/2208.08722}{{\ttfamily 2208.08722}}].

\bibitem{Ostrik2003}
V.~Ostrik, \emph{{Module categories, weak Hopf algebras and modular invariants}}, \href{https://doi.org/10.1007/s00031-003-0515-6}{\emph{Transformation Groups} {\bfseries 8} (2003) 177} [\href{https://arxiv.org/abs/math/0111139}{{\ttfamily math/0111139}}].

\bibitem{Kapustin:2016jqm}
A.~Kapustin, A.~Turzillo and M.~You, \emph{{Topological Field Theory and Matrix Product States}}, \href{https://doi.org/10.1103/PhysRevB.96.075125}{\emph{Phys. Rev. B} {\bfseries 96} (2017) 075125} [\href{https://arxiv.org/abs/1607.06766}{{\ttfamily 1607.06766}}].

\bibitem{Mackey1951}
G.W.~Mackey, \emph{On induced representations of groups}, \href{https://doi.org/10.2307/2372309}{\emph{American Journal of Mathematics} {\bfseries 73} (1951) 576}.

\bibitem{Greenough_2010}
J.~Greenough, \emph{Monoidal 2-structure of bimodule categories}, \href{https://doi.org/https://doi.org/10.1016/j.jalgebra.2010.06.018}{\emph{Journal of Algebra} {\bfseries 324} (2010) 1818}.

\bibitem{Perez-Garcia:2006nqo}
D.~P\'erez-Garc\'ia, F.~Verstraete, M.M.~Wolf and J.I.~Cirac, \emph{{Matrix product state representations}}, \href{https://doi.org/10.26421/QIC7.5-6-1}{\emph{Quant. Inf. Comput.} {\bfseries 7} (2007) 401} [\href{https://arxiv.org/abs/quant-ph/0608197}{{\ttfamily quant-ph/0608197}}].

\bibitem{gordon1995coherence}
R.~Gordon, A.J.~Power and R.~Street, \emph{{Coherence for tricategories}}, vol.~558, American Mathematical Soc. (1995).

\bibitem{schommerpries2014classification}
C.J.~Schommer-Pries, \emph{The classification of two-dimensional extended topological field theories},  \href{https://arxiv.org/abs/1112.1000}{{\ttfamily 1112.1000}}.

\end{thebibliography}\endgroup

\end{document}